\documentclass[11pt]{article}

\usepackage{jheppub}

\RequirePackage{lmodern}      
\RequirePackage{amssymb}
\RequirePackage{amsfonts}
\RequirePackage{amsmath}

\usepackage{hyperref}
\usepackage[usenames,dvipsnames]{xcolor} 
\numberwithin{equation}{section}
\usepackage{shuffle}
\usepackage{bm}
\usepackage{latexsym}
\usepackage{esint}
\usepackage{graphicx}
\usepackage{soul}
\usepackage{cancel}

\usepackage{tikz}
\usepackage{tikz-cd}
\usetikzlibrary{decorations.markings,arrows.meta }
\usetikzlibrary{arrows,shapes.geometric,positioning}
\usepackage{adjustbox}

\usepackage{subcaption}
\usepackage{extarrows}
\usepackage{bbm}
\usepackage{mathtools,slashed}
\usepackage{hyperref}
\usepackage{tablefootnote}
\usepackage{diagbox}
\usepackage{multirow}
\usepackage{graphbox}
\usepackage{fnpct}
\usepackage{dsfont}
\usepackage{stmaryrd} 
\usepackage{relsize}
\usepackage{orcidlink}

\usepackage[most]{tcolorbox}
\tcbset{
    enhanced jigsaw, 
    colback=gray!5!white,
    colframe=gray!20!white,
    coltitle=black,
    fonttitle=\bfseries
}
\newtcolorbox[auto counter,number within=section]{mybox }[2][]{%
fonttitle=\bfseries,
breakable,
title=Box \thetcbcounter: #2,#1
}

\newcommand{\myblue}{RoyalBlue}
\newcommand{\myred}{BrickRed}

\hypersetup{
    colorlinks=true,
    linkcolor=\myblue,
    citecolor=\myred   
}

\definecolor{darkCyan}{RGB}{0, 139, 139}
\definecolor{darkMagenta}{RGB}{139, 0, 139}

\renewcommand{\d}{\mathrm{d}}
\newcommand{\res}{\text{Res}}
\newcommand{\disc}{\text{Disc}}

\newcommand{\vep}{\varepsilon}
\newcommand{\vphi}{\varphi}

\newcommand{\ra}{\rangle}
\newcommand{\la}{\langle}
\newcommand{\dlog}{\mathrm{dlog}}

\newcommand{\G}{\mathcal{G}}
\newcommand{\g}{{\mathfrak{g}}}
\newcommand{\h}{{\mathfrak{h}}}
\newcommand{\f}{{\mathfrak{f}}}
\newcommand{\T}{\mathcal{T}}

\newcommand{\E}{\mathcal{E}}
\newcommand{\V}{\mathcal{V}}
\newcommand{\B}{\mathcal{B}}
\newcommand{\Csf}{{\mathsf{C}}}
\newcommand{\csf}{\mathsf{c}}

\newcommand{\Rsf}{\mathsf{R}}
\newcommand{\rsf}{\mathsf{r}}
\newcommand{\nn}{\nonumber}

\newcommand{\id}{{\mathds{1}}}

\newcommand{\R}{\mathcal{R}}

\newcommand{\mbf}[1]{\mathbf{#1}}
\newcommand{\bs}[1]{\boldsymbol{#1}}
\newcommand{\mat}[1]{\underline{\bs{#1}}}

\newcommand{\be}{\begin{equation}\begin{aligned}}
\newcommand{\ee}{\end{aligned}\end{equation}}

\DeclareMathOperator{\sgn}{sgn}

\newcommand{\solidEdge}[1]{
    \adjustbox{valign=c}{
        \begin{tikzpicture}[scale=#1]
            \coordinate (A) at (0,0);
            \coordinate (B) at (1/2,0);
            \coordinate (C) at (1,0);
            \coordinate (D) at (3/2,0);
            \draw[thick,double] (A) -- (B);
            \fill[black] (A) circle (2pt);
            \fill[black] (B) circle (2pt);
        \end{tikzpicture}
    }
}
\newcommand{\directedEdge}[1]{
    \adjustbox{valign=c}{
        \begin{tikzpicture}[scale=#1]
            \coordinate (A) at (0,0);
            \coordinate (B) at (1/2,0);
            \draw[thick,-stealth] (A) -- (1/3,0);
            \draw[thick] (1/3,0) -- (B);
            \fill[black] (A) circle (2pt);
            \fill[black] (B) circle (2pt);
        \end{tikzpicture}
    }
}
\newcommand{\ddirectedEdge}[1]{
    \adjustbox{valign=c}{
        \begin{tikzpicture}[scale=#1]
            \coordinate (A) at (0,0);
            \coordinate (B) at (1/2,0);
            \draw[thick,-stealth] (B) -- (1/6,0);
            \draw[thick] (1/6,0) -- (A);
            \fill[black] (A) circle (2pt);
            \fill[black] (B) circle (2pt);
        \end{tikzpicture}
    }
}
\newcommand{\brokenEdge}[1]{
    \adjustbox{valign=c}{
        \begin{tikzpicture}[scale=#1]
            \coordinate (A) at (0,0);
            \coordinate (B) at (1/2,0);
            \coordinate (C) at (1,0);
            \coordinate (D) at (3/2,0);
            \draw[thick,dotted] (A) -- (B);
            \fill[black] (A) circle (2pt);
            \fill[black] (B) circle (2pt);
        \end{tikzpicture}
    }
}
\newcommand{\brEdge}[1]{
    \adjustbox{valign=c}{
        \begin{tikzpicture}[scale=#1]
            \coordinate (A) at (0,0);
            \coordinate (B) at (1/2,0);
            \draw[thick,dotted] (A) -- (B);
            \fill[Orange] (A) circle (2pt);
        \end{tikzpicture}
    }
}
\newcommand{\outEdge}[1]{
    \adjustbox{valign=c}{
        \begin{tikzpicture}[scale=#1]
            \coordinate (A) at (0,0);
            \coordinate (B) at (1/2,0);
            \coordinate (AB) at (3/8,0);
            \draw[thick,-stealth] (A) -- (AB);
            \draw[thick] (A) -- (B);
            \fill[Orange] (A) circle (2pt);
        \end{tikzpicture}
    }
}
\newcommand{\inEdge}[1]{
    \adjustbox{valign=c}{
        \begin{tikzpicture}[scale=#1]
            \coordinate (A) at (0,0);
            \coordinate (B) at (1/2,0);
            \coordinate (AB) at (1/8,0);
            \draw[thick,-stealth] (B) -- (AB);
            \draw[thick] (A) -- (B);
            \fill[Orange] (A) circle (2pt);
        \end{tikzpicture}
    }
}

\newcommand{\pinch}{\mathrm{pinch}}
\newcommand{\antipinch}{\mathrm{antipinch}}

\newcommand{\twoChaing}{\includegraphics[scale=.5]{figs/2chain_gs}}
\newcommand{\twoChaingg}{\includegraphics[scale=.5]{figs/2chain_gr}}
\newcommand{\twoChainggg}{\includegraphics[scale=.5]{figs/2chain_gl}}
\newcommand{\twoChaingggg}{\includegraphics[scale=.5]{figs/2chain_gb}}

\newcommand{\ar}{\tikz \filldraw[scale=0.1, rotate=-90] (0,0) -- (1,0) -- (0.5,0.866) -- cycle;}

\newcommand{\coactionurl}{https://frwcoaction.ca/}

\newcommand{\coactionsymbolAlt}{%
     \includegraphics[align=c, scale=1.5]{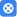}
}

\newcommand{\webappAlt}{\href{\coactionurl}{\coactionsymbolAlt}}

\usepackage{fontawesome5}

\makeatletter
\newcommand{\github}{%
   {\href{https://github.com/ASphericalCow?tab=repositories}{\faGithub}}%
}
\makeatother

\preprint{QMUL-PH-26-22}

\title{ A Graphical Coaction for FRW Integrals 
from Partial/Relative Twisted (Co)homology 
}

\author[1]{Andrew J.~McLeod\orcidlink{0000-0001-7685-2929},}\emailAdd{andrew.mcleod@ed.ac.uk}
\author[2]{Andrzej Pokraka\orcidlink{0000-0003-1186-4624},}\emailAdd{a.m.pokraka@uva.nl}
\author[3]{and Lecheng Ren\orcidlink{0000-0002-0846-8017}}\emailAdd{lecheng.ren@qmul.ac.uk}

\affiliation[1]{%
    Higgs Centre for Theoretical Physics, 
    School of Physics and Astronomy, 
    The University of Edinburgh 
    \\ Edinburgh EH9 3FD, Scotland, UK
}
\affiliation[2]{%
        Institute of Physics, University of Amsterdam, 
        Amsterdam, 1098 XH, The Netherlands
}
\affiliation[3]{%
    Centre for Theoretical Physics, 
    Department of Physics and Astronomy, 
    \\ Queen Mary University of London, E1 4NS, UK
}

\abstract{%
We construct a graphical coaction for Friedmann-Robertson-Walker (FRW) integrals at all loop orders in conformally-coupled scalar theories with non-conformal polynomial interactions. Our construction makes use of intersection theory in the context of (partial/relative) twisted (co)homology, which we use to decompose FRW integrals (and their discontinuities and derivatives) into building blocks that can be represented as decorations of the original Feynman diagram. This facilitates a purely graphical description of the coaction, up to rational prefactors that can be read off from the graph. Our construction provides a comprehensive combinatorial framework for dissecting the analytic properties of cosmological observables; in particular, we demonstrate that the combinatorics of the differential equations that govern FRW integrals---their so-called kinematic flow---is a natural consequence of our coaction. 
\\[1em]
We have also developed a user-friendly web application that computes the graphical coaction of any graph:  \webappAlt\!. 
Whenever possible, the web application also computes the differentials and discontinuities. 
A \texttt{Mathematica} notebook with the same functionality is also hosted at the following repository: \github.
}

\begin{document}
\maketitle
\allowdisplaybreaks

\newpage 
\section{Introduction}
\label{sec:intro}

In recent years, cosmology has emerged as a fertile ground for the application of ideas from algebraic geometry, combinatorics, and number theory. These mathematical perspectives have yielded new insight into the structure of cosmological observables, and---increasingly---the ways in which this structure is dictated by basic physical principles. 
The result has been a rich interplay of ideas that continue to reveal deep mathematical principles at the heart of quantum field theory.

Many of these developments in cosmology have drawn inspiration from 
advances in our understanding of scattering amplitudes, whose mathematical structure has been under intense investigation for over half a century~\cite{ELOP}, from the classes of special functions amplitudes evaluate to~\cite{Broadhurst:1998rz,Bogner:2007mn,Bourjaily:2022bwx}, to the ways in which their analytic structure is constrained by physical principles~\cite{Landau:1959fi,Steinmann,Steinmann2,pham,Hannesdottir:2024cnn,Caron-Huot:2019bsq} and the discovery of unexpected number-theoretic symmetries~\cite{Arkani-Hamed:2012zlh,Schlotterer:2012ny,Brown:2015fyf,Panzer:2016snt,Schnetz:2017bko,Caron-Huot:2019bsq,Gurdogan:2020ppd,Dixon:2021tdw,Dixon:2022xqh}. One important facet of this research has focused on amplitudes as \emph{motivic periods}~\cite{Gonch3,FBThesis,2015arXiv151206410B}, namely as (sums of) integrals whose analytic and algebraic properties can be investigated with the use of geometric tools from homology and cohomology. 
In particular, 
the \emph{motivic coaction} is used to decompose an integral into simpler building blocks that faithfully capture the original integral's analytic properties. 
By iteratively applying the coaction, one can often decompose Feynman integrals into tensor products of one-fold integrals that can be evaluated in closed form (into functions such as logarithms). The upshot is that identities between the (highly nontrivial) special functions that arise in scattering amplitudes can be understood in terms of relations between these simpler functions. Using this technology, the analytic structure of amplitudes involving even billions of terms can be made comprehensible (see for instance~\cite{Caron-Huot:2018dsv,Dixon:2022rse,Dixon:2023kop,Basso:2024hlx,He:2025tyv}).

While the motivic coaction first appeared in the scattering amplitudes literature in connection with multiple polylogarithms (MPLs)~\cite{Chen,G91b,Goncharov:1998kja,Remiddi:1999ew,Borwein:1999js,Moch:2001zr,DelDuca:2009au,DelDuca:2010zg,Gonch2,Goncharov:2010jf,Brown:2011ik,Brown1102.1312,Frost:2023stm,Frost:2025lre}, coactions have been constructed/conjectured for more general classes of functions. These include elliptic polylogarithms~\cite{Broedel:2018iwv, tapuskovic2023cosmic, brown2011multiple}, hypergeometric functions~\cite{Brown:2019jng,Abreu:2019wzk}, and iterated integrals over holomorphic Eisenstein series~\cite{Kleinschmidt:2025dtk}. Intriguingly, a \emph{diagrammatic} coaction has also been defined for one-loop Feynman integrals~\cite{Abreu:2017enx,Abreu:2017mtm}, which decomposes the integral associated with a given Feynman diagram into a basis of integrals associated with contracted and cut versions of the same graph---thereby translating the nontrivial algebraic and analytic properties of these integrals into simple graph-theoretic relations. Effort has gone into extending this diagrammatic coaction to two-loop Feynman integrals~\cite{Abreu:2021vhb}, although a general formulation that works beyond one loop is not yet known.

In this paper, we define a similar graphical coaction on the integrals in cosmology that contribute to the Friedmann-Robertson-Walker (FRW) wavefunction of the universe, for any number of external edges and to all loop orders in conformally-coupled scalar theories with non-conformal polynomial interactions. In particular, we demonstrate that the mathematical objects that appear in the coaction of these integrals are in one-to-one correspondence with decorated graphs $\g$ involving pinched, broken, and directed edges ($\adjustbox{valign=c}{\begin{tikzpicture}[scale=1]
    \coordinate (A) at (0,0);
    \coordinate (B) at (1/2,0);
    \coordinate (C) at (1,0);
    \coordinate (D) at (3/2,0);
    \draw[thick,double] (A) -- (B);
    \fill[black] (A) circle (2pt);
    \fill[black] (B) circle (2pt);
\end{tikzpicture}}$, 
$\adjustbox{valign=c}{\begin{tikzpicture}[scale=1]
    \coordinate (A) at (0,0);
    \coordinate (B) at (1/2,0);
    \coordinate (C) at (1,0);
    \coordinate (D) at (3/2,0);
    \draw[thick,dotted] (A) -- (B);
    \fill[black] (A) circle (2pt);
    \fill[black] (B) circle (2pt);
\end{tikzpicture}}$,
and 
$\adjustbox{valign=c}{\begin{tikzpicture}[scale=1]
    \coordinate (A) at (0,0);
    \coordinate (B) at (1/2,0);
    \coordinate (C) at (1,0);
    \coordinate (D) at (3/2,0);
    \draw[thick] (A) -- node{\ar} (B);
    \fill[black] (A) circle (2pt);
    \fill[black] (B) circle (2pt);
\end{tikzpicture}}$), such that the directed edges do not form an oriented cycle. While the pinched and broken edges play a similar role to contracted and cut edges in the graphical coaction on Feynman integrals, the additional appearance of directed edges reflects the fundamental role that time ordering plays in cosmological perturbation theory.
This time ordering restricts which singularities actually occur in the FRW wavefunction (out of the ones that generically arise, for general integration contours).

To construct the graphical coaction, we adopt a geometric perspective in which FRW wavefunction coefficients---the building blocks of the FRW wavefunction---are expressed as sums of \emph{twisted periods}.
Twisted periods are common in quantum field theory computations, and arise when factors in the integrand appear raised to general complex powers. 
In this context, the FRW integrals that we encounter are of the form
\be \label{eq:generalized_hypergeometrics}
    F(\bs{\beta};\mat{l}\bs{c})
    :=\int_{\Delta_m} \frac{
        z_1^{\beta_1} \cdots z_m^{\beta_m} 
        (1-z_1-z_2- \cdots - z_m)^{\beta_{m+1}}
    }{
        (\bs{l}_1\cdot \bs{z}-c_1)
        (\bs{l}_2\cdot \bs{z}-c_2)
        \cdots
        (\bs{l}_m\cdot \bs{z} - c_m)
    }
    \d^m\bs{z}
\ee
where $\Delta_m$ is the standard $m$-simplex, each $\bs{l}_i$ is an $m$-component vector with entries drawn from $\{0,\pm1\}$, and the powers $\beta_i$ are complex numbers.
The argument $\mat{l}\bs{c}:=\bs{l}_1 \cdots \bs{l}_n \bs{c}$ is an $(n{+}1) \times m$ matrix whose columns are constituted by the vectors $\bs{l}$ or $\bs{c}$.

Conveniently, there exists a natural (albeit conjectural) coaction  on twisted periods that reduces the study of their analytic structure to knowledge of the corresponding twisted homology and cohomology groups, their Poincar\'e duals, and the associated set of positive geometries. Similar to the coactions that have been developed for Feynman integrals, this coaction for FRW integrals breaks each wavefunction coefficient into a sum over simpler building blocks. 
In particular, when applied to the FRW integral associated with a graph $\g$, the coaction can be phrased graphically as
\be \label{eq:schemCosCoaction}
    \Delta (\g)
    =\sum_{\text{compatible } \h}
    C_{\h\h}^{-1}\; 
    \h \otimes 
    \disc_{\h}[\g] 
    \,,
\ee
where the $C_{\h\h}$ are self-intersection numbers that evaluate to rational functions of the parameters of one's theory (which can be pulled out of the tensor product), and the decorated graphs $\h$ represent different (often simpler) FRW integrals.
Importantly, the FRW integrals in the first component of this tensor product (multiplied by rational prefactors) constitute the derivatives of the original integral, while the integrals in the second component constitute the discontinuities of the original integral (modulo factors of $i \pi$).
When iterated, the coaction produces tensor products of arbitrary length (with each component other than the first interpreted modulo $i\pi$). 

Before introducing the mathematical details that underlie this construction, let us illustrate it for the two-site chain graph. 
First, we decompose the wavefunction coefficient associated with this graph into our basis of decorated graphs (each of which represents an FRW integral). In this case, we get
\be \label{eq:two_site_decomposition}
    \includegraphics[scale=.7,align=c]{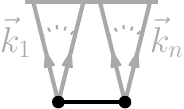}  
    &= 
        \includegraphics[scale=.7,align=c]{figs/2chain_gr}  
        + \includegraphics[scale=.7,align=c]{figs/2chain_gl} 
        -\includegraphics[scale=.7,align=c]{figs/2chain_gb}
    \,.
\ee
Applying the coaction $\Delta$ to these decorated graphs yields
\be \label{eq:Delta2ChainIntro}
    \Delta\left(
        \includegraphics[scale=.7,align=c]{figs/2chain_ext}
    \right)
    &= 
    \Delta\left(
        \includegraphics[scale=.7,align=c]{figs/2chain_gr}  
    \right)
    + \Delta\left(
        \includegraphics[scale=.7,align=c]{figs/2chain_gl} 
    \right)
    - \Delta\left(
        \includegraphics[scale=.7,align=c]{figs/2chain_gb}
    \right)
    \\
    &=
    \frac{\alpha_1 \alpha_2}{\alpha_1 + \alpha_2} \cdot
    \includegraphics[scale=.7,align=c]{figs/2chain_gs}  
    \otimes 
    \includegraphics[scale=.7,align=c]{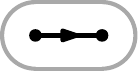}
    + 
    1 \cdot
    \includegraphics[scale=.7,align=c]{figs/2chain_gr}  
    \otimes 
    \includegraphics[scale=.7,align=c]{figs/2chainCr}
    \\&+
    \frac{\alpha_1 \alpha_2}{\alpha_1 + \alpha_2} \cdot
    \includegraphics[scale=.7,align=c]{figs/2chain_gs}  
    \otimes 
    \includegraphics[scale=.7,align=c]{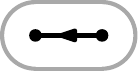}
    + 
    1 \cdot 
    \includegraphics[scale=.7,align=c]{figs/2chain_gl}  
    \otimes 
    \includegraphics[scale=.7,align=c]{figs/2chainCl}
    \\&-
    1 \cdot
    \includegraphics[scale=.7,align=c]{figs/2chain_gb}  
    \otimes 
    \includegraphics[scale=.7,align=c]{figs/2chainCb}
    \,,
\ee
where we have separated the parts of the expression that are generated by different FRW integrals onto different lines. 
First, let us discuss the first component of these tensor products. 
The graphs $\h$ that are compatible with the original graph $\g$ in~\eqref{eq:schemCosCoaction} are those graphs that can be generated by replacing a subset of the oriented edges ($\includegraphics[scale=.6,align=c]{figs/2chain_gr}$) with pinched edges ($\includegraphics[scale=.6,align=c]{figs/2chain_gs}$), where we discard any graphs that have an oriented cycle after all pinched edges are contracted. 
In the decomposition in~\eqref{eq:two_site_decomposition}, the first two decorated graphs are compatible with a pair of graphs (one either replaces the single oriented edge or not), while the third decorated graph is only compatible with itself (since it contains no directed edges that can be replaced). 
To each of these compatible graphs $\h$, we associate a unique cut graph that appears in the second component of the tensor product. These cuts are denoted by the {\color{gray} gray} tubes, which identify the maximal set of denominator factors that are compatible with the decorated graph in the first entry, and indicate that these denominators have all been placed on shell. We will provide a complete description of how to construct these sets of tubings below; for now,  we just note that
\begin{itemize}
    \item no tubes cross pinched lines,
    \item no tubes encircle broken lines,
    \item tubes only cross directed edges where the directed edge points away from the tube.
\end{itemize}
Finally, the rational prefactors involving $\alpha_1$ and $\alpha_2$ arise from the self-intersection numbers $C_{\h\h}$, and only depend on the details of the underlying QFT and FRW spacetime. 
We stress that no knowledge of intersection theory is needed to use the coaction, since we provide explicit formulas for all intersection numbers $C_{\h\h}$. 

We again emphasize that each of the graphs in~\eqref{eq:Delta2ChainIntro} represents an FRW integral, which evaluates to a generalized hypergeometric function of the type shown in~\eqref{eq:generalized_hypergeometrics}. However, the FRW coaction~\eqref{eq:schemCosCoaction} is also compatible with the more familiar coaction on MPLs, in the sense that it commutes with expanding FRW integrals in limits that evaluate (order by order) to MPLs. That is, one gets the same result by applying the coaction~\eqref{eq:schemCosCoaction} and then expanding all tensor factors in terms of MPLs, or by first expanding the original FRW integral in terms of MPLs and then computing the MPL coaction of each term.

More formally, to construct the coaction in~\eqref{eq:schemCosCoaction}, we associate to each decorated graph $\g$ four (co)homological objects: a pair of cycles $\gamma_\g$ and $\check{\gamma}_\g$, and a pair of cocycles $\vphi_\g$ and $\check{\vphi}_\g$. Roughly speaking:
\begin{itemize}
    \item[$\gamma_\g$:] 
    \emph{Elements of the FRW homology}: integration contours for FRW integrals and their cuts (which encode the discontinuities of these integrals). 
    \item[$\vphi_\g$:] 
    \emph{Elements of the FRW cohomology}: the rational differential forms of FRW integrals, which can have poles where propagators go on-shell, or at the branch loci of the twist. 
    \item[$\check{\gamma}_\g$:] 
    \emph{Elements of the dual homology}: 
    these objects do not appear directly as integration contours, but are (Poincaré) dual to the FRW integrands $\vphi_\g$.
    \item[$\check\vphi_\g$:] 
    \emph{Elements of the dual cohomology}: differential forms that are (Poincar\'e) dual to the $\gamma_\g$. These objects are used to compute the coefficients $C_{\h\h}$ in \eqref{eq:schemCosCoaction}. 
\end{itemize}
As we will show, the graph $\g$ encodes all the information needed to build $\gamma_\g$, $\vphi_\g$, $\check{\gamma}_\g$, and $\check\vphi_\g$, as well as the set of compatible graphs $\h$ that are needed to compute the coaction~\eqref{eq:schemCosCoaction}.
The pairings between these four objects (and the canonical-form and intersection maps that relate them) are summarized in figure~\ref{fig:pairings} of section~\ref{sec:coactionTwPer}.

The remainder of this paper is structured as follows.
Section \ref{sec:FRWIntegrals} reviews the computation of FRW wavefunction coefficients as twisted periods.  This includes a discussion of the peculiarities of the geometry underlying FRW integrals, and the development of combinatorial tools for organizing this geometry.
In section \ref{sec:coaction}, we review the properties and utility of the coaction as well as how it is constructed from twisted (co)homology and intersection theory.
Sections \ref{sec:FRWIntegrals} and \ref{sec:coaction} provide all the needed machinery for understanding the examples in section \ref{sec:examples} without delving into the details of twisted (co)homology or intersection theory.
Readers interested in understanding the full geometric construction of the FRW (co)homology and their duals in the partial/relative twisted setting can find this material in section \ref{sec:cohomDets}.
There, the details are spelled out with the two-site chain graph as a pedagogical example.

\section{FRW integrals}
\label{sec:FRWIntegrals}

In this work, we study conformally-coupled scalar field theories in FRW cosmologies with non-conformal polynomial interactions. The action in $(d+1)$ spacetime dimensions is 
\begin{align} 
    \mathcal{S} = \int \d^d x \, \d\eta \sqrt{-g} \left[-\frac{1}{2} g^{\mu \nu} \partial_{\mu} \phi \partial_{\nu} \phi - \frac{d-1}{8d} R \phi^2 - \sum_{p \geq 3} \frac{\lambda_p}{p!} \phi^p \right]~,
    \label{eq:actioninFRW}
\end{align}
where the FRW metric is $\d s^2 = a^2(\eta) \left[-\d \eta^2 + \d x_i \d x^i\right]$ in comoving coordinates with conformal time $\eta \in (-\infty, 0]$, and $i$ should be understood as running over all spatial coordinates $i=1,\dots,d$. The scale factor is chosen to be a power-law characterized by the cosmological parameter $\varepsilon$, 
\begin{equation}
    a(\eta) = (\eta/\eta_{0})^{-(1+\varepsilon)}\, .
\end{equation}
This power-law specializes to several well-studied cosmological scenarios, such as de Sitter ($\varepsilon = 0$), flat ($\varepsilon = -1$), radiation-dominated ($\varepsilon = -2$), and matter-dominated ($\varepsilon = -3$) universes~\cite{De:2023xue, Arkani-Hamed:2023kig}; moreover, for $0<\varepsilon\ll1$ this model is expected to capture the essential dynamics of near de Sitter inflationary cosmology.

The main quantities of interest in these theories are equal-time correlation functions---these functions encode the quantum fluctuations present at the end of inflation, which ultimately seed the temperature and density variations observed today. 
Conveniently, for theories with the action in~\eqref{eq:actioninFRW}, the perturbative contributions to these correlation functions take a universal form. Namely, the integral that we associate with each Feynman diagram in the flat-space theory can be upgraded to any power-law FRW spacetime with $\vep \notin\mathbb{Q}$ through the inclusion of a twist factor $u_\G$:
\begin{align} \label{eq:psiphys}
    \psi_\G = 
    \int_0^\infty 
    u_\G\  \vphi_\G \, ,
\end{align}
where 
\begin{align}
    \vphi_\G = 
    \hat{\Omega}_\G (\mbf{x}+\mbf{X},\mbf{Y})\;
    \d^n\mbf{x}
\end{align}
and $\hat{\Omega}_\G (\mbf{X},\mbf{Y})$
is the flat-space wavefunction coefficient of the flat-space integrand associated with the (truncated) Feynman graph $\G$. It depends on the energies $X_v$ and $Y_e$ that enter each vertex $v$ and flow through each edge $e$, respectively. (We denote the collection of each set of variables by $\mbf{X}$ and $\mbf{Y}$.) These energies all take positive values in the physical region. Denoting the set of vertices in $\G$ by $\V$, the twist factor
\begin{align}
    u_\G= \prod_{v\in \V} x_v^{\alpha_v}
\end{align}
is a multi-valued function whose exponents $\alpha_v$ are determined by the underlying power-law cosmology (as specified by $\vep$), the number of spatial dimensions $d$, and the valency $p_v$ of each vertex $v$:  \begin{align}
    \alpha_v = d + \vep (d + 1) + \frac{1}{2} p_v (1 + \vep)(1 - d)
    \,.
\end{align} 
Our assumption that $\vep \notin \mathbb{Q}$ guarantees that $\alpha_v \notin \mathbb{Z}$, which places us in the context of twisted cohomology. 

In more detail, the flat-space 
integrand $\vphi_\G$ 
is a rational function of the variables $x_v$, $X_v$, and $Y_e$ that inherits a
simple combinatorial definition from the flat space wavefunction $\hat{\Omega}_\G$, namely 
\begin{align} \label{eq:Psi}
    \vphi_\G
    = \d^{|\V|} \mbf{x} \; \sum_{T_\G} \prod_{\tau \in T_\G} 
    \frac{ 1 }{ B_\tau } \, ,
\end{align}
where $T_\G$ denotes the set of complete tubings of $\G$. The denominator factors $B_\tau$ are the basic atoms of the on-shell variety and appear throughout the paper; we collect their definition in box~\ref{box:tubePoly}.

\begin{mybox}[label={box:tubePoly}]{Tubes and tube polynomials }
A \emph{tube} $\tau$ on $\G$ is a simple closed curve encircling a subset of $\G$, specified by a pair of sets
\[
    \tau = \{\mathcal{V}_\tau,\, \mathcal{E}_\tau\}\,,
\]
where $\mathcal{V}_\tau$ is the set of vertices of $\G$ encircled by $\tau$ and $\mathcal{E}_\tau$ is the set of edges of $\G$ that $\tau$ crosses. A \emph{complete tubing} is a maximal set of non-intersecting tubes; the set of complete tubings of $\G$ is denoted $T_\G$. To each tube $\tau$ we associate the \emph{tube polynomial}
\begin{align} \label{eq:singular_divisors}
    B_\tau \;=\; \sum_{v\in \mathcal{V}_\tau} \left( x_v +  X_v \right) \;+\; \sum_{e\in \mathcal{E}_\tau} Y_e
    \,.
\end{align}
These polynomials are the denominators in~\eqref{eq:Psi}; their zero loci $\{B_\tau = 0\}$ make up the on-shell variety $\B$ and govern the cut structure used throughout the paper.
\end{mybox}

As an example, consider the three-site chain graph 
\begin{align} \label{eq:3chainEx1}
    \includegraphics[scale=1.5]{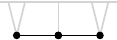}\, .
\end{align}
It has the following tubes and associated tube polynomials:
\begin{align} \label{eq:3chainEx2}
\begin{tabular}{cl|cl}
     $\tau$ &  \qquad $B_\tau$
     & $\tau$ & \qquad $B_\tau$
     \\
     \hline
     \includegraphics[align=c,scale=.7]{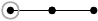}
     & $x_1+X_1+Y_{12}$
     & \includegraphics[align=c,scale=.7]{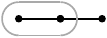}
     & $x_1+x_2+X_1+X_2+Y_{23}$
     \\
     \includegraphics[align=c,scale=.7]{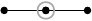}
     & $x_2+X_2+Y_{12}+Y_{23}$
     & \includegraphics[align=c,scale=.7]{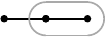}
     & $x_2+x_3+X_2+X_3+Y_{12}$
     \\
     \includegraphics[align=c,scale=.7]{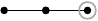}
     & $x_3+X_3+Y_{23}$
     & \includegraphics[align=c,scale=.7]{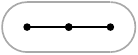}
     & $x_1+x_2+x_3+X_1+X_2+X_3$
\end{tabular}
\end{align}
The full integrand for this diagram can be constructed in terms of these tube polynomials by summing over the two complete tubings
\begin{equation}
    T_\G=\left\{
    \includegraphics[align=c,scale=1]{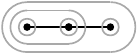}
    ,
    \includegraphics[align=c,scale=1]{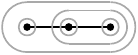}
    \right\}\, .
\end{equation}
For more details, see~\cite{Arkani-Hamed:2017fdk}.

More formally, we think of the integral~\eqref{eq:psiphys} as a period, namely the pairing of a \emph{partially twisted cycle} $[\gamma_\G|$
with a \emph{partially twisted cocycle}
$\vert \vphi_\G\ra$:
\begin{align} \label{eq:integral_family}
   \psi_\G :=
    \int_{0}^\infty u_\G\; \vphi_\G
    =: [\gamma_\G \vert \vphi_\G \ra
    \,.
\end{align}
The bracket types in $[\gamma_\G\vert\vphi_\G\ra$ are not decorative: square and angle brackets each encode whether the corresponding slot holds a homology or cohomology class. The three pairings used throughout the paper all follow this rule, and we collect them in box~\ref{box:braket} for reference.

\begin{mybox}[label={box:braket}]{Bra-ket conventions}
\textbf{Rule.} In any bracket pairing, a \emph{square} bracket marks a (co)homology slot containing a \emph{cycle} (homology class), while an \emph{angle} bracket marks a slot containing a \emph{cocycle} (cohomology class). The argument inside each bra or ket is one representative of its equivalence class.

\medskip
\noindent The three pairings that appear in this paper, together with their short names, are:
\vspace{-.5em}
\begin{center}
\renewcommand{\arraystretch}{1.35}
\footnotesize
\begin{tabular}{@{}l@{\hspace{1.2em}}l@{\hspace{1.2em}}l@{}}
\textbf{Pairing} & \textbf{Notation} & \textbf{Short name} \\
\hline
period (integration) & $[\gamma_\g\vert\vphi_\h\ra = \int_{\gamma_\g} u_\G\,\vphi_\h$ & $P_{\g\h}$ \\
cohomology intersection & $\la\check\vphi_\g\vert\vphi_\h\ra$ & $C_{\g\h}$ \\
homology intersection & $[\gamma_\g\vert\check\gamma_\h]$ & 
\end{tabular}
\end{center}
\end{mybox}


As we will see, $\vphi_\G$ has poles outside of the vanishing loci of the twist $u_\G$, which is why we refer to it as \emph{partially} twisted. 
In this language, it becomes natural to extend our attention to the complete set of integrals that can be constructed from the geometry that defines $\vphi_\G$ and $u_\G$. More specifically, given any Feynman graph $\G$ we can construct a pair of varieties: 
\begin{itemize}
    \item The \textit{twisted} variety, $\mathcal{T} = V(x_1 \cdots x_n)$. 
    This is the locus where the twist $u_\G$ vanishes. 
    \item The \textit{on-shell} variety, $\mathcal{B} = V(B_{\tau_1} \cdots B_{\tau_N})$; the loci where the propagators $B_\tau$ (the hyperplanes that appear in the denominator of $\vphi_\G$) vanish.
\end{itemize}
Given these varieties, we define a pair of topological spaces
\begin{align}
    \mathcal{M}:= M \setminus \mathcal{B}
    \,,
    \qquad
    M := \mathbb{C}^n \setminus \mathcal{T}
    \,,
\end{align}
in which the relevant hyperplanes have been excised. 
We then consider the set of partially twisted cycles $[\gamma_{\g_i} |$ that do not intersect $\B$, and that end on $\T$. 
These cycles can be paired with a dual set of cocycles $\vert \vphi_{\g_j}\ra$
to define the larger set of periods
\begin{align} \label{eq:periodpairing}
    P_{\g_1, \g_2} := \int_{\gamma_{\g_1}} u_\G\; \vphi_{\g_2}
    =: [\gamma_{\g_1} 
    \vert \vphi_{\g_2}\ra
    \,.
\end{align}
As we will see later, each of the $\vert \vphi_{\g_j}\ra$ can be chosen to be the canonical form of one of the bounded chambers in $\mathcal{M}$ (meaning that, like $|\vphi_\G \rangle$, these forms can have poles on both $\mathcal{B}$ and $\T$).

It has been shown that all kinematic derivatives of the integral~\eqref{eq:psiphys} can be expressed in terms of a strict subset of the family of periods that take the form~\eqref{eq:periodpairing}~\cite{Glew:2025ypb,Arkani-Hamed:2023kig, Baumann:2025qjx, He:2024olr}. 
That is, the number of periods that are actually relevant to physics is generally much smaller than the number of elements in the full (co)homology. The relevant cohomological subspace is spanned by the physical wavefunction coefficient $\vphi_\G$ and its derivatives,
\begin{align} \label{eq:physical_space}
    H^n_\mathrm{phys}:= 
    \mathrm{Span}\{
        \vphi_\G, 
        \partial_{X_\bullet}^\mathbb{N}\partial_{Y_\bullet}^\mathbb{N}\vphi_\G
    \}
    \,,
\end{align}
while the homological subspace $H_n^\mathrm{phys}$ is spanned by contours that take cuts which are compatible with the elements of $H^n_\mathrm{phys}$. We take advantage of this fact, by constructing a graphical coaction that only invokes elements of the (co)homology that appear in this physical sector.\footnote{More specifically, while the coaction can be straightforwardly extended to include any period of the form~\eqref{eq:periodpairing}, the \emph{graphical} coaction that we construct is coextensive with just the physical subspace.} 
In particular, we describe how to construct a basis for the physical subspace of the partially twisted cycles $[\gamma_{\g_1} \vert$  and cocycles $ \vert \vphi_{\g_2}\ra$ that can be put in one-to-one correspondence with certain decorations of the graph $\G$.

In the remainder of this section, we introduce the minimal background material needed to construct the graphical coaction.
First, in section~\ref{sec:degenArr}, we identify where degenerate configurations of the cosmological hyperplane arrangements $\B$ occur, and how this affects the partially twisted (co)homology of $\mathcal{M}$.
We then describe, in section~\ref{sec:ResolveDegen}, how this degeneracy can be resolved in the physical sector by considering certain decorations of the graph $\G$. 
In particular, these decorated graphs are in one-to-one correspondence with the set of cuts (or residues) that are nonzero for the wavefunction coefficient. In section~\ref{sec:cutGeom}, we describe how to construct a positive geometry for each of these cuts by restricting the twisted hyperplanes $\T$ to the cut.
As we will see, each physical cut and its associated positive geometry corresponds to a basis element of the physical FRW homology. 
Moreover, each physical cut also defines an element in the \emph{dual} homology whose canonical form provides a basis element for the FRW cohomology.

\subsection{Degenerate hyperplane arrangements}
\label{sec:degenArr}

The tube polynomials $B_\tau$ defined in~\eqref{eq:singular_divisors} are not linearly independent; whenever a pair of tubes overlap, the corresponding $B_\tau$ satisfy a linear relation. 
In such cases, the corresponding hyperplane arrangement $\B$ is said to be degenerate.%
\footnote{%
	On the other hand, the hyperplanes defining $\mathcal{T}$ are always in a generic arrangement; they meet transversely. 
} 
As a result, the naive sequential residues are redundant. Thus, before we can assign a basis of residue operators to a graph, we have to resolve these redundancies---this is the goal of the subsection.

Whenever a pair of tubes $\tau_1$ and $\tau_2$ intersect, there exists a relation
\begin{align}\label{eq:linRel}
    B_{\tau_1} + B_{\tau_2} 
    = B_{\tau_1 \cup \tau_2} + B_{\tau_1\cap\tau_2}
    \,.
\end{align}
Thus, such a relation arises whenever we have partially-overlapping tubes; for instance, for the following pair of partially-overlapping two-site tubes, we generate the relation: 
\begin{align} \label{eq:3chainEx3}
    \underset{
        \includegraphics[align=c,scale=.7]{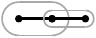}
    }{\underbrace{
        B_{\includegraphics[align=c,scale=.5]{figs/3chainT12.pdf}}
        + B_{\includegraphics[align=c,scale=.5]{figs/3chainT23.pdf}}
    }}
    = \underset{
        \includegraphics[align=c,scale=.7]{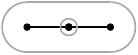}
    }{\underbrace{
        B_{\includegraphics[align=c,scale=.5]{figs/3chainT2.pdf}}
        + B_{\includegraphics[align=c,scale=.5]{figs/3chainT123.pdf}}
    }}
    \,.
\end{align}

Because of the relations generated by \eqref{eq:linRel}, the intersection of $|K|$ hyperplanes $\smash{\{\B_{k}\}_{k\in K}}$ does not necessarily define a codimension-\hspace{-.04cm}$|K|$ hypersurface. 
For example, for the three-site chain, equation~\eqref{eq:3chainEx3} implies that setting any three of the polynomials
\begin{equation} \label{eq:degenerate_tube_polynomials}
\{
    B_{\includegraphics[align=c,scale=.5]{figs/3chainT123.pdf}},
    B_{\includegraphics[align=c,scale=.5]{figs/3chainT12.pdf}},
    B_{\includegraphics[align=c,scale=.5]{figs/3chainT23.pdf}},
    B_{\includegraphics[align=c,scale=.5]{figs/3chainT2.pdf}}
\}
\end{equation}
to zero automatically sets the fourth to zero as well. 
Geometrically, this means that the four hyperplanes 
defined by these polynomials intersect at a point in the (three-dimensional) space of integration variables $\{x_1, x_2, x_3\}$.
This non-genericity generates identities between differential forms and sequential residues; for instance, one can check that the following identity holds at the level of $\dlog$-forms:
\be \label{eq:dlogID}
    0=&
    \dlog B_{\includegraphics[align=c,scale=.5]{figs/3chainT123.pdf}}
    \wedge \dlog B_{\includegraphics[align=c,scale=.5]{figs/3chainT12.pdf}}
    \wedge \dlog B_{\includegraphics[align=c,scale=.5]{figs/3chainT23.pdf}}
    \\
    & -
    \dlog B_{\includegraphics[align=c,scale=.5]{figs/3chainT123.pdf}}
    \wedge \dlog B_{\includegraphics[align=c,scale=.5]{figs/3chainT12.pdf}}
    \wedge \dlog B_{\includegraphics[align=c,scale=.5]{figs/3chainT2.pdf}}
    \\
    & +
    \dlog B_{\includegraphics[align=c,scale=.5]{figs/3chainT123.pdf}}
    \wedge \dlog B_{\includegraphics[align=c,scale=.5]{figs/3chainT23.pdf}}
    \wedge \dlog B_{\includegraphics[align=c,scale=.5]{figs/3chainT2.pdf}}
    \\ 
    & -
    \dlog B_{\includegraphics[align=c,scale=.5]{figs/3chainT12.pdf}}
    \wedge \dlog B_{\includegraphics[align=c,scale=.5]{figs/3chainT23.pdf}}
    \wedge \dlog B_{\includegraphics[align=c,scale=.5]{figs/3chainT2.pdf}}
    \,.
\ee
Similarly, at the level of sequential residues, it is easy to see that 
\begin{equation} \label{eq:residue_relation_example}
\res_{
        \includegraphics[align=c,scale=.5]{figs/3chainT123.pdf}, \includegraphics[align=c,scale=.5]{figs/3chainT12.pdf}, \includegraphics[align=c,scale=.5]{figs/3chainT2.pdf} 
    }
    = \res_{
        \includegraphics[align=c,scale=.5]{figs/3chainT123.pdf}, \includegraphics[align=c,scale=.5]{figs/3chainT12.pdf}, \includegraphics[align=c,scale=.5]{figs/3chainT23.pdf}
    } \, ,
\end{equation} 
since~\eqref{eq:3chainEx3} implies that 
$
    B_{\includegraphics[align=c,scale=.5]{figs/3chainT23.pdf}}
    \vert_{
        B_{\includegraphics[align=c,scale=.3]{figs/3chainT123.pdf}} 
        = B_{\includegraphics[align=c,scale=.3]{figs/3chainT12.pdf}}
        = 0
    }
    = B_{\includegraphics[align=c,scale=.5]{figs/3chainT2.pdf}}
    \vert_{
        B_{\includegraphics[align=c,scale=.3]{figs/3chainT123.pdf}} 
        = B_{\includegraphics[align=c,scale=.3]{figs/3chainT12.pdf}}
        = 0
    }
$
and consequently that $\res_{\includegraphics[align=c,scale=.5]{figs/3chainT2.pdf}}$ and $\res_{\includegraphics[align=c,scale=.5]{figs/3chainT23.pdf}}$ are equivalent once we have computed $\smash{\res_{\includegraphics[align=c,scale=.5]{figs/3chainT123.pdf},\includegraphics[align=c,scale=.5]{figs/3chainT23.pdf}}}$. Note that our convention for sequential residues is
\be\res_{\tau_1,\tau_2,\cdots, \tau_m} := \res_{\B_{\tau_m,\dots,\tau_1}} \circ \cdots \circ \res_{\B_{\tau_1\tau_2}} \circ \res_{\B_{\tau_1}}
\ee 
where each residue operator acts to the right, and we recall that $\B_{\tau_1,\dots,\tau_m} = V(B_{\tau_1}) \cap \cdots \cap V(\B_{\tau_m})$ is the intersection of the hyperplanes generated by $B_{\tau_1}, \dots, B_{\tau_m}$.
Denoting residues by the corresponding hyperplane intersections---which are coordinate independent---removes any coordinate ambiguity. 

To construct a coaction, we would like to resolve these degeneracies by identifying sets of non-redundant sequential residues. To do so, we make use of the following result:
\begin{mybox}[label={box:spanningRes}]{Spanning Residues}
    The set of all sequential residue operators is spanned by those that respect a chosen global ordering of the tubes.
\end{mybox}
\noindent 
This statement is obvious when the residue hyperplanes are generic (namely, when the corresponding tube polynomials are linearly independent).
In this case, sequential residue operators are fully anti-symmetric in the sense that $\res_{i,j,k,\dots} = \text{sign}(\sigma) \res_{\sigma(i,j,k,\dots)}$ for any permutation $\sigma$. 
Therefore, one can span the space of sequential residue operators using only those that respect a global ordering.
Conversely, in non-generic arrangements, some sequential residue operators satisfy linear relations and will not be totally antisymmetric. Specifically, given any minimal collection of linearly dependent tube polynomials $\{B_{\tau_j}\}_{j=1}^J$ (meaning all subsets with cardinality $|J|-1$ are linearly independent) equations \eqref{eq:dlogID} and \eqref{eq:residue_relation_example} generalize to  
\begin{align}
    \label{eq:genDlogID}
    &\sum_{j = 1}^J (-1)^{|J|-j}
    \bigwedge_{k\in J \setminus \{j\}}
    \dlog B_k = 0
    \,,
    \\
    \label{eq:genResID}
    &\res_{\tau_1,\cdots, \tau_{|J|-2}, \tau_{|J|-1}}
    = \res_{\tau_1,\cdots, \tau_{|J|-2},\tau_{|J|}}
    \,,
\end{align}
where the order of tubes $\tau_1, \dots, \tau_{|J|}$ is compatible with the chosen global tube order. 
Since the existence of the additional relations~\eqref{eq:genResID} can only reduce the number of independent operators needed to span the full set, the statement in Box~\ref{box:spanningRes} must still hold.

When considering the wavefunction of the universe, some orderings are better than others. To illustrate this, consider the following random ordering of the tubes that appear in the three-site chain:
$    (
         \includegraphics[align=c,scale=.5]{figs/3chainT123.pdf}, \includegraphics[align=c,scale=.5]{figs/3chainT2.pdf}, \includegraphics[align=c,scale=.5]{figs/3chainT12.pdf},
         \includegraphics[align=c,scale=.5]{figs/3chainT23.pdf} 
    )
$.
From \eqref{eq:genResID}, we have that two of the non-crossed tubings with this ordering are equal, namely
\begin{align}
    \res_{
        \includegraphics[align=c,scale=.5]{figs/3chainT123.pdf}, \includegraphics[align=c,scale=.5]{figs/3chainT2.pdf}, \includegraphics[align=c,scale=.5]{figs/3chainT12.pdf}
    }
    = \res_{
        \includegraphics[align=c,scale=.5]{figs/3chainT123.pdf}, \includegraphics[align=c,scale=.5]{figs/3chainT2.pdf},
        \includegraphics[align=c,scale=.5]{figs/3chainT23.pdf} 
    }
    \,.
\end{align}
Moreover, this pair of non-crossed residues annihilate the physical form, since
\begin{align}
    \res_{
        \includegraphics[align=c,scale=.5]{figs/3chainT123.pdf}, \includegraphics[align=c,scale=.5]{figs/3chainT2.pdf}, \includegraphics[align=c,scale=.5]{figs/3chainT12.pdf}
    }[\vphi_{\includegraphics[scale=.1]{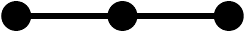}}]
    &\propto 
    \res_{
        \includegraphics[align=c,scale=.5]{figs/3chainT123.pdf}, \includegraphics[align=c,scale=.5]{figs/3chainT2.pdf}, \includegraphics[align=c,scale=.5]{figs/3chainT12.pdf}
    }\left[\frac{
            B_{\includegraphics[align=c,scale=.5]{figs/3chainT12.pdf}}
            + B_{\includegraphics[align=c,scale=.5]{figs/3chainT23.pdf}}
        }{
            B_{\includegraphics[align=c,scale=.5]{figs/3chainT2.pdf}}
            B_{\includegraphics[align=c,scale=.5]{figs/3chainT12.pdf}}
            B_{\includegraphics[align=c,scale=.5]{figs/3chainT23.pdf}}
            B_{\includegraphics[align=c,scale=.5]{figs/3chainT123.pdf}}
    }\right]
    \nn\\
    &= \res_{
            \includegraphics[align=c,scale=.5]{figs/3chainT12.pdf}
    }\left[
        \frac{
            B_{\includegraphics[align=c,scale=.5]{figs/3chainT12.pdf}}
            + B_{\includegraphics[align=c,scale=.5]{figs/3chainT23.pdf}}
        }{
            B_{\includegraphics[align=c,scale=.5]{figs/3chainT12.pdf}}
            B_{\includegraphics[align=c,scale=.5]{figs/3chainT23.pdf}}
         }
    \right]_{ 
            B_{\includegraphics[align=c,scale=.4]{figs/3chainT2.pdf}}
            = B_{\includegraphics[align=c,scale=.4]{figs/3chainT123.pdf}}
            = 0
        }    
    = 0 
    \,,
\end{align}
where in the last line above we have used \eqref{eq:3chainEx3} to set the numerator to zero. At the same time, for one of the crossed tubings, we have that
\begin{align}
    \res_{
        \includegraphics[align=c,scale=.5]{figs/3chainT2.pdf}, \includegraphics[align=c,scale=.5]{figs/3chainT12.pdf}, \includegraphics[align=c,scale=.5]{figs/3chainT23.pdf}
    }[\vphi_{\includegraphics[scale=.1]{figs/3chain/3chain.pdf}}]
    &= \res_{
            \includegraphics[align=c,scale=.5]{figs/3chainT23.pdf}
    }\left[
        \frac{
            1
        }{
            B_{\includegraphics[align=c,scale=.5]{figs/3chainT123.pdf}}
         }
    \right]_{ 
            B_{\includegraphics[align=c,scale=.4]{figs/3chainT2.pdf}}
            = B_{\includegraphics[align=c,scale=.4]{figs/3chainT12.pdf}}
            = 0
        }    
    \\&= \res_{
            \includegraphics[align=c,scale=.5]{figs/3chainT23.pdf}
    }\left[
        \frac{
            1
        }{
            B_{\includegraphics[align=c,scale=.5]{figs/3chainT23.pdf}}
         }
    \right]_{ 
            B_{\includegraphics[align=c,scale=.4]{figs/3chainT2.pdf}}
            = B_{\includegraphics[align=c,scale=.4]{figs/3chainT12.pdf}}
            = 0
        } 
    = \frac{1}{Y_{12} Y_{23}}
    \,.
\end{align}
We conclude that this choice of ordering spoils the distinguished role that non-crossed tubings play in the construction of $\vphi_\G$ (recall that each term in the wavefunction $\vphi_\G$ is associated with a maximal \emph{non-crossing} tubing~\cite{Arkani-Hamed:2017fdk}).

It turns out that a preferred ordering, in which the residues associated with crossed tubings can be ignored, does exist.%
\footnote{%
    Note that changing the order of the tubes can only alter \eqref{eq:dlogID} by an overall sign. Thus, the identity among forms \eqref{eq:genDlogID} is less sensitive to ordering changes than the residue operators \eqref{eq:genResID}.
}
The key is to order tubes that encircle more vertices before tubes that encircle fewer vertices; we break ties by ordering tubes that encircle the largest number of edges first, and then lexicographically by the minimal non-shared vertex.%
\footnote{%
    One could instead choose the opposite ordering, in which tubes that encircle fewer vertices come first. 
    However, we prefer to compute the residues associated to 
    larger tubes first. 
    With this ordering, computing residues of the wavefunction reduces it to an amplitude or several sub-amplitudes \cite{Arkani-Hamed:2017fdk}. 
} 
Then, the sequential residue associated to each crossed tubing is either:
\begin{itemize}
    \item equivalent to some linear combination of non-crossed residue operators, or \item annihilates the physical FRW-form $\vphi_\G$.
\end{itemize}
Adopting this convention automatically reduces the number of residue operators---and, as demonstrated below, the number of elements of the (co)homology---needed to construct the coaction of $\vphi_\G$.
We emphasize that this set of residue operators is designed to span the space relevant to the wavefunction coefficient (as outlined in~\eqref{eq:physical_space}), rather than the complete space of objects that can be defined on $\mathcal{M}$. 

Putting everything together, and adopting a lexicographical ordering to order tubes of the same size, our tube ordering is given as follows:

\begin{mybox}[label={box:tubeOrd}]{Tube Ordering}
We always use sequential residues where the tubes are ordered such that 
\begin{itemize}
    \item[(1)] tubes that encircle more vertices come before tubes that encircle fewer vertices; 
    \item[(2)] tubes that encircle more edges come before tubes that encircle fewer edges; 
    \item[(3)] tubes that encircle the same number of vertices and edges are ordered lexicographically via the minimal vertex not shared by both. 
\end{itemize}
In this ordering, the space of sequential residue operators is spanned by the set of operators associated with non-crossing tubings.
\end{mybox}

To close this subsection, we highlight that we can diagonalize the space of degenerate residues with respect to the physical FRW-form by a rotation of the independent residue operators. 
For instance, in the degenerate sector of the three-site chain, we define
\begin{align} \label{eq:residue_rotation_example}
        \res_{-} 
        := 
        \res_{\includegraphics[scale=.5, align=c]{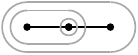}}
        - \res_{\includegraphics[scale=.5, align=c]{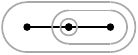}}
        \,,
        \qquad\mathrm{and}\qquad
        \res_{+} 
        :=
        \res_{\includegraphics[scale=.5, align=c]{figs/3Chain123_12_2.pdf}}
        + \res_{\includegraphics[scale=.5, align=c]{figs/3Chain123_23_2.pdf}}
        \,,
\end{align}
after which the space of residues is spanned by $\res_{\pm}$ and $\res_{\includegraphics[scale=.5, align=c]{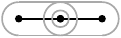}}$. 
Moreover, in the new basis, only $\res_-$ does not annihilate $\vphi_\G$:
\begin{align}
      \res_{-}[\vphi_{\includegraphics[scale=.1]{figs/3chain/3chain.pdf}}] 
      = \frac{1}{4Y_{12}Y_{23}}
      \,,
      \qquad
      \res_{+}[\vphi_{\includegraphics[scale=.1]{figs/3chain/3chain.pdf}}] 
      = 0
      \,,
      \qquad
      \res_{\includegraphics[scale=.5, align=c]{figs/3Chain12_23_2.pdf}}[\vphi_{\includegraphics[scale=.1]{figs/3chain/3chain.pdf}}] 
      =0
      \,.
\end{align}
This is a general feature; one can always organize the space of residue operators in a degenerate sector so that all but one of these operators annihilates $\vphi_\G$. 
In the next section, we will develop a combinatorial algorithm for constructing this distinguished operator. We can then safely ignore all other residue operators in this sector, as they are unphysical.

\subsection{Cut degeneracy and decorated graphs}
\label{sec:ResolveDegen}

Using the ideas from the last section, it is already possible to enumerate an independent basis of sequential residue operators that take into account the linear relations generated by~\eqref{eq:linRel}, for any FRW integral $\vphi_\G$. 
However, we would like to automate this construction as much as possible. 
For this purpose, we consider different ways of decorating the graph $\G$. 
As we will show, a basis of residue operators that do not annihilate the FRW form can be chosen that are in one-to-one correspondence with the \textit{acyclic minors} of $\G$.

The acyclic minors of a graph $\G$ are constructed by assigning decorations to each of its edges:
\begin{mybox}[label={box:edgeDecorations}]{Edge decorations}
\vspace{-.5em}
\begin{center}
\renewcommand{\arraystretch}{1.2}
\begin{tabular}{@{}c@{\hspace{1.5em}}p{0.75\textwidth}@{}}
\textbf{Decoration} & \textbf{Bulk physics interpretation} \\
\hline
\includegraphics[scale=.7, align=c]{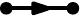} \;\; \emph{oriented}
    & Energy (or time) flows in the direction of the arrow. \\
\includegraphics[scale=.7, align=c]{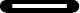} \;\; \emph{pinched}
    & The connected vertices merge into a single contact interaction. \\
\includegraphics[scale=.7, align=c]{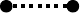} \;\; \emph{broken}
    & No particles are exchanged between the two vertices. 
\end{tabular}
\end{center}
\vspace{-1em}
\end{mybox}
\noindent
Since the resulting graph is (partially) oriented, we refer to these graphs as {\it decorated orientations} of $\G$. We denote the set of all such decorated orientations of $\G$ by $\text{Dec}(\G)$. 
It is clear that $|\text{Dec}(\G)|=4^{|\E_\G|}$, where $\E_\G$ is the set of edges of $\G$.
A decorated orientation $\g \in \text{Dec}(\G)$ is {\it acyclic} if the oriented graph produced by deleting all broken edges and contracting all pinched edges does not contain any cycles.
We denote the set of all {\it acyclic minors} by $\text{aDec}(\G) \subseteq \text{Dec}(\G)$.%
\footnote{%
    The pinching (contracting) and breaking (removing) edges of $\G$ corresponds to minors of graph $\G$ in the usual mathematical sense; the only difference is that we have given each minor an acyclic orientation.
} 
For tree level graphs we simply have $\text{aDec}(\G) = \text{Dec}(\G)$.


It turns out that we can associate a set of ``physical'' cut tubings $\Csf_\g$ to each acyclic decorated graph $\g \in \text{aDec}(\G)$, according to    the following rules:
\begin{mybox}[label={box:acyclicToTubings}]{From acyclic decorated graphs to ``physical'' cut-tubings}
    Each tubing $\csf\in\Csf_\g$ is a maximal collection of non-crossing tubes where 
    \vspace{-.5em}\begin{itemize}
    \item[(1)] no $\tau \in \csf$ crosses a pinched edges of $\g$,
    \item[(2)] no $\tau \in \csf$ encircles a broken edge of $\g$,
    \item[(3)] any $\tau \in \csf$ crosses an edge $e$ only if 
        \vspace{-.5em}\begin{itemize}
            \item[(a)] $e$ is a broken edge of $\g$, 
            \item[(b)] or $e$ is an oriented edge \includegraphics[align=c,scale=.3]{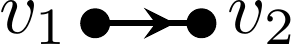} with $v_1 \in \V_\tau$ and $v_2 \not\in \V_\tau$.
        \end{itemize}   
    \end{itemize}
\end{mybox}
\noindent
For many acyclic minors, there is a unique tubing that satisfies the conditions in Box~\ref{box:acyclicToTubings}. 
As an example, consider the following acyclic minors of the four-site chain and the corresponding set of cut-tubings
\begin{align}\begin{aligned}
    \Csf_{\includegraphics[align=c, scale=.3]{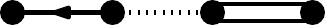}}
    = \left\{
        \includegraphics[align=c, scale=.4]{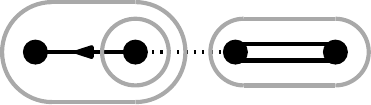}
    \right\}
    \,,
    \qquad
    \Csf_{\includegraphics[align=c, scale=.3]{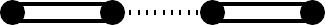}}
    = \left\{
        \includegraphics[align=c, scale=.4]{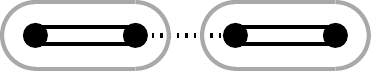}
    \right\}
    \,.
\end{aligned}\end{align}
There are also many acyclic decorated graphs that have multiple tubings $\csf$, each of which satisfy the rules of Box~\ref{box:acyclicToTubings}. 
Some examples for the four-site chain are 
\begin{align}
    \Csf_{\includegraphics[align=c, scale=.3]{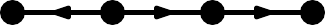}}
    &= \left\{
        \includegraphics[align=c, scale=.4]{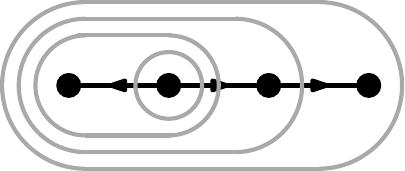}
        , 
        \includegraphics[align=c, scale=.4]{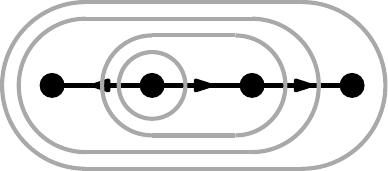}
        , 
        \includegraphics[align=c, scale=.4]{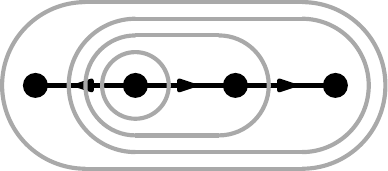}
    \right\}
    \,,
    \\
    \Csf_{\includegraphics[align=c, scale=.3]{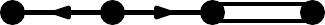}}
    &= \left\{
        \includegraphics[align=c, scale=.4]{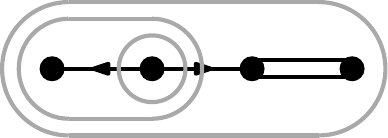}
        , 
        \includegraphics[align=c, scale=.4]{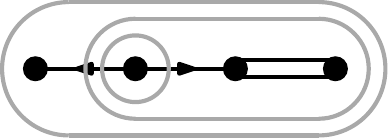}
    \right\}
    \,,
\end{align}
There are 60 more acyclic decorated orientations for the four-site chain, each of which gives rise to its own set of cut tubings. 

Next, we associate a distinguished sequential residue operator to each $\g \in \text{aDec}(\G)$. 
For decorations that give rise to a unique cut-tubing, this correspondence is clear---we just construct a sequential residue operator out of the set of tubes that appear in this cut-tubing, where the order of residues is dictated by our chosen tube ordering in Box \ref{box:tubeOrd}. 
Note that, in general, these sequential residue operators will have different lengths; for instance, consider the following pair of examples:
\begin{align}
    \res\Csf_{\includegraphics[align=c, scale=.3]{figs/sbs.pdf}}
    &= \res_{
       \includegraphics[align=c, scale=.3]{figs/Csbs.pdf}
    }
    := \res_{
        \includegraphics[align=c, scale=.3]{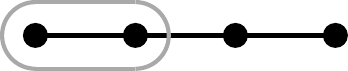}
        ,\includegraphics[align=c, scale=.3]{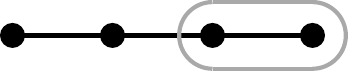}
    }
    \,,
    \\
    \res\Csf_{\includegraphics[align=c, scale=.3]{figs/lbs.pdf}}
    &= \res_{
       \includegraphics[align=c, scale=.3]{figs/Clbs.pdf}
    }
    := \res_{
        \includegraphics[align=c, scale=.3]{figs/4ChainB12.pdf}
        ,\includegraphics[align=c, scale=.3]{figs/4ChainB34.pdf}
        ,\includegraphics[align=c, scale=.3]{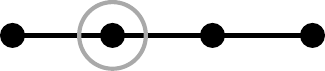}
    }
    \,.
\end{align}
Only two tubes are associated with $\adjustbox{valign=c}{\includegraphics[align=c, scale=.3]{figs/sbs.pdf}}$, 
while three tubes are associated with $\adjustbox{valign=c}{\includegraphics[align=c, scale=.3]{figs/lbs.pdf}}$.

The distinguished sequential residue operator that we associate with each acyclic decorated graph that has multiple cut-tubings is slightly more complicated. 
To construct these residue operators, we first note that all the cut-tubings associated to an acyclic decorated graph specify the same variety. 
Namely, for any pair $\csf, \csf^\prime \in \Csf$ that are given by $\csf=\{\tau_1, \dots, \tau_{k}\}$ and $\csf^\prime = \{\tau_1^\prime, \dots, \tau_k^\prime\}$, we have that 
\begin{equation}
    \B_{\csf} := V(B_{\tau_1}) \cap \cdots \cap V(B_{\tau_k}) = V(B_{\tau_1^\prime}) \cap \cdots \cap V(B_{\tau_k^\prime}) =: \B_{\csf^\prime} \, .
\end{equation}
In other words, each $\csf \in \Csf_\g$ just corresponds to a different way of writing the same linear system of polynomial equations, using identities of the form \eqref{eq:linRel}.
As a result, we associate a unique \emph{cut space} to each decorated graph $\g \in \text{aDec}(\G)$:
\begin{align} \label{eq:cutSpace1}
    M_\g := M\cap \B_{\g}
    = (\mathbb{C}^{|\V_\G|}\setminus\T) \cap \B_\g
    \,,
\end{align}
where we have adopted the notation $\B_\g := \B_\csf \;\forall\; \csf \in \Csf_\g$.
This codimension-$|\csf|$ space is the complement of $\mathcal{T}$ on the intersection of the hyperplanes $\B_\g$. 
Since the polynomials $B_{\tau_i}$ that appear in $\B_\g$ are all linear and (by construction) correspond to non-crossing tubes, they identify a single bounded chamber in $M_\g$.%
\footnote{%
    In fact, the bounded chamber on $M_\g = M \cap \B_{\csf \in \Csf_\g}$ is a Cartesian product of simplices. 
    This is not the case for cut spaces $ M \cap \B_{\tau}$ that are defined by collections of tubings $\tau$ that cannot be associated to an acyclic decorated graph via the rules outlined in Box~\ref{box:acyclicToTubings}. 
} 
(We will have more to say about this bounded chamber in section~\ref{sec:cutGeom}.)
It follows that the twisted (co)homology on $M_\g$ is one-dimensional. This matches what was indicated at the end of section~\ref{sec:degenArr}, namely that the space of inequivalent sequential residue operators that localize us to $\B_\g$ can always be rotated so that all but one of the operators in this sector annihilate $\vphi_\G$. Therefore, without loss of generality, we associate the unique operator that acts nontrivially on $\vphi_\G$ with the acyclic minor $\{\res_\csf\}_{\csf\in\Csf_\g}$. 
It can be shown that this distinguished residue operator always takes the form 
\begin{align} \label{eq:degenRes}
    \res\,\Csf_\g 
    &= \sum_{\csf \in \Csf_\g}
        \sgn_\csf
        \res_\csf
        \,,
\end{align}
where $\sgn_\csf$ is determined by comparing the signature of some ordered sets. 
(While the explicit formula for $\sgn_\csf$ is not complicated, it depends on notation that distracts from the current exposition; it is given later in~\eqref{eq:sgn}.)
Some simple examples include 
\begin{align}
    \res\Csf_{\includegraphics[align=c, scale=.3]{figs/lrr.pdf}}
    &= - \res_{
        \includegraphics[align=c, scale=.3]{figs/Clrr_1.pdf}
    }
    + \res_{
        \includegraphics[align=c, scale=.3]{figs/Clrr_2.pdf}
    }
    - \res_{
        \includegraphics[align=c, scale=.3]{figs/Clrr_3.pdf}
    }
    \,,
    \\
    \res\Csf_{\includegraphics[align=c, scale=.3]{figs/lrs.pdf}}
    &= \res_{
        \includegraphics[align=c, scale=.3]{figs/Clrs_1.pdf}
    }
    - \res_{
        \includegraphics[align=c, scale=.3]{figs/Clrs_2.pdf}
    }
    \,.
\end{align}
The operator $\res_{-}$, as defined in~\eqref{eq:residue_rotation_example} for the three-site chain, constitutes another example.

We end this section by describing the inverse of the map in box \ref{box:acyclicToTubings}, by means of which we can construct the acyclic decorated graph $\g_{\csf}$ associated with any physical tubing $\csf$. 
Namely, for each edge \includegraphics[align=c,scale=.3]{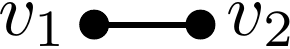}
 in the graph $\G$, we assign the following decoration:
\begin{itemize}
\item[(1)] oriented as  \includegraphics[align=c,scale=.3]{figs/v1_to_v2.pdf}, 
    if all the tubes in $\csf$ crossing it are encircling $v_1$;
\item[(2)] oriented as  \includegraphics[align=c,scale=.3]{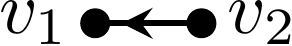},     
    if all the tubes in $\csf$ crossing it are encircling $v_2$;
\item[(3)] broken \includegraphics[align=c,scale=.3]{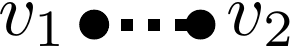},     
     if $v_1$ and $v_2$ are both encircled by the tubes crossing the edge;
\item[(4)] pinched  \includegraphics[align=c,scale=.3]{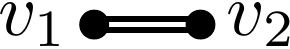}, 
    if there is no tube in $\csf$ crossing it.
\end{itemize}
It can be checked that, with these assignments,  all the elements in $\Csf_{\g}$ map back to $\g$ itself.

\subsection{The positive geometry of physical cuts}
\label{sec:cutGeom}

Having associated a set of tubings with each acyclic decorated graph $\g$, we now also introduce a notion of graph regions:

\begin{mybox}[label={box:regions}]{Regions of an acyclic decorated graph}
    To every 
    acyclic decorated graph $\g$, we assign a set of graph regions $\Rsf_\g := \{\rsf_\bullet\}$ that partition the vertices in $\g$ so that each vertex is assigned to a unique region.
    Each region $\rsf$ is a connected subgraph of $\g$ that either consists of only pinched edges (\includegraphics[scale=.6,align=c]{figs/2chain_gs.pdf}), or a single vertex that is not connected to a pinched edge.
\end{mybox}

\noindent 
The regions $\Rsf_\g$ of $\g$ do not match the sets of vertices that are enclosed by the tubes in $\Csf_\g$. To illustrate this, we consider the following example:
\be \label{eq:g_dblbox}
    \g = \includegraphics[align=c,scale=2]{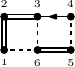}
    \,.
\ee
This acyclic graph has three regions, which we assign different colors for convenience:
\be
    \rsf_{1} &:= \includegraphics[align=c,scale=2]{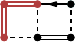}
    \,,
    &
    \rsf_{2} &:= \includegraphics[align=c,scale=2]{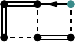}
    \,,
    &
    \rsf_{3} &:= \includegraphics[align=c,scale=2]{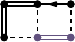}
    \,,
\ee
Overlaying these regions on top of this graph's (unique) cut tubing, we have:
\begin{center}
    \begin{minipage}{.2\textwidth}
        \includegraphics[align=c,width=\textwidth]{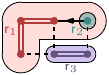}
    \end{minipage}
    \begin{minipage}{.78\textwidth}
        \be \label{eq:regionExample}
            B_{\includegraphics[align=c,scale=.7]{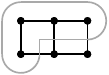}} 
                &= x_1 {+} x_2 {+} x_3 {+} x_4 {+} X_1 {+} X_2 {+} X_3 {+} X_4 {+} Y_{16} {+} Y_{36} {+} Y_{45}
            \,,
            \\
            B_{\includegraphics[align=c,scale=.7]{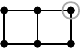}}
                &= x_4 {+} X_4 {+} Y_{34} {+} Y_{45}
            \,,
            \\
            B_{\includegraphics[align=c,scale=.7]{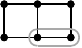}} 
                &= x_5 {+} x_6 {+} X_5 {+} X_6 {+} Y_{16} {+} Y_{36} {+} Y_{45}
            \,. 
        \ee
    \end{minipage}
\end{center}
Thus, while the regions do not correspond to the tubes appearing in the cut tubing directly, the set of vertices in each region are bounded by these tubes. While in some cases, the vertices in a region are bounded by a single outer tube (in this example, $\rsf_2$ and $\rsf_3$), others are also bounded by inner tubes ($\rsf_1$). 

If we rearrange the on-shell conditions that come from setting all three of the tube polynomials in~\eqref{eq:regionExample} to zero, we see that
\be\label{eq:partialRegion}
    \partial\rsf_{1}
    = B_{\includegraphics[align=c,scale=.7]{figs/dblBoxB1234.pdf}}  
        - B_{\includegraphics[align=c,scale=.7]{figs/dblBoxB4.pdf}} 
    = 0 
    &\implies  
    x_1 {+} x_2 {+} x_3 {+}  
        X_1 {+} X_2 {+} X_3 {+} Y_{16} {+} Y_{36} {-} Y_{34}
    = 0
    \,,
    \\
    \partial\rsf_{2} 
    = B_{\includegraphics[align=c,scale=.7]{figs/dblBoxB4.pdf}}   
    = 0
    &\implies  
    x_4 {+} X_4 {+} Y_{34} {+} Y_{45} 
    = 0
    \,,
    \\
    \partial\rsf_{3} 
    = B_{\includegraphics[align=c,scale=.7]{figs/dblBoxB56.pdf}}  
    = 0 
    &\implies 
    x_5 {+} x_6 {+} X_5 {+} X_6 {+} Y_{16} {+} Y_{36} {+} Y_{45} 
    = 0
    \,.
\ee
Written in this form, it becomes clear that the on-shell conditions place a single linear constraint on the sum of variables $\sum x_i$ associated with each region. 
Thus, the regions $\Rsf_\g$ can be thought of as identifying the geometry of the cut space $M_\g$ defined in~\eqref{eq:cutSpace1}, which is obtained by setting all of the tube polynomials in any $\csf\in\Csf_\g$ to zero. 
More explicitly, 
\be
    M_\g = \{ 
        (x_1,\dots,x_{|\V_\g|}) \in \mathbb{C}^{|\V_\G|} \setminus \T 
    \,\big|\, 
        \partial r_1 = \cdots = \partial r_{\Rsf_\g} = 0 
    \}  
    \,,
\ee
where $\partial r_j$ represents the boundary of region $r_j$, as illustrated in~\eqref{eq:partialRegion}.
The cut spaces $M_\g$ provide us with important topological building blocks when constructing the FRW homology, as we will see in sections \ref{sec:hom} and \ref{sec:dualCohom}.

\begin{figure} 
    \centering
    \includegraphics[align=c, scale=.6]{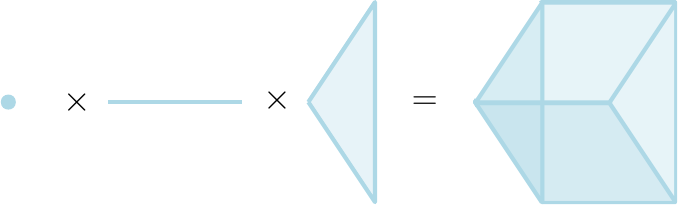}
    \qquad
    \includegraphics[align=c, scale=.4]{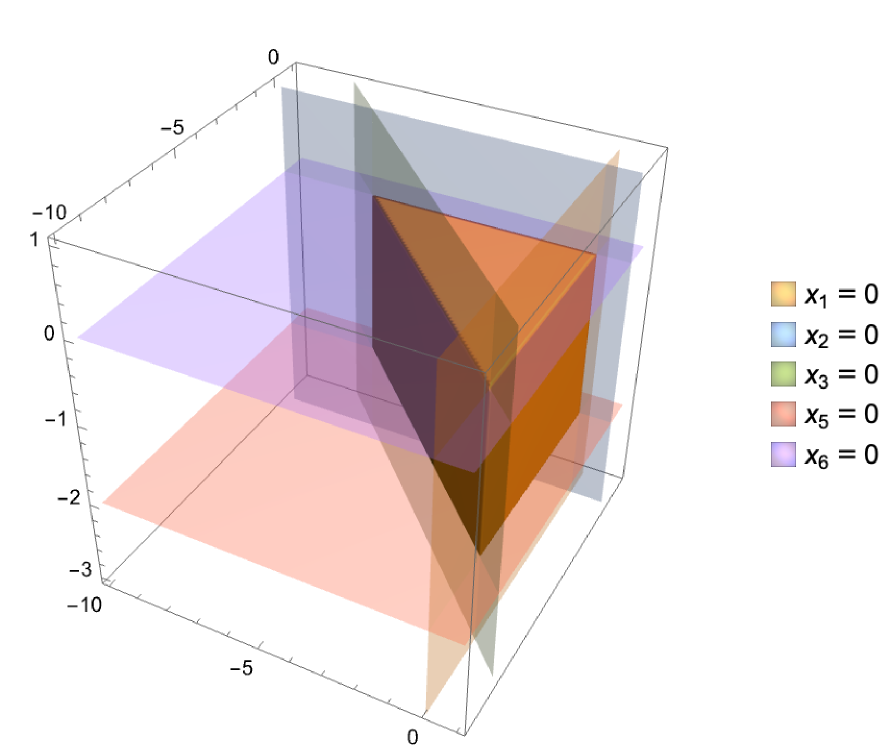}
    \caption{%
    Left: The unique bounded chamber $\Delta_\g$ on $M_\g$ expressed as a Cartesian product of region simplices. 
    Right: Plot of the hyperplanes $x_1=x_2=x_6=0$,
    $x_3 = - f_{\rsf_1} - x_1 - x_2 = 0$, 
    and $x_5 = - f_{\rsf_3} - x_6 = 0$, where we have eliminated $x_3$, $x_4$, and $x_5$ using \eqref{eq:partialRegion}.
    Note that the on-shell conditions fix $x_4 = -f_{\rsf_2}$, and therefore the twisted plane $x_4=0$ does not appear in the right plot. 
    The bounded chamber corresponds to the solid orange bounded region. 
    \label{fig:toblerone}
    }
\end{figure}

In the cut space $M_\g$, the twisted variety $\T \cap \B_\g$ carves out a single bounded chamber denoted by $\Delta_\g$. 
In the example graph from~\eqref{eq:g_dblbox}, this chamber is given by 
\be
    \Delta_\g := \{ (x_1, x_2, x_3) \in \mathbb{R}^3_- : \partial\rsf_{123}=0\} \times \{ x_4 \in \mathbb{R}_- : \partial\rsf_{4}=0\} \times \{ (x_5, x_6) \in \mathbb{R}^2_- : \partial\rsf_{56}=0\}
    \,,
\ee
and is illustrated in figure \ref{fig:toblerone}.
Here, each piece of the Cartesian product corresponds to a region, and constitutes a 2-, 1- or 0-simplex in the corresponding negative orthant.

This construction of graph regions generalizes straightforwardly to any acyclic decorated graph $\g$, and shows that the on-shell conditions do not depend on the details of individual cut tubings $\csf$. 
In terms of the kinematic variables $\{X_\bullet,Y_\bullet\}$, the on-shell conditions $\partial\rsf_i =0$ are 
\begin{align} \label{eq:partialRg}
    \partial\rsf 
    &= \sum_{v\in\V_\rsf} x_v + f_\rsf
    \,,
    &
    f_\rsf 
    &= \sum_{v\in\V_\rsf^{{\color{Orange}\bullet}}} X_v 
        + \sum_{e\in\E_\rsf^{\brEdge{.8}}} Y_e 
        + \sum_{e\in\E_\rsf^{\outEdge{.8}}} Y_e 
        - \sum_{e\in\E_\rsf^{\inEdge{.8}}} Y_e 
    \,,
\end{align}
where $\E_\rsf^{\brEdge{.8}}$, $\E_\rsf^{\outEdge{.8}}$ and $\E_\rsf^{\inEdge{.8}}$ are the broken, incoming and outgoing edges attached to the vertices of $\rsf$.
In this way, the regions determine the \emph{letters} $f_\rsf$ that appear in the differential equation \cite{Glew:2025ypb}.

In summary, the regions of $\g$ partition the vertices of $\G$. 
Using this partition of vertices, we decompose the space $(x_1, \cdots, x_{|\V_\g|}) \in \mathbb{C}^{|\V_\g|}$ into a Cartesian product of the spaces associated with each region $\rsf$, namely $\mathbb{C}_\rsf := \{(x_{v_1},x_{v_2}, \dots,x_{v_{|\rsf}|}) \in \mathbb{C}^{|\rsf|} : v_{i=1,\dots,|\rsf|} \in \rsf \}$. 
The on-shell conditions take the form $\{ \partial\rsf=0 : \rsf \in \R_\g \}$, as in equation \eqref{eq:partialRg},  
and each hyperplane $\partial \rsf = 0$ bounds an $(|\rsf|{-}1)$-dimensional simplex in the negative orthant of the region space $\mathbb{C}_\rsf$ (for physical kinematics $X_\bullet,Y_\bullet\geq0$). 
We denote the corresponding simplex by $\Delta_\rsf$.
Taking the Cartesian product of these region simplices produces the bounded chamber on the cut $M_\g$: 
\be\label{eq:Delta_g}
    \text{${\Delta_\g} :=$ $ {\bigtimes}_{\rsf \in \Rsf_\g} \Delta_\rsf$}
\ee
The bounded chamber constructed in this way is guaranteed to be a polytope, and therefore, a positive geometry (since all polytopes are positive geometries). 
All physical (co)cycles are constructed starting from these positive geometries.

\subsection{The two-chain (part 1): a pedagogical running example}
\label{sec:2chainInto}

The two-site chain, 
\be
    \G_{\text{2-chain}} = \begin{gathered}
        \begin{tikzpicture}
            \node[label={180:$1$}] (X1) at (0,0) {};
            \node[label={0:$2$}] (X2) at (1,0) {};
            \draw[thick] (0,0) -- (1,0);
            \fill[black] (X1) circle (2pt);
            \fill[black] (X2) circle (2pt);
        \end{tikzpicture}
    \,,
    \end{gathered}
\ee
serves as the running pedagogical example throughout the paper.
It first appears here (section~\ref{sec:2chainInto}), where we build the basis of decorated graphs, regions, and cut contours. 
It is continued in section~\ref{sec:2chainCoaction_full}, where we compute the full coaction of a 2-chain period using the general rules of section~\ref{sec:quickStart}. 
Optional follow-ups appear in section~\ref{sec:cohomDets}, where the (co)homological derivations of the formulas used here are spelled out.

The FRW integral associated with the two-site chain is
\begin{equation} \label{eq:complement_space_two_site}
    \psi_{\G_{\text{2-chain}}} = \int_0^\infty 
    u_{ \includegraphics[align=b, scale=.7]{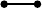} } \;
    \vphi_{ \includegraphics[align=b, scale=.7]{figs/2chainNoLabels.pdf} } 
    \,,
    \qquad
    \vphi_{ \includegraphics[align=b, scale=.7]{figs/2chainNoLabels.pdf} } 
    = \frac{
        \d^2\bs{x}
    }{
        B_{\includegraphics[align=c, scale=.4]{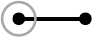}}
        B_{\includegraphics[align=c, scale=.4]{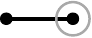}}
        B_{\includegraphics[align=c, scale=.4]{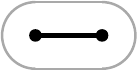}}
    }
    \,,
\end{equation}
where the relevant tube polynomials are
\begin{equation} \label{eq:2chainBs}
    \begin{aligned}
        & B_{\includegraphics[align=c, scale=.4]{figs/2ChainB1.pdf}} 
        = x_1 + X_1 + Y_{12} 
        \,, \hspace{.8em} 
        B_{\includegraphics[align=c, scale=.4]{figs/2ChainB2.pdf}}  
        = x_2 + X_2 + Y_{12}
        \,, \hspace{.8em} 
        B_{\includegraphics[align=c, scale=.4]{figs/2ChainB12.pdf}} 
        = x_1 + x_2 + X_1 + X_2  
        \,.
    \end{aligned}
\end{equation} 
These polynomials define the on-shell variety $\B$ for this graph. Simultaneously, the twisted variety $\T$ is generated by the set of coordinate hyperplanes $\{x_i=0\}$. Graphically, these two varieties can be depicted as
\begin{align}\label{eq:2chainTvarBvar}
    \mathcal{T} &= \includegraphics[align=c,scale=.7]{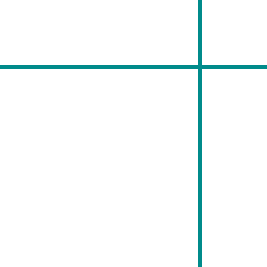}
    \,,
    &
    \mathcal{B} &= \includegraphics[align=c,scale=.7]{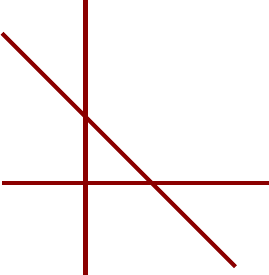}
    \,.
\end{align}
We think of the shifted flat space wavefunction coefficient 
$\vphi_{ \includegraphics[align=b, scale=.7]{figs/2chainNoLabels.pdf} }$
as a differential form on the complement of $\B$ and $\T$ in $\mathbb{C}^2$:
\be \label{eq:2chainTopSpace}
    \mathbb{C}^2 \setminus (\mathcal{T}\cup\mathcal{\B}) =
    \includegraphics[align=c,scale=.8]{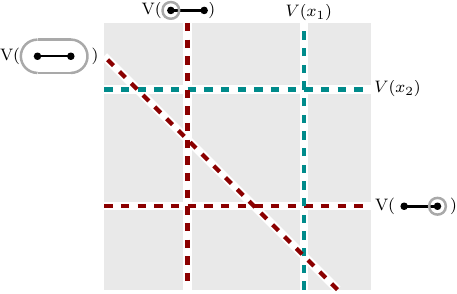}
    \,,
\ee
where we have drawn $\B$ and $\T$ as dashed lines to emphasize these varieties have been removed. 
The derivatives/discontinuities of the FRW integral defined by the geometry~\eqref{eq:2chainTvarBvar} also correspond to forms/contours that live in this space.

To construct the elements that will enter the graphical coaction of $\psi_{\G_{\text{2-chain}}}$, we consider the four ways we can decorate this graph to get an acyclic minor: 
\begin{align}\label{eq:2chaing}
    \g_1 &= \includegraphics[scale=.7,align=c]{figs/2chain_gs}
    \,,
    &
    \g_2 &= \includegraphics[scale=.7,align=c]{figs/2chain_gr}
    \,,
    &
    \g_3 &= \includegraphics[scale=.7,align=c]{figs/2chain_gl}
    \,,
    &
    \g_4 &= \includegraphics[scale=.7,align=c]{figs/2chain_gb}
    \,.
\end{align}
In this example, each acyclic minor gives rise to just a single cut tubing:
\begin{align}\label{eq:2chainCg}
    \Csf_{\g_1} &= \includegraphics[align=c,scale=.7]{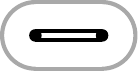}
    \,,
    &
    \Csf_{\g_2} &= \includegraphics[align=c,scale=.7]{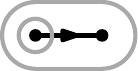}
    \,,
    &
    \Csf_{\g_3} &= \includegraphics[align=c,scale=.7]{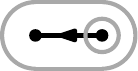}
    \,,
    &
    \Csf_{\g_4} &= \includegraphics[align=c,scale=.7]{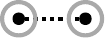}
    \,.
\end{align}
If we first consider the cut tubing for $\g_1$, we see that it involves just a single tube, so the corresponding cut space must have codimension one; in fact, it is given by a line with two points removed, as seen here:
\be
    M_{\includegraphics[scale=.5]{figs/2chain_gs}}
    &= \underset{M}{\underbrace{ 
            \includegraphics[scale=1,align=c]{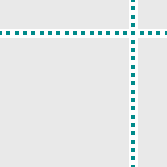} 
        }}
        \quad\bigcap\quad
        \underset{
            \B_{\includegraphics[scale=.3]{figs/2chain_gs}}
        }{\underbrace{ 
            \includegraphics[scale=1,align=c]{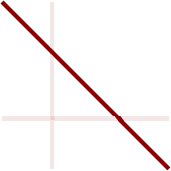}
        }}
    = \includegraphics[scale=1,align=c]{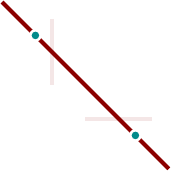}
\ee
Conversely, the cut space identified by the other acyclic minors---whose cut tubings each involve two tubes---are points:
\be
    M_{\twoChaingg}
    &= \includegraphics[align=c,scale=.8]{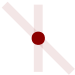}
    \,,
    &
    M_{\twoChainggg}
    &= \includegraphics[align=c,scale=.8]{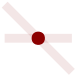}
    \,,
    &
    M_{\twoChaingggg}
    &= \includegraphics[align=c,scale=.8]{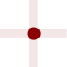}
    \,.
\ee
We already see that each acyclic graph identifies a single bounded region. More generally, though, we can identify these bounded regions by identifying the regions of each acyclic graph, as described in section~\ref{sec:cutGeom}:
\begin{align}\label{eq:2chainRg}
    \Rsf_{\twoChaing} &= \left\{ 
        \includegraphics[align=c,scale=.7]{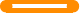}
    \right\}
    \,,
    &
    \Rsf_{\twoChaingg} &= \left\{ 
        \includegraphics[align=c,scale=.7]{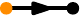}
        ,\includegraphics[align=c,scale=.7]{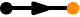}
    \right\}
    \,,
    \\
    \Rsf_{\twoChainggg} &= \left\{ 
        \includegraphics[align=c,scale=.7]{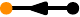}
        ,\includegraphics[align=c,scale=.7]{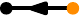}
    \right\}
    \,,
    &
    \Rsf_{\twoChaingggg} &= \left\{ 
        \includegraphics[align=c,scale=.7]{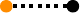}
        ,\includegraphics[align=c,scale=.7]{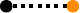}
    \right\}
    \,.
\end{align}
In terms of these regions, the relevant bounded chamber $\Delta_\g$ (in {\color{Orange} orange}) can be computed as the Cartesian product of all region-simplices $\Delta_\rsf$ associated to $\g$:
\be\label{eq:2chainDeltag}
    \Delta_{\twoChaing} 
    &= \Delta_{\includegraphics[align=c,scale=.5]{figs/2chain_Rs1}}
    = \includegraphics[align=c,scale=.8]{figs/2chainM12.pdf}
    \,,
    &
    \Delta_{\twoChaingg} 
    &= \Delta_{\includegraphics[align=c,scale=.5]{figs/2chain_Rr1}}
    \times \Delta_{\includegraphics[align=c,scale=.5]{figs/2chain_Rr2}}
    = \includegraphics[align=c,scale=.8]{figs/2ChainM12_1}
    \,,
    \\
    \Delta_{\twoChainggg} 
    &= \Delta_{\includegraphics[align=c,scale=.5]{figs/2chain_Rl1}}
    \times \Delta_{\includegraphics[align=c,scale=.5]{figs/2chain_Rl2}}
    = \includegraphics[align=c,scale=.8]{figs/2ChainM12_2}
    \,,
    &
    \Delta_{\twoChaingggg} 
    &= \Delta_{\includegraphics[align=c,scale=.5]{figs/2chain_Rb1}}
    \times \Delta_{\includegraphics[align=c,scale=.5]{figs/2chain_Rb2}}
    = \includegraphics[align=c,scale=.8]{figs/2ChainM1_2}
    \,.
\ee
These bounded regions constitute essential building blocks of the FRW homology, as we will describe in section \ref{sec:hom}.
Roughly speaking, the cycles $\gamma_\g$ that appear in this homology correspond to wrapping $\Delta_\g$ in a set of infinitesimal ``residue''-like tubes. 
For the two-site chain, these contours (derived in section~\ref{sec:hom}) can be depicted as follows:
\be
    \label{eq:2chain_contours}
    \allowdisplaybreaks
    \gamma_{\twoChaing}
    &= \delta_{\includegraphics[align=c, scale=.3]{figs/2ChainB12.pdf}}
    \left( \includegraphics[align=c,scale=.6]{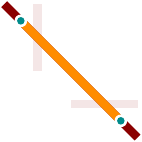} \right) 
    &\qquad\qquad
    \gamma_{\twoChaingg}
    &= \delta_{\includegraphics[align=c, scale=.3]{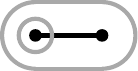}}
    \left( \includegraphics[align=c,scale=.6]{figs/2chainM12_1.pdf} \right) 
    \\
    &= \includegraphics[align=c,scale=.6]{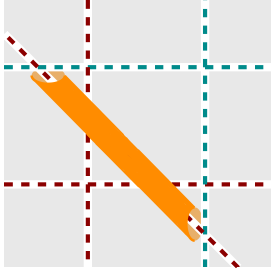}
    \,,
    &\qquad
    &= \includegraphics[align=c,scale=.6]{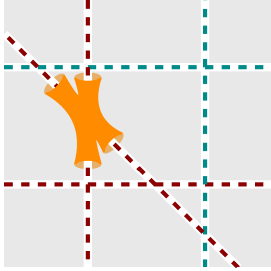}  \,,
    \\[1em]
    \gamma_{\twoChainggg}
    &= \delta_{\includegraphics[align=c, scale=.3]{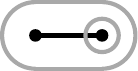}}
    \left( \includegraphics[align=c,scale=.6]{figs/2ChainM12_2} \right) 
    &\qquad\qquad
    \gamma_{\twoChaingggg}
    &= \delta_{\includegraphics[align=c, scale=.3]{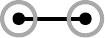}}
    \left( \includegraphics[align=c,scale=.6]{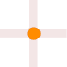} \right) 
    \\
    &= \includegraphics[align=c,scale=.6]{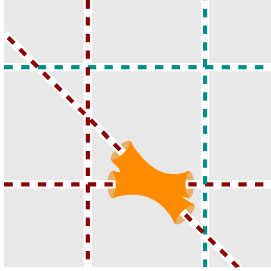}
    \,,
    &\qquad\qquad
    &= \includegraphics[align=c,scale=.6]{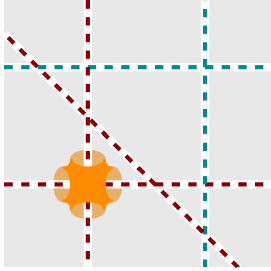}
    \,.     
\ee
where $\delta_{\csf} := \delta_{\tau_{|\csf|}} \circ \cdots \circ \delta_{\tau_1}$ is Leray's iterated coboundary. 
Namely, each $\delta_{\tau}$ represents a tube that encircles the locus $B_\tau=0$ and---when inserted in an integration contour---corresponds to computing a (sequential) residue of the integrand:
\be
    \int_{\delta_{\csf}(\Delta_\g)} u_{ \includegraphics[align=b, scale=.7]{figs/2chainNoLabels.pdf} }\; \vphi_{\h}
    = \int_{\Delta_\g} 
    \res_{\csf}[u_{ \includegraphics[align=b, scale=.7]{figs/2chainNoLabels.pdf} }\; \vphi_{\h}]
    \,.
\ee

Next, we would like to associate a set of differential forms with the minors~\eqref{eq:2chaing}. In section \ref{sec:FRWquickStart}, we provide a general formula for the form we associate with each acyclic minors; for now,  we simply quote the two-site chain result:
\be \label{eq:vphi_2chain}
    \vphi_{\twoChaing}
    &= \dlog B_{\includegraphics[scale=.3]{figs/2chainB12}}
    \wedge \dlog \frac{x_2}{x_1}
    \,,
    \\
    \vphi_{\twoChaingg}
    &= \dlog B_{\includegraphics[scale=.3]{figs/2chainB1}}
    \wedge \dlog B_{\includegraphics[scale=.3]{figs/2chainB12}}
    = - \dlog B_{\includegraphics[scale=.3]{figs/2chainB12}}
    \wedge \dlog B_{\includegraphics[scale=.3]{figs/2chainB1}}
    \,,
    \\
    \vphi_{\twoChainggg}
    &= \dlog B_{\includegraphics[scale=.3]{figs/2chainB12}}
    \wedge \dlog B_{\includegraphics[scale=.3]{figs/2chainB2}}
    \,,
    \\
    \vphi_{\twoChaingggg}
    &= \dlog B_{\includegraphics[scale=.3]{figs/2chainB1}}
    \wedge \dlog B_{\includegraphics[scale=.3]{figs/2chainB2}}
    \,.
\ee
These forms pair nicely with the cut/residue contours in \eqref{eq:2chain_contours};
in fact, their maximal cuts precisely correspond to the canonical forms of the bounded chambers $\Delta_\g$, in the sense that
\be \label{eq:maxCutOf2chain}
    \res_\g[\vphi_\g] = \Omega[\Delta_\g]
    \,.
\ee
These forms are the canonical forms of dual relative cycles $\check{\gamma}_\g$, which correspond to unbounded chambers with $\Delta_\g$ on their maximal boundary. Using the properties of the canonical form map this translates into equation \eqref{eq:maxCutOf2chain}. 
Furthermore, because the \emph{homological} intersection matrix  $[\gamma_\g\vert\check{\gamma}_\h] \propto \delta_{\g\h}$ is  diagonal when paired with the cut/residue contours, it follows that the cohomology intersection matrix in the coaction is also diagonal. 




In the physical region, namely $X_i > Y_{12} > 0$, we can compute the period integral that arises for each pairing of one of the contours $\gamma_{\g_a}$ with the form $\vphi_{\g_b}$. The corresponding twisted period matrix is 
\begin{align} \label{eq:2chainPer}
    P_{\g_a \g_b} := \int_{\gamma_{\g_a}} u\; \vphi_{\g_b}
    &= \begin{pmatrix}
        \frac{\Gamma(\alpha_1) \Gamma(\alpha_2) }{\Gamma (\alpha_1+\alpha_2)}
        f_{\includegraphics[scale=.4]{figs/2chain_Rs1}}^{\alpha_1+\alpha_2}
        & P_{\g_1 \g_2}
        & P_{\g_1 \g_3}
        & 
    \\
        {}
        & f_{\includegraphics[scale=.4]{figs/2chain_Rr1}}^{\alpha_1}
            f_{\includegraphics[scale=.4]{figs/2chain_Rr2}}^{\alpha_2}
        & 
        & 
    \\
        {}
        & 
        & f_{\includegraphics[scale=.4]{figs/2chain_Rl1}}^{\alpha_1}
            f_{\includegraphics[scale=.4]{figs/2chain_Rl2}}^{\alpha_2}
        & 
    \\
        {}
        & 
        & 
        & f_{\includegraphics[scale=.4]{figs/2chain_Rb1}}^{\alpha_1}
            f_{\includegraphics[scale=.4]{figs/2chain_Rb2}}^{\alpha_2}
    \end{pmatrix}
\end{align}
where all omitted entries are zero, and
\begin{align}
    \frac{ P_{\g_1 \g_2} }{ P_{\g_1 \g_1} }
    &= 
    \frac{
        \alpha_1\alpha_2 
        \Gamma(\alpha_1{+}\alpha_2)
    }{
        \Gamma(2{+}\alpha_1{+}\alpha_2)
    } 
    \frac{
        f_{\includegraphics[scale=.4]{figs/2chain_Rs1}}
    }{
        f_{\includegraphics[scale=.4]{figs/2chain_Rr1}}
    }
    \; 
    {{}_2{F}_1}\left(
        1,
        \alpha_1{+}1;
       \alpha_1{+}\alpha_2{+}2;
        \frac{
            f_{\includegraphics[scale=.4]{figs/2chain_Rs1}}
        }{
            f_{\includegraphics[scale=.4]{figs/2chain_Rr1}}
        }
    \right)
    \,,
    \nn\\&
    = \frac{
        \alpha_1\alpha_2 
        \Gamma(\alpha_1{+}\alpha_2)
    }{
        \Gamma(2{+}\alpha_1{+}\alpha_2)
    } 
    \frac{
        f_{\includegraphics[scale=.4]{figs/2chain_Rs1}}
    }{
        f_{\includegraphics[scale=.4]{figs/2chain_Rr2}}
    }
    \;
    {{}_2{F}_1}\left(
        1,
        \alpha_2{+}1;
        \alpha_1{+}\alpha_2{+}2;
        \frac{
            f_{\includegraphics[scale=.4]{figs/2chain_Rs1}}
        }{
            f_{\includegraphics[scale=.4]{figs/2chain_Rr2}}
        }
    \right)
    \,,
    \\
    \frac{ P_{\g_1 \g_3} }{ P_{\g_1 \g_1} }
    &= 
    \frac{
        \alpha_1\alpha_2 
        \Gamma(\alpha_1{+}\alpha_2)
    }{
        \Gamma(2{+}\alpha_1{+}\alpha_2)
    }
    \frac{
            f_{\includegraphics[scale=.4]{figs/2chain_Rs1}}
        }{
            f_{\includegraphics[scale=.4]{figs/2chain_Rl2}}
        }
    \; 
    {{}_2F_1}\left(
        1,
        \alpha_1{+}1;
       \alpha_1{+}\alpha_2{+}2;
        \frac{
            f_{\includegraphics[scale=.4]{figs/2chain_Rs1}}
        }{
            f_{\includegraphics[scale=.4]{figs/2chain_Rl2}}
        }
    \right)
    \,,
    \nn\\&
    = \frac{
        \alpha_1\alpha_2 
        \Gamma(\alpha_1{+}\alpha_2)
    }{
        \Gamma(2{+}\alpha_1{+}\alpha_2)
    } 
    \frac{
        f_{\includegraphics[scale=.4]{figs/2chain_Rs1}}
    }{
        f_{\includegraphics[scale=.4]{figs/2chain_Rl1}}
    }
    \;
    {{}_2{F}_1}\left(
        1,
        \alpha_2{+}1;
        \alpha_1{+}\alpha_2{+}2;
        \frac{
            f_{\includegraphics[scale=.4]{figs/2chain_Rs1}}
        }{
            f_{\includegraphics[scale=.4]{figs/2chain_Rl1}}
        }
    \right)
    \,.
\end{align}
Here, the $f_\rsf$ are the letters that appear in the differential equations for the two-site graph, given by \eqref{eq:partialRg}. Explicitly:
\be
    f_{\includegraphics[scale=.4]{figs/2chain_Rs1}}
    &= X_1 + X_2  
    \,,
    &
    f_{\includegraphics[scale=.4]{figs/2chain_Rr1}}
    &= X_1 + Y_{12}
    = f_{\includegraphics[scale=.4]{figs/2chain_Rb1}}
    \,,
    &
    f_{\includegraphics[scale=.4]{figs/2chain_Rr2}}
    &= X_2 - Y_{12}
    \,,
    \\
    f_{\includegraphics[scale=.4]{figs/2chain_Rl1}}
    &= X_1 - Y_{12}
    \,,
    &
    f_{\includegraphics[scale=.4]{figs/2chain_Rl2}}
    &= X_2 + Y_{12}
    = f_{\includegraphics[scale=.4]{figs/2chain_Rb2}}
    \,.
\ee
Note also that $P_{\g_1 \g_3} = - P_{\g_1 \g_2}\vert_{\alpha_1 \leftrightarrow \alpha_2, X_1 \leftrightarrow X_2}$, while the ratio $P_{\g_1\g_{i}}/P_{\g_1\g_1}$ expands into weight zero functions if the transcendental weight of the expansion parameters $\alpha_i$ is taken to be -1.\footnote{%
    For readers unfamiliar with the term, transcendental weight is introduced in the next section.
}

The period matrix \eqref{eq:2chainPer} has a block diagonal decomposition 
\begin{align} \label{eq:p_two_chain}
    \mat{P}_{\text{2-chain}}
    = \mat{P}[\mathcal{Z}_{\twoChaing}] 
    \oplus \mat{P}[\mathcal{Z}_{\twoChaingggg}]
    \,,
\end{align}
where 
\begin{align}
    \mat{P}[\mathcal{Z}_{\twoChaing}] &: = 
    \begin{pmatrix}
        \frac{\Gamma(\alpha_1) \Gamma(\alpha_2) }{\Gamma (\alpha_1+\alpha_2)}
        f_{\includegraphics[scale=.4]{figs/2chain_Rs1}}^{\alpha_1+\alpha_2}
        & P_{\g_1 \g_2}
        & P_{\g_1 \g_3}
        & 0
    \\
        0
        & f_{\includegraphics[scale=.4]{figs/2chain_Rr1}}^{\alpha_1}
            f_{\includegraphics[scale=.4]{figs/2chain_Rr2}}^{\alpha_2}
        & 0
        & 0
    \\
        0
        & 0
        & f_{\includegraphics[scale=.4]{figs/2chain_Rl1}}^{\alpha_1}
            f_{\includegraphics[scale=.4]{figs/2chain_Rl2}}^{\alpha_2}
        & 0
    \\
        0
        & 0
        & 0
        & 0
    \end{pmatrix}
    \,,
    \\
    \mat{P}[\mathcal{Z}_{\twoChaingggg}] &: = 
    \begin{pmatrix}
        0
        & 0
        & 0
        & 0
    \\
        0
        & 0
        & 0
        & 0
    \\
        0
        & 0
        & 0
        & 0
    \\
        0
        & 0
        & 0
        & f_{\includegraphics[scale=.4]{figs/2chain_Rb1}}^{\alpha_1}
            f_{\includegraphics[scale=.4]{figs/2chain_Rb2}}^{\alpha_2}
    \end{pmatrix}
    \,.
\end{align}
While this direct sum decomposition looks rather trivial in the case of the two-site chain graph, it generalizes and becomes more dramatic for more complicated graphs. 
This block triangular decomposition of the period matrix is also inherited by the differential equations for these periods. 
Since the periods associated to different blocks cannot interact, the periods and differential equations of each block can be studied independently of all other blocks.

\begin{figure}
    \centering
    \includegraphics[width=0.5\linewidth]{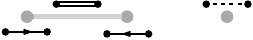}
    \caption{%
        The Zonotopes for the 2-chain graph: $\mathcal{Z}_{\twoChaing}$ (left) and $\mathcal{Z}_{\twoChaingggg}$ (right). 
    }
    \label{fig:2chainZonotope}
\end{figure}

The block decomposition of the period matrices corresponds to a certain combinatorial structure, specifically graphical zonotopes, in the on-shell variety $\B$.
In general, the residues of $\vphi_\G$ are canonical forms of graphical zonotopes (which will be discussed further in section \ref{sec:partialFracAndZono}), where each face of the zonotope corresponds to a basis element of the physical (co)homology.
For the two-site chain, we have two graphical zonotopes, labeled by their highest-dimensional face, $\mathcal{Z}_{\twoChaing}$ and $\mathcal{Z}_{\twoChaingggg}$, as shown in figure \ref{fig:2chainZonotope}.

It is important to highlight that the physical FRW form $\vphi_{ \includegraphics[align=b, scale=.7]{figs/2chainNoLabels.pdf} }$ and the physical contour $\gamma_{ \includegraphics[align=b, scale=.7]{figs/2chainNoLabels.pdf} } = (x_1,x_2) \in \mathbb{R}^2_+$ do not appear directly in our cohomology or homology basis. 
Instead, they are linear combinations of our basis elements. 
The physical FRW form is
\be \label{eq:2chainPartialFraction}
    \vphi_{ \includegraphics[align=b, scale=.7]{figs/2chainNoLabels.pdf} }
    = \vphi_{\twoChaingg} + \vphi_{\twoChainggg} 
    - \vphi_{\twoChaingggg}
    \,.
\ee
while the physical contour is
\be \label{eq:2chainContourDecomp}
    \gamma_{\includegraphics[align=b, scale=.7]{figs/2chainNoLabels.pdf} }
    &= \frac{
        -\pi e^{- \pi i (\alpha_1{+}\alpha_2)} 
    }{
        \sin\big(\pi (\alpha_1{+}\alpha_2) \big)
    }
    \gamma_{\g_1}    
    + \frac{
        \pi^2 e^{- \pi i \alpha_1} 
        }{
            \sin\big(\pi \alpha_2 \big) 
            \sin\big(\pi (\alpha_1{+}\alpha_2) \big) 
        }
        \gamma_{\g_2}
    \\& 
    + \frac{
        \pi^2 e^{- \pi i \alpha_2}  
    }{
        \sin(\pi \alpha_1) 
        \sin\big(\pi (\alpha_1{+}\alpha_2) \big)
    }
    \gamma_{\g_3}
    + \frac{
        \pi^2 e^{- \pi i (\alpha_1{+}\alpha_2)} 
    }{
        \sin(\pi \alpha_1) \sin(\pi \alpha_2)
    }
    \gamma_{\g_4}
    \,.
\ee
Given equations \eqref{eq:2chainPartialFraction} and \eqref{eq:2chainContourDecomp}, the decomposition of the wavefunction coefficient in terms of the periods is
\be
    \psi_{\includegraphics[align=b, scale=.7]{figs/2chainNoLabels.pdf}} 
    &= \int_{\gamma_{\includegraphics[align=b, scale=.7]{figs/2chainNoLabels.pdf}} } u_{\includegraphics[align=b, scale=.7]{figs/2chainNoLabels.pdf}}\; \vphi_{ \includegraphics[align=b, scale=.7]{figs/2chainNoLabels.pdf} }
    \\
    &= \frac{
            -\pi e^{- \pi i (\alpha_1{+}\alpha_2)} 
        }{
            \sin\big(\pi (\alpha_1{+}\alpha_2) \big)
        }
        (P_{\g_1\g_2}{+}P_{\g_1\g_3})
    + \frac{
            \pi^2 e^{- \pi i \alpha_1} 
        }{
            \sin\big(\pi \alpha_2 \big) 
            \sin\big(\pi (\alpha_1{+}\alpha_2) \big) 
        } 
        P_{\g_2\g_2}
    \\&\quad 
    + \frac{
            \pi^2 e^{- \pi i \alpha_2}  
        }{
            \sin(\pi \alpha_1) 
            \sin\big(\pi (\alpha_1{+}\alpha_2) \big)
        } 
        P_{\g_3\g_3}
    - \frac{
        \pi^2 e^{- \pi i (\alpha_1{+}\alpha_2)} 
    }{
        \sin(\pi \alpha_1) \sin(\pi \alpha_2) 
    } 
    P_{\g_4\g_4}
    \,.
\ee
Once in this form, we can easily apply the graphical coaction (as we will see in section~\ref{sec:quickStart}).

\subsection{Graphical building blocks for any $\G$}
\label{sec:FRWquickStart}

This graphical decomposition of the two-site graph integral is just one instance of a general set of formulas that hold for any graph $\G$.
In this subsection, we collect some of the general formulas that describe the basis of physical homology and cohomology (the FRW cycles $\gamma_\g$ and cocycles $\vphi_\g$), the corresponding periods $P_{\g\h}$, the intersection numbers $C_{\g\h}$ that appear in the coaction, and the decomposition of $\vphi_\G$ in terms of the basis of forms.
We defer the derivations of these formulas to section \ref{sec:cohomDets}.

\paragraph{FRW cycles.}

As argued in section \ref{sec:ResolveDegen}, the degeneracy of the on-shell variety $\B$ introduces relations between residue contours and $\dlog$-forms. 
Due to this, we must consider distinguished linear combinations of residue contours and $\dlog$-forms to find a minimal spanning set (or basis) of cycles that appear in $\vphi_\G$ and its kinematic derivatives. 
In fact, the formula for our chosen basis of residue operators was already presented in equation \eqref{eq:degenRes}. 
Here, we provide the definition of $\sgn_\csf$, which compares the ordering described in box \ref{box:tubeOrd} to the natural lexicographic ordering (for more details, see \cite{Glew:2025ypb}).

To compute $\sgn_\csf$, we define a map  $\la v \ra_\csf$ from each vertex to one of the tubes in the cut tubing; specifically, $\la v \ra_\csf$ is the smallest tube $\tau\in\csf$ that contains the region $\rsf$ which also contains the vertex $v$:%
\be
    \la v \ra_\csf 
    &:= \tau \in \csf 
    \text{ such that }
    v \in \V_\rsf 
    \text{ and } 
    |\V_\rsf| > |\V_{\rsf^\prime}| 
    \;\forall\; \rsf^\prime \in \tau 
    \text{ with } \rsf^\prime \neq \rsf
    \,.
\ee
Note that the map $\la v \ra_\csf : \V_\g \to \csf$ is many-to-one whenever the associated minor $\g$ contains pinched edges; that is, some regions contain more than one vertex. 
If we let $\sigma$ be the permutation that orders the tubes $\tau\in\csf$ according to box \ref{box:tubeOrd}, we have
\begin{align} \label{eq:sgn}
    \sgn_\csf &:= 
    \frac{
        \mathrm{sig}\left(
            \la \bar{\rsf}_1 \ra_\csf, 
            \dots, 
            \la \bar{\rsf}_{|\Rsf_\g|} \ra_\csf
        \right) 
    }{
        \mathrm{sig}\left(
            \sigma\left( \la \bar{\rsf}_{1} \ra_\csf \right), 
            \dots, 
            \sigma\left( \la \bar{\rsf}_{|\Rsf_\g|} \ra_\csf \right)
        \right) 
    }
    \,,
\end{align}
where $\bar{\rsf}_i := \min \V_\rsf$ is minimal vertex of the region $\rsf_i$. This sign can be derived by thinking about the number of pairwise swaps (of linearly independent tubes) that are needed to bring $\csf$ into the chosen global ordering (the swaps among linearly dependent tubes, supplied by \eqref{eq:genResID}, contribute no extra sign). 

Having defined $\sgn_\csf$, we can make the cut contours $\gamma_\g$ and FRW-forms $\vphi_\g$ explicit for any $\g$.
The set of FRW-contours indexed by acyclic decorated graphs $\g$ forms a basis of the physical homology 
\begin{align} \label{eq:FRWContours_quick}
    \gamma_\g 
    &= \delta_{\Csf_\g} ( \Delta_\g )
        \,,
    &
    \delta_{\Csf_\g}
    &= \sum_{\csf \in \Csf_\g}
        \sgn_\csf
        \delta_{\csf}
        \,,
\end{align}
where $\Delta_{\g}:= {\bigtimes}_{\rsf \in \Rsf_\g} \Delta_\rsf$ is the bounded chamber on $M_\g$. 
Here, $\delta_\tau$ should be thought of as the residue contour that localizes to $B_\tau=0$, and 
$\delta_\csf := \delta_{\tau_{|\csf|}} \circ \cdots \circ \delta_{\tau_1}$ with $\tau_i\in\csf$ is the sequential residue contour. 
Thus, for any FRW-form $\vphi$, 
\be
    \int_{\delta_{\Csf_\g}(\Delta_\g) } u_\G\;\vphi
    &:= \int_{\Delta_\g} \res_{\Csf_\g}[u_\g\;\vphi]
    \,,
\ee
where the remaining integral is integrated over the region $\Delta_\g$ defined in section \ref{sec:cutGeom}.
Note that, like sequential residue operators, $\delta_\csf$ depends on the ordering of its indices.

\paragraph{FRW cocycles.}

For a basis of physical cocycles, we choose the canonical forms of special unbounded regions $\check{\Gamma}_\g$ (see section \ref{sec:dualhom} and \ref{sec:FRWCohom}). 
Explicitly, we have \cite{Glew:2025ypb} 
\be \label{eq:Omega[CheckGamma]}
    \vphi_\g := 
    \Omega[\check{\Gamma}_\g]
    = \left[ 
        \sum_{\csf \in \Csf_\g} \sgn_\csf
        \left(\bigwedge_{\tau\in\csf} \dlog B_\tau\right)
    \right]
    \wedge
    \left[
        \bigwedge_{\rsf\in\Rsf_\g : |\V_\g|>1}
        \tilde{\Omega}[\Delta_\rsf]
    \right]
    \,,
\ee
where the $\dlog B_\bullet$'s are ordered according to box \ref{box:tubeOrd}, and 
\be
    \tilde{\Omega}[\Delta_\rsf]
    := \bigwedge_{ v \in \V_\rsf \setminus \bar{\rsf} } 
    \dlog\frac{ x_{v} }{ x_{\bar{\rsf}} } 
    \,.
\ee
Since $\tilde{\Omega}[\Delta_\rsf]$ becomes the canonical form of $\Delta_\rsf$ once restricted to the cut, it is easy to see that the maximal cut of $\vphi_\g$ is the canonical form  $\Omega[\Delta_\g]$ of $\Delta_\g$. 
That is,
\be
    \res_{\Csf_\g}[\vphi_\g]
    = |\Csf_\g| \; \Omega[\Delta_\g]
    \,,
\ee
where 
\be \label{eq:OmegaR}
    \Omega[\Delta_\g] 
    &:= \bigwedge_{\rsf\in\Rsf_\g : |\V_\g|>1}
    \Omega[\Delta_\rsf]
    \,,
    \\
    \Omega[\Delta_\rsf]
    &:= \left.
    \bigwedge_{ v \in \V_\rsf \setminus \bar{\rsf} } 
        \dlog\frac{ x_{v} }{ x_{\bar{\rsf}} } 
    \right\vert_{\Csf_\g}
    = \bigwedge_{ v \in \V_\rsf \setminus \bar{\rsf} } 
        \dlog\frac{ x_{v} }{ f_\rsf +\sum_{v\in\V_\rsf\setminus\bar{\rsf}}x_v} 
    \,.
\ee
Here, $\vert_{\Csf_\g}$ denotes the restriction to the cut $M_\g$. 
Because $\Delta_\g$ is the Cartesian product of region simplices $\Delta_\rsf$, the canonical form of $\Delta_\g$ is the wedge product of the canonical form of the region simplices.
Also, to get an explicit representation for $\Omega[\Delta_\rsf]$, we have used the cut condition $\partial \rsf = 0$ to eliminate $x_{\bar{r}}$ for each region.

\paragraph{FRW periods.}

The set of FRW periods associated to a graph $\G$ are the twisted periods 
\begin{align}
    [ \gamma_\g \vert \vphi_{\h} \ra
    := \int_{\gamma_\g} u_\G \; \vphi_{\h} 
    =: P_{\g\h} 
    \,.
\end{align}
where both $\g$ and $\h$ correspond to acyclic minors of $\G$. In general, up to simple prefactors, these periods evaluate to specialized Aomoto hypergeometric functions 
\be
    F(\bs{\beta};\mat{l}\bs{c})
    :=\int_{\Delta_m} \frac{
        z_1^{\beta_1} \cdots z_m^{\beta_m} 
        (1-z_1-z_2- \cdots - z_m)^{\beta_{m+1}}
    }{
        (\bs{l}_1\cdot \bs{z}-c_1)
        (\bs{l}_2\cdot \bs{z}-c_2)
        \cdots
        (\bs{l}_m\cdot \bs{z} - c_m)
    }
    \d^m\bs{z}
\ee
(or products thereof), where $\Delta_m$ is the standard $m$-simplex and $\bs{l}_i \in \{\pm1,0\}^m$.
Here, the $\beta_i$ are linear combinations of the $\alpha_i$ and $z_i$ are the remaining unfixed $x_i$ after taking all cuts.
Also, $\mat{l}\bs{c}:=\bs{l}_1 \cdots \bs{l}_m \bs{c}$ is a matrix whose columns are the $\bs{l}_i$ and $\bs{c}$.
Only in special cases do these evaluate to named classical hypergeometric functions like the Appell-Lauricella functions.

\paragraph{Cohomology intersection numbers.}

The cohomology intersection matrix is diagonal and evaluates to the following rational functions of the $\alpha_\bullet$ parameters:
\be \label{eq:Cggp_quick}
    C_{\g\h} 
    := \la \check{\vphi}_{\g_1}
    \vert \vphi_{\g_2} \ra 
    = \delta_{\g\h}\;  
        |\Csf_\g| \; 
        \prod_{\rsf \in \Rsf_\g} 
        \frac{
            \sum_{v\in\V_\rsf} \alpha_v
        }{
            \prod_{v\in\V_\rsf} \alpha_v
        }
\ee
where $\delta_{\g\h}$ is the Kronecker delta symbol and $|\Csf_\g|$ is the cardinality of the set $\Csf_\g$.

\paragraph{Partial fractions formula for the wavefunction.}

The physical differential form $\vphi_\G$ has a closed form expansion in terms of the $\vphi_\g$
\begin{align} \label{eq:partialFractionedPsi_quick}
    \vphi_\G &= \frac{1}{\prod_{e\in\E_\G} Y_e}
    \left(\sum_{\substack{
        \g \in \mathrm{aDec}(\G)  
        \\ 
        \text{pinched edges}(\g) = \varnothing
    }} 
    (-1)^{ \#(\text{broken edges of }\g) }
    \vphi_{\g} \right)
    \,,
\end{align}
This form is fully partial-fractioned with respect to the $x_i$ variables (so, each differential form in the sum is projective) \cite{Glew:2025ypb}.

\section{The coaction from twisted (co)homology and intersection theory}
\label{sec:coaction}

As we have already seen, when studying cosmological correlators in perturbation theory (and, similarly, flat-space scattering amplitudes), one encounters twisted period integrals that evaluate to complicated transcendental functions. Often, these integrals are too difficult to evaluate directly, so we are interested in studying their analytic properties in order to systematically exploit these structures in our computations. For example, it is useful to distinguish physical branch points that actually appear in amplitudes/correlators from spurious branch points that are only present in intermediate steps. 
The coaction helps reveal this type of structure, by breaking periods up into their simplest building blocks. 

In simple examples (for instance, one-loop amplitudes in dimensional regularization and pure de Sitter correlators), perturbation theory produces integrals that evaluate to functions called multiple polylogarithms (MPLs). 
This family of functions includes the logarithm, classical polylogarithms, and more general iterated integrals that involve logarithmic integration kernels.
The properties of MPLs are well studied in both the mathematics and physics literature \cite{Chen,G91b,Goncharov:1998kja,Remiddi:1999ew,Borwein:1999js,Moch:2001zr}. 
In particular, the coaction on the space of MPLs is well known \cite{Gonch2,Goncharov:2010jf,Brown:2011ik,Brown1102.1312,Duhr:2012fh,Duhr:2011zq}. For an excellent introductory treatment, see \cite{Duhr:2014woa}. 

More generally, Standard Model amplitudes and the FRW wavefunction coefficients considered in this work can be thought of as hypergeometric functions that depend on the (non-integer) spacetime dimension $d=\mathbb{N}-2\vep$, or on the (non-integer) cosmological parameters $\alpha_\bullet$. 
While hypergeometric functions are not directly related to MPLs, 
the $\alpha_\bullet$ or $\vep$ expansion of hypergeometric functions that arise in physics often have MPL coefficients. 
In fact, the diagrammatic coaction that has been developed for dimensionally-regularized Feynman integrals is compatible with the coaction of MPLs, when the latter is applied order-by-order in the dimensional regularization parameter $\epsilon$~\cite{Abreu:2014cla,Abreu:2017ptx,Abreu:2017enx,Abreu:2017mtm,Brown:2019jng,Abreu:2019wzk,Abreu:2019xep,Abreu:2021vhb,Gardi:2022wro}. 

The properties of the MPL coaction are reviewed in section~\ref{sec:coactionMPLs}. 
Section~\ref{sec:coactionTwPer} constructs the coaction for twisted periods specialized to FRW integrals. 
The two-site chain (section~\ref{sec:2chainCoaction_full}) illustrates this concretely, and section~\ref{sec:quickStart} distills the result into the four-step graphical recipe of box~\ref{box:graphical_rule}---the form used throughout the rest of the paper.




\subsection{The coaction on MPLs}
\label{sec:coactionMPLs}

Multiple polylogarithms are defined as iterated integrals over $\dlog$ integration kernels,
\be \label{eq:MPL_def}
    G(a_1, \dots, a_n; z) 
    := \int_{0}^z G(a_2, \dots, a_n;t)\; \dlog(t-a_1) \, ,
\ee
with the base case $G(;z)=1$. 
Here, $(a_1, \dots ,a_n)  = \bs{a}\in\mathbb{C}^n$ is a vector of constants or parameters, while $z$ is a complex variable. Note that care is needed when the last $m$ entries of $\bs{a}$ vanish; to regulate these divergent cases, the definition $G(\bs{0}_n;z) := \frac{1}{n!} \log^n(z)$ is usually adopted. 

The vector space of MPLs, which we denote by $\mathcal{A}$, comes equipped with a grading referred to as \emph{transcendental weight} (or just \emph{weight}), which corresponds to the length of the vector $\bs{a}$. 
This notion of weight provides us with an important organizing principle, insofar as (conjecturally) there are no relations between MPLs of different weights.
The space of MPLs also forms an algebra, via which the product of weight-$w_1$ and weight-$w_2$ MPLs can always be expressed as linear combinations of weight-$(w_1{+}w_2)$ MPLs (see~\cite{Duhr:2014woa} for more details).

Clearly, the weight of an MPL is tied to the number of integrations that enter its definition; consequently, the derivative of a weight-$w$ MPL is expressible as a linear combination of weight-$(w{-}1)$ MPLs. 
Schematically, this means the total differential of a weight-$w$ MPL $g^{(w)}$ takes the form
\be \label{eq:mpl_deriv}
    \d g^{(w)}
    = 
    \sum_{i} f^{(w-1)}_i\; \d\log (s_i)
    \,,
\ee
where the $\smash{f^{(w-1)}_i}$ are MPLs of weight $w{-}1$, and the arguments $s_i$ are rational functions of the argument $z$ and the indices $a_j$ that appear in the definition~\eqref{eq:MPL_def}. In other words, the derivative lowers the weight of an MPL by one.

The \textit{coaction} can be thought of as taking~\eqref{eq:mpl_deriv} a step further. While the total differential splits off individual $\d\log$ factors from the rest of the integrations in $g^{(w)}$, the coaction splits up the full set of (ordered) $\d\log$ integrations into two sets in all possible ways. Schematically, this takes the form:
\be 
    \Delta g^{(w)} = \sum_{w'=0}^w \sum_{i,j} f_i^{(w')} \otimes f_j^{(w-w')}
    \,.
\ee
In simple cases, such as for powers of logarithms 
\be
    \log^n(z) = n! \, G({\underset{n}{\underbrace{0,\cdots,0}}};z)
    \,
\ee 
or the classical polylogarithms
\be
    \mathrm{Li}_n(z) = - G(\underset{n{-}1}{\underbrace{0,\cdots,0}},1;z)
    \,,
\ee 
the coaction can be written in a simple closed form:  
\be \label{eq:DeltaClassicalPL}
    \Delta\Big(\log^n(z)\Big) &= \sum_{k=0}^n \binom{n}{k} \log^{n-k}(z) \otimes \log^k(z)
    \,,
    \\    
    \Delta\Big(\mathrm{Li}_n(z)\Big) &= 1 \otimes \mathrm{Li}_n(z) + \sum_{k=0}^{n-1} \mathrm{Li}_{n-k} \otimes \frac{\log^k(z)}{k!}    
    \,.
\ee
For a general formula for the coaction on MPLs, see~\cite{Duhr:2014woa}. In general, one can think of the functions that appear in the left factor as coming from carrying out the first $w'$ integrations in the definition of the original function, while the functions that appear in the right factor come from carrying out the remaining $w{-}w'$ integrations. 
The only subtlety is that---since the functions that appear in the left and right factors are formally defined in terms of pairings involving different types of homology---the right factor should be interpreted modulo factors of $i \pi$. 
Thus, constants such as $\zeta_2 = (i\pi)^2/6$, which naturally appear as special limits of MPLs, must be dropped any time they appear in the right entry.


The utility of the coaction largely comes from the fact that it cleanly separates information about the discontinuities and the derivatives of MPLs. Namely, discontinuities only act on the left entry, while derivatives only act on the right entry:
\begin{align} 
    \Delta \circ \d &= (\id \otimes \d) \circ \Delta
    \,, \label{eq:diffCoaction} \\
    \Delta \circ \disc &= (\disc \otimes \id) \circ \Delta
    \,. \label{eq:DiscCoaction}
\end{align}
This remains true even if we apply the coaction multiple times, to break an MPL into more than two parts; derivatives only act on the rightmost entry, while discontinuities only act on the leftmost one. In fact, if we apply the coaction a maximal number of times to break down a weight-$w$ MPL into a $w$-fold tensor product populated by logarithms, we arrive at the \emph{symbol} of an MPL~\cite{Goncharov:2010jf}. The symbol proves particularly useful for understanding the analytic structure of multiple polylogarithms, because (modulo boundary constants proportional to higher-weight transcendental constants), it faithfully exposes all of the logarithmic branch cuts that are encoded in the function (including where these branch cuts live in the global Riemann sheet structure). By virtue of keeping track of all these branch cuts, it also faithfully manifests all functional identities between polylogarithms (which now reduce to simple logarithmic identities). For this reason, it is common to work with the symbol of MPLs directly, and only re-upgrade them to full functions at the end of a calculation (see for instance~\cite{Caron-Huot:2020bkp}).

More generally---in the context of not only MPLs, but also the more general classes of functions that appear in FRW integrals---the coaction has a number of properties that make it useful to work with. First, it is coassociative, meaning that $(\Delta \otimes \id)\circ\Delta = (\id \otimes \Delta)\circ\Delta$. This ensures that its repeated application always gives rise to the same result, and correspondingly that there exists a unique way to split an element into three or more pieces of a given set of weights.  
Additionally, the tensor product is bilinear in its arguments:
\be
    (a + b) \otimes c
    = a \otimes c + b \otimes c
    \,,
    \qquad
    a \otimes (b + c) 
    = a \otimes b + a \otimes c
    \,,
\ee
for $a,b,c \in \mathcal{A}$, while any non-transcendental $\mathbb{Q}$-number $K$ can be brought out of the tensor product:
\be
    (K a) \otimes b = a \otimes (K b) = K ( a \otimes b)  
    \,.
\ee
Lastly, multiplication and comultiplication on $\mathcal{A}$ are compatible, in the sense that for functions $f$ and $g$ with $\Delta(f) = \sum_{i} a_i \otimes b_i$ and $\Delta(g) = \sum_{j} c_j \otimes d_j$, one has
\be
    \Delta(f) \cdot \Delta(g) = \Delta(f\cdot g), 
    \text{ with } \Delta(f) \cdot \Delta(g) := \sum_{i,j} a_i c_j \otimes b_i d_j, \\
\ee
All of this guarantees that the object we get by acting with the coaction (possibly more than once) is well-defined, and faithfully encodes the salient analytic properties of the original function. 

\subsection{The coaction on twisted periods}
\label{sec:coactionTwPer}

FRW integrals only evaluate to MPLs in special limits, such as when $\varepsilon \to 0$. More generally, these integrals evaluate to generalized Hypergeometric functions of the type shown in~\eqref{eq:generalized_hypergeometrics}. However, a coaction can also be constructed for this class of Hypergeometric functions, which reduces to the MPL coaction in the appropriate limits. To construct this coaction for general FRW integrals, we make use of relative/partially twisted (co)homology.

\begin{figure}
\begin{equation*}
    \begin{tikzcd}
        \substack{
            \text{ \normalsize Relative twisted cohomology }
        \\
            \mathbin{\rotatebox[origin=c]{90}{$\in$}}
        \\
            { 
                \color{\myblue} 
                \la\check{\vphi}_\g := \check{\Omega}[\gamma_\g] \vert 
            }
        }
        \arrow[r, leftrightarrow, Purple, "{\color{Purple}\la\bullet\vert\bullet\ra}"]
        & 
        \substack{
            \text{\normalsize Partially twisted cohomology }
        \\
            \mathbin{\rotatebox[origin=c]{90}{$\in$}}
        \\
            {
                \color{\myred} 
                \vert 
                \vphi_\g:= \Omega[\check{\gamma}_\g]
                \ra
            }
        }
        \\[1em]
        \substack{
            {\vert\check{\gamma}_\g]}
        \\
            \mathbin{\rotatebox[origin=c]{-90}{$\in$}}
        \\
            \text{ \normalsize Relative twisted homology }
        }
        \arrow[r, leftrightarrow, lightgray, "{\color{lightgray}[\bullet\vert\bullet]}", swap]
        \arrow[u, leftrightarrow, lightgray, "{\color{lightgray}\la\bullet\vert\bullet]}"]
        \arrow[ur, BrickRed, swap, "\Omega" near end]
        & 
        \substack{
            {[\gamma_\g\vert}
        \\
            \mathbin{\rotatebox[origin=c]{-90}{$\in$}}
        \\
            \text{\normalsize Partially twisted homology }
        }
        \arrow[u, leftrightarrow, "{[\bullet\vert\bullet\ra}", swap]
        \arrow[ul, crossing over, MidnightBlue, "\check{\Omega}" near end]
    \end{tikzcd}
\end{equation*}
\caption{%
    \label{fig:pairings}
    Relationships between the partially twisted (co)homology and the dual relative twisted (co)homology. 
    The pairings relevant to the coaction are emphasized in color (while those less relevant are de-emphasized in {\color{gray}gray}).
    The vertical pairings $[\bullet\vert\bullet\ra$ and $\color{lightgray}\la\bullet\vert\bullet]$ are simply integration, and result in twisted periods (cosmological integrals). 
    The horizontal pairings are the cohomology and homology intersection numbers $\color{Purple}\la \bullet \vert \bullet \ra$ and $\color{lightgray}[\bullet\vert\bullet]$. 
    These pairings function as inner products on the twisted (co)homology vector spaces. 
    The diagonal maps ${\color{MidnightBlue}\check{\Omega}}$ and ${\color{BrickRed}{\Omega}}$ are the (dual) canonical form maps. 
    These are essentially a realization of Poincar\'e duality, which is necessary for formulating the coaction. 
}
\end{figure}
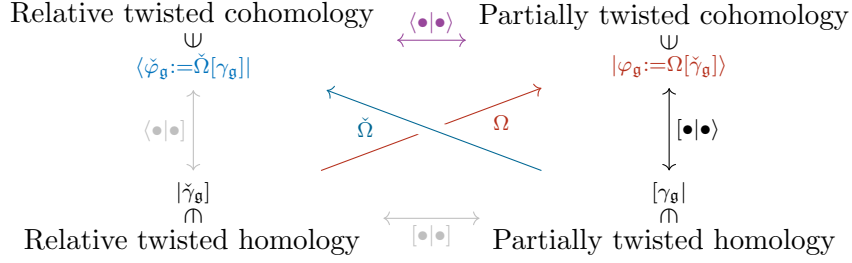

Roughly speaking, partially twisted (co)homology characterizes the space of ``interesting'' contours ($\gamma_\g$) and integrands ($u_\G\; \vphi_\g$) associated to FRW integrals. 
These spaces are fixed once the underlying topological space ($\mathbb{C}^n\setminus(\T\cup\B)$) and twist ($u_\G$) are chosen.  
The (Poincar\'e) dual space to partially twisted (co)homology is the relative twisted cohomology on $\mathbb{C}^n\setminus\T$ with boundaries in $\B$, and twist $\check{u}_\G=1/u_\G$.
In total there are four related spaces: two cohomologies and two homologies. 
The various relations between these spaces are summarized in figure \ref{fig:pairings}. 
We will only make use of the pairings colored in {\color{\myblue}blue ($\check{\Omega}$)}, 
{\color{\myred}red ($\Omega$)}, {\color{Purple}purple ($\la\bullet\vert\bullet\ra$)}, and black ($[\bullet\vert\bullet\ra$) to construct the coaction \cite{Abreu:2019wzk, Brown:2019jng}:
\begin{align} \label{eq:coactionMaster}
    \Delta \Big[ \gamma \Big| \varphi \Big\ra
    &= \sum_{\g,\mathfrak{h}} 
    {\color{Purple}C_{\g\mathfrak{h}}^{-1}}
    \Big[ 
        \gamma 
    \Big\vert 
       {\color{BrickRed}
            \vphi_\g
            := \Omega
            [\gamma_{\g}]
        }
    \Big\ra 
    \otimes \Big[
        \gamma_{\mathfrak{h}}
    \Big\vert 
        \varphi
    \Big\ra 
\end{align}
where
\begin{align}
    {\color{Purple} C_{\g\h} }
    &:= \Big\la 
        {\color{MidnightBlue}
            \check{\vphi}_\g
            := \check{\Omega}
            [\gamma_\g] 
        }
    \Big\vert 
        {\color{BrickRed}
            \vphi_\mathfrak{h} 
            := \Omega
            [\check{\gamma}_{\mathfrak{h}}]
        }
    \Big\ra
    \,.
\end{align}
Note that while we only sum over acyclic minors, equation \eqref{eq:coactionMaster} can be extended to the full twisted (co)homology, by summing over a basis for the full twisted (co)homology, instead of the physical subspace.

The colored diagonal lines in figure \ref{fig:pairings}, ${\color{MidnightBlue}\check{\Omega}}$ and ${\color{BrickRed}{\Omega}}$, represent the canonical form and dual canonical form maps.  
Specifically, ${\color{BrickRed}{\Omega}}$ maps a relative twisted (dual) cycle to a partially twisted FRW form while ${\color{MidnightBlue}\check{\Omega}}$ maps a partially twisted FRW cycle to a relative twisted (dual) form. 
That is, these maps make Poincar\'e duality explicit. 
Equation \eqref{eq:coactionMaster} is only valid when the forms and contours are related in this way. 
We provided the explicit recipe for constructing $\gamma_\g$ and $\vphi_\g$ in section \ref{sec:FRWquickStart}; additional details can be found in section \ref{sec:cohomDets}. 

The vertical black line in figure \ref{fig:pairings} represents the twisted period pairing that is computed via integration,
\begin{align}
    P_{\g\mathfrak{h}}
    := \int_{\gamma_\g} u_\G\; \vphi_{\mathfrak{h}}
    =: [ \gamma_{\g} \vert \vphi_{\mathfrak{h}} \ra 
    \,. \label{eq:graphical_basis_periods}
\end{align}
To make the notation more graphical for the periods $P_{\g\h}=[\gamma_\g\vert\vphi_\h\ra$, whenever possible, we superimpose the tubes associated to the cut $\Csf_\g$ on the acyclic graph $\h$. 
For example, in this graphical notation, the non-trivial periods of the two-site chain graph (section \ref{sec:2chainInto}) are represented by
\begin{align} \label{eq:graphicalPeriods_2chain}
    P_{{\twoChaing} , {\twoChaing} }
    & = [ \gamma_{\twoChaing} \vert \vphi_{\twoChaing} \ra 
    = \includegraphics[align=c, scale=.7]{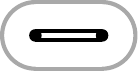}
    \,,
    &
    P_{{\twoChaing} , {\twoChaingg}}
    &= [ \gamma_{\twoChaing} \vert \vphi_{\twoChaingg} \ra 
    = \includegraphics[align=c, scale=.7]{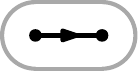}
    \,,
    \nn\\
    P_{{\twoChaing} , {\twoChainggg}}
    &= [ \gamma_{\twoChaing} \vert \vphi_{\twoChainggg} \ra 
    = \includegraphics[align=c, scale=.7]{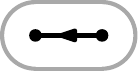}
    \,,
    &
    P_{{\twoChaingg} , {\twoChaingg}}
    &= [ \gamma_{\twoChaingg} \vert \vphi_{\twoChaingg} \ra 
    = \includegraphics[align=c, scale=.7]{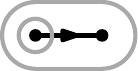}
    \,,
    \\
    P_{{\twoChainggg} , {\twoChainggg}}
    &= [ \gamma_{\twoChainggg} \vert \vphi_{\twoChainggg} \ra 
    = \includegraphics[align=c, scale=.7]{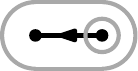}
    \,,
    &
    P_{{\twoChaingggg} , {\twoChaingggg}}
    &= [ \gamma_{\twoChaingggg} \vert \vphi_{\twoChaingggg} \ra 
    = \includegraphics[align=c, scale=.7]{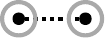}
    \,.
    \nn
\end{align}
We will use this graphical representation of period integrals when these objects cannot be confused for cut tubings.

This graphical notation also makes it clear which periods vanish. 
Whenever a tube crosses a pinched edge, an oriented edge that points the wrong way, or encircles a broken edge, the integral vanishes: 
\be
    0&=\includegraphics[align=c,scale=.7]{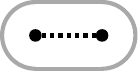}
    = \includegraphics[align=c,scale=.7]{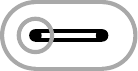}
    = \includegraphics[align=c,scale=.7]{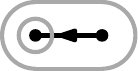}
    = \includegraphics[align=c,scale=.7]{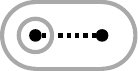}
    = \includegraphics[align=c,scale=.7]{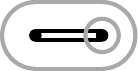}
    \\&\quad
    = \includegraphics[align=c,scale=.7]{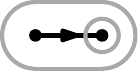}
    = \includegraphics[align=c,scale=.7]{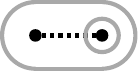}
    = \includegraphics[align=c,scale=.7]{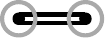}
    = \includegraphics[align=c,scale=.7]{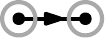}
    = \includegraphics[align=c,scale=.7]{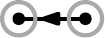}
    \,.
\ee
This is the graphical translation of the fact that 
\be \label{eq:vanishingCuts}
    \res_\mathfrak{\g}[\vphi_\h] = 0 
    \;\forall\; 
    \g \notin \mathrm{pinch}(\h)
    \,,
\ee
where $\mathrm{pinch}(\h)$ is the set of all acyclic minors obtained from $\h$ by turning any number of oriented edges into pinched edges (including changing no edges). 
In other words, $P_{\g\h}\neq0$ if and only if $\g$ is a pinch of $\h$.

The only other pairing that enters the coaction \eqref{eq:coactionMaster} is the intersection pairing between relative and partially twisted cohomology
\be \label{eq:intersection_numbers}
    C_{\g\mathfrak{h}} 
    := \la \check{\vphi}_\g \vert \vphi_\mathfrak{h} \ra 
    = \la 
        \check{\Omega}[\gamma_\g] 
    \vert 
        \Omega[\check{\gamma}_\mathfrak{h}] 
    \ra
    = \text{rational function of } \alpha_i 
    \quad \Big( \text{see } \eqref{eq:Cggp_quick} 
    \text{ or } \eqref{eq:Cggp} \Big)
    \,.
\ee
It functions as an inner product on the partially twisted cohomology, and can be used to project any FRW-form onto our basis. 
%
Moreover, the intersection number $\la \check{\Omega}[\gamma_\g] \vert \Omega[\check{\gamma}_\mathfrak{h}] \ra$ is the leading contribution to the period $[\gamma_\g\vert\vphi_\mathfrak{h}\ra$. 
Therefore, 
\be
    \sum_{\mathfrak{h}} 
    {C}^{-1}_{\g\mathfrak{h}} 
    [\gamma_\mathfrak{h} \vert \vphi_{\g'}\ra
    \bigg\vert_{\alpha_i \to \alpha' \alpha_i}
    = \delta_{\g\g'} 
    + \mathcal{O}(\alpha')
    \,,
\ee
where our (co)homology bases have been built so that the intersection matrix is diagonal, namely $\la \check{\vphi}_{\g} \vert \vphi_{\g'} \ra \propto \delta_{\g\g'}$.%
\footnote{%
    The homology intersection matrix of our choice of homology bases is also diagonal $[ \gamma_{\g} \vert \check{\gamma}_{\g'} ] \propto \delta_{\g\g'}$. 
}

Whenever we consider an FRW integral $F$ that has a series expansion around small values of $\alpha_i$ that evaluates to MPLs, whose coaction takes the form $\Delta F = \sum_{i}c_i (f_i \otimes g_i)$, equations \eqref{eq:diffCoaction} and \eqref{eq:DiscCoaction} generalize to 
\begin{align}
        \label{eq:deqHyperGeo}
        & \d F = \sum_{i} c_i \;  f_i \; \d W_1^\mathrm{MPL,\alpha}[g_i] \,, \\
        \label{eq:discHyperGeo}
        & \disc F = \sum_{i} c_i \; \disc\left( W_1^\mathrm{MPL,\alpha}[f_i] \right) \; g_i \,.
\end{align}
When computing these expansions, it is useful assign weight $-1$ to the expansion parameters $\alpha_i$. 
Then, each term in the MPL expansion of the hypergeometric function has the same, uniform weight.

\subsection{The two-chain (part 2): the coaction}
\label{sec:2chainCoaction_full}

In this section, we pick up where section~\ref{sec:2chainInto} left off. The basis of decorated graphs $\{\twoChaing,\twoChaingg,$ $\twoChainggg,\twoChaingggg\}$ and the cycles $\gamma_\g$ were constructed there; the cocycles $\vphi_\g$ and the period matrix $P_{\g\h}$ were written down in \eqref{eq:vphi_2chain} and \eqref{eq:2chainPer}. The point of this subsection is to apply the master coaction formula~\eqref{eq:coactionMaster} to those data and check that the results support the graphical recipe of box~\ref{box:graphical_rule}.


Consider the period $[\gamma_{\twoChaing} \vert \vphi_{\twoChaingg} \ra = \includegraphics[align=c, scale=.4]{figs/2chainP12.pdf}$. Implementing \eqref{eq:coactionMaster}, we find 
\be \label{eq:graphical_coaction_example_1}
    \Delta \includegraphics[align=c, scale=.7]{figs/2chainP12.pdf}
    &:= \sum_{\mathfrak{h}} 
    \frac{
         1
    }{
        \la \check{\vphi}_\mathfrak{h} \vert \vphi_\mathfrak{h} \ra
    }
    [\gamma_{\twoChaing} \vert \vphi_{\mathfrak{h}} \ra
    \otimes 
    [\gamma_{\mathfrak{h}} \vert \vphi_{\mathfrak{\twoChaingg}} \ra
    \\
    &= \frac{\alpha_1\alpha_2}{\alpha_1+\alpha_2} 
    \includegraphics[align=c, scale=.7]{figs/2chainP11.pdf}
    \otimes 
    \includegraphics[align=c, scale=.7]{figs/2chainP12.pdf}
    + \includegraphics[align=c, scale=.7]{figs/2chainP12.pdf}
    \otimes 
    \includegraphics[align=c, scale=.7]{figs/2chainP22.pdf}
\ee
where $
\includegraphics[align=c, scale=.5]{figs/2chainP32.pdf}
= \includegraphics[align=c, scale=.5]{figs/2chainP14.pdf}
= \includegraphics[align=c, scale=.5]{figs/2chainP42.pdf}
= 0
$ and
\be
    \la \check{\vphi}_{\twoChaing} \vert \vphi_{\twoChaing} \ra 
    &= \frac{\alpha_1 + \alpha_2}{\alpha_1 \alpha_2} 
    \,,
    \\
    \la \check{\vphi}_{\twoChaingg} \vert \vphi_{\twoChaingg} \ra 
    &= \la \check{\vphi}_{\twoChainggg} \vert \vphi_{\twoChainggg} \ra 
    = \la \check{\vphi}_{\twoChaingggg} \vert \vphi_{\twoChaingggg} \ra 
    = 1
    \,,
\ee
from \eqref{eq:Cggp_quick} or \eqref{eq:Cggp}.
Note that $\res_{\includegraphics[align=c,scale=.4]{figs/2chainCl.pdf}}$ and $\res_{\includegraphics[align=c,scale=.4]{figs/2chainCb.pdf}}$ in the second entry annihilate the form $\vphi_{\twoChaingg}$. We can think of this being a consequence of the zonotopal structure of the sequential residues (as described in section \ref{sec:partialFracAndZono}). 
The coaction only receives contribution from terms where $\mathfrak{h}$ can be obtained from $\twoChaingg$ by pinching a number oriented edges. 
This is analogous to the pinching rules of the  graphical coaction on Feynman integrals \cite{Abreu:2017ptx, Abreu:2018sat}.

Computing the graphical coaction for the remaining nontrivial two-site chain periods, we find
\begin{align} \label{eq:graphical_coaction_example_2}
    \Delta 
    \left[\includegraphics[align=c, scale=.7]{figs/2chainP12.pdf}\right]
    &= \frac{\alpha_1\alpha_2}{\alpha_1+\alpha_2} 
    \includegraphics[align=c, scale=.7]{figs/2chainP11.pdf}
    \otimes 
    \includegraphics[align=c, scale=.7]{figs/2chainP12.pdf}
    + 
    \includegraphics[align=c, scale=.7]{figs/2chainP12.pdf}
    \otimes 
    \includegraphics[align=c, scale=.7]{figs/2chainP22.pdf}
    \,, 
\\
    \Delta 
    \left[\includegraphics[align=c, scale=.7]{figs/2chainP13.pdf}\right]
    &= \frac{\alpha_1\alpha_2}{\alpha_1+\alpha_2} 
    \includegraphics[align=c, scale=.7]{figs/2chainP11.pdf}
    \otimes 
    \includegraphics[align=c, scale=.7]{figs/2chainP13.pdf}
    + 
    \includegraphics[align=c, scale=.7]{figs/2chainP13.pdf}
    \otimes 
    \includegraphics[align=c, scale=.7]{figs/2chainP33.pdf}
    \,, 
\\
    \Delta 
    \left[\includegraphics[align=c, scale=.7]{figs/2chainP22.pdf}\right]
    &= 
    \includegraphics[align=c, scale=.7]{figs/2chainP22.pdf}
    \otimes 
    \includegraphics[align=c, scale=.7]{figs/2chainP22.pdf}
    \,, 
\\
    \Delta 
    \left[\includegraphics[align=c, scale=.7]{figs/2chainP33.pdf}\right]
    &= 
    \includegraphics[align=c, scale=.7]{figs/2chainP33.pdf}
    \otimes 
    \includegraphics[align=c, scale=.7]{figs/2chainP33.pdf}
    \,, 
\\
    \Delta 
    \bigg[\includegraphics[align=c, scale=.7]{figs/2chainP44.pdf}\bigg]
    &= 
    \includegraphics[align=c, scale=.7]{figs/2chainP44.pdf}
    \otimes 
    \includegraphics[align=c, scale=.7]{figs/2chainP44.pdf}
    \,,  \label{eq:graphical_coaction_example_n}
\end{align}
where (as always) the second entries should be interpreted $\mathrm{mod}\; i\pi$.
As a first check of this formalism, we use these results and the weight-one term of $\alpha_\bullet$-expansion of $\mat{P}_{\mathrm{2-chain}} \mod i\pi$ (as defined in~\eqref{eq:p_two_chain}) to recover the differential equations that govern the two-site chain integrals:
\begin{align}
    \mat{A}
    &= \d \left(
        \mat{C}^{-1}\cdot W_1^\mathrm{MPL,\alpha}\left[\mat{P}_{\mathrm{2-chain}}
        \;\mathrm{mod}\; i\pi \right]
    \right)^\top 
    \,,
    \\
    &= \begin{pmatrix}
        (\alpha_1{+}\alpha_2)\;
        \dlog f_{\includegraphics[scale=.4]{figs/2chain_Rs1}}
        & 
        & 
        & 
    \\
        \frac{\alpha_1{+}\alpha_2}{\alpha_1\alpha_2}
        \dlog\frac{f_{\includegraphics[scale=.4]{figs/2chain_Rr2}}}{f_{\includegraphics[scale=.4]{figs/2chain_Rr1}}}
        & \substack{\mathlarger{
            \alpha_1\; 
            \dlog f_{\includegraphics[scale=.4]{figs/2chain_Rr2}}
        } \\ \mathlarger{
            \qquad {+} \alpha_2\; 
            \dlog f_{\includegraphics[scale=.4]{figs/2chain_Rr1}}
        }}
        & 
        & 
    \\
        \frac{\alpha_1{+}\alpha_2}{\alpha_1\alpha_2}
        \dlog\frac{f_{\includegraphics[scale=.4]{figs/2chain_Rl1}}}{f_{\includegraphics[scale=.4]{figs/2chain_Rl2}}}
        & 
        & \substack{\mathlarger{
            \alpha_1\; \dlog f_{\includegraphics[scale=.4]{figs/2chain_Rl2}}
        } \\ \mathlarger{
            \qquad {+} \alpha_2\; 
            \dlog f_{\includegraphics[scale=.4]{figs/2chain_Rl1}}
        }}
        &
    \\
        {}
        & 
        & 
        & \substack{\mathlarger{
            \alpha_1\; \dlog f_{\includegraphics[scale=.4]{figs/2chain_Rb2}}
        } \\ \mathlarger{
            \qquad {+} \alpha_2\; \dlog f_{\includegraphics[scale=.4]{figs/2chain_Rb1}}
        }}
    \end{pmatrix}
   \,.
    \nn
\end{align}
This matches the results of \cite{De:2023xue, Arkani-Hamed:2023kig, Glew:2025arc, Baumann:2025qjx} (up to a constant rotation of the basis). 
A general formula for the weight-one terms that arise in our basis integrals is derived in appendix \ref{app:weight1}.

The graphical rules described above are also compatible with any choice of contour. 
The most directly interesting examples are those integrals that involve the physical contour. 
After writing the differential form in our basis, we recover the formula teased in \eqref{eq:Delta2ChainIntro}:
\begin{align}
    \Delta [\psi_{\includegraphics[align=b, scale=.7]{figs/2chainNoLabels.pdf}}]
    &= \Delta \int_{0}^\infty u_{\includegraphics[align=b, scale=.7]{figs/2chainNoLabels.pdf}}\;  \vphi_{\includegraphics[align=b, scale=.7]{figs/2chainNoLabels.pdf}}
    = \Delta \int_{0}^\infty u_{\includegraphics[align=b, scale=.7]{figs/2chainNoLabels.pdf}}\;  \vphi_{\includegraphics[align=b, scale=.5]{figs/2chain_gr.pdf}}
    + \Delta \int_{0}^\infty u_{\includegraphics[align=b, scale=.7]{figs/2chainNoLabels.pdf}}\;  \vphi_{\includegraphics[align=b, scale=.5]{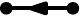}}
    - \Delta \int_{0}^\infty u_{\includegraphics[align=b, scale=.7]{figs/2chainNoLabels.pdf}}\;  \vphi_{\includegraphics[align=b, scale=.5]{figs/2chain_gb.pdf}}
    \nn\\
    &= \frac{\alpha_1 \alpha_2}{\alpha_1 + \alpha_2} \cdot
    \int_{0}^\infty u_{\includegraphics[align=b, scale=.7]{figs/2chainNoLabels.pdf}}\;  \vphi_{\includegraphics[align=c, scale=.5]{figs/2chain_gs}}
    \otimes 
    \includegraphics[scale=.7,align=c]{figs/2chain_gs_gr}
    + 
    1 \cdot
    \int_{0}^\infty u_{\includegraphics[align=b, scale=.7]{figs/2chainNoLabels.pdf}}\;  \vphi_{\includegraphics[align=c, scale=.5]{figs/2chain_gr.pdf}}
    \otimes 
    \includegraphics[scale=.7,align=c]{figs/2chainCr}
    \nn\\&+
    \frac{\alpha_1 \alpha_2}{\alpha_1 + \alpha_2} \cdot
    \int_{0}^\infty u_{\includegraphics[align=b, scale=.7]{figs/2chainNoLabels.pdf}}\;  \vphi_{\includegraphics[align=c, scale=.5]{figs/2chain_gs}}
    \otimes 
    \includegraphics[scale=.7,align=c]{figs/2chain_gs_gl}
    + 
    1 \cdot 
    \int_{0}^\infty u_{\includegraphics[align=b, scale=.7]{figs/2chainNoLabels.pdf}}\;  \vphi_{\includegraphics[align=c, scale=.5]{figs/2chain_gl}}
    \otimes 
    \includegraphics[scale=.7,align=c]{figs/2chainCl}
    \nn\\&-
    1 \cdot
    \int_{0}^\infty u_{\includegraphics[align=b, scale=.7]{figs/2chainNoLabels.pdf}}\;  \vphi_{\includegraphics[align=c, scale=.5]{figs/2chain_gb}}
    \otimes 
    \includegraphics[scale=.7,align=c]{figs/2chainCb}
    \,.
\end{align}
Because we have not first decomposed the physical integration contour into our chosen graphical basis, this result is not as naturally compact as what we found in~\eqref{eq:graphical_coaction_example_1} or~\eqref{eq:graphical_coaction_example_2} through~\eqref{eq:graphical_coaction_example_n}. For this reason, we generally decompose physical FRW integrals into the basis of periods introduced in~\eqref{eq:graphical_basis_periods} before computing the coaction. This allows us to leverage the full graphical power of our notation, which greatly simplifies the analysis (as we now expand upon).

\subsection{The graphical coaction}
\label{sec:quickStart}

Having presented the general coaction formula~\eqref{eq:coactionMaster}, which can be applied to any FRW period, we finish this section by presenting a simplified formula for the coaction of elements in our chosen (graphical) basis of periods. Namely, for any period $[ \gamma_\g \vert \vphi_\h \ra$, the coaction formula simplifies to  
\begin{align} \label{eq:basis_simplifications_graphical_coaction}
    \Delta [ \gamma_\g \vert \vphi_\h \ra
    = \sum_{
        \f \in \pinch(\g,\h)
    }
    C_{\f\f}^{-1}
    [\gamma_\g \vert \vphi_\f \ra 
    \otimes 
    [\gamma_\f \vert \vphi_\h \ra 
    \,,
\end{align}
where the sum is now just over pinched graphs $\f$, namely those in the set
\be
     \pinch(\g,\h) = \{
        \f \in \mathrm{aDec}(\G) : 
        \g \in \pinch(\f) \text{ and } \f \in \pinch(\h)
    \}
    \,,
\ee
where, as stated earlier, 
\begin{align}
    \pinch(\h)=\left\{\substack{
        \text{set of all acyclic minors obtained from $\h$ by turning any number} 
        \\
        \text{of oriented edges into pinched edges  (including changing no edges) }
    }\right\}
    \,.
\end{align} 
This restriction on the sum follows from the fact that  $\res_\mathfrak{\g}[\vphi_\h] \neq 0 \iff \g \in \pinch (\h)$, as highlighted in~\eqref{eq:vanishingCuts}. Additionally, the simplified formula~\eqref{eq:basis_simplifications_graphical_coaction} takes advantage of the fact that the intersection matrix in this chosen basis is diagonal.

In practice, we can equivalently think of constructing the coaction of $[\gamma_\g \vert \vphi_\h \ra$ using the following combinatorial recipe:
\begin{mybox}[label={box:graphical_rule}]{Graphical coaction} To construct the coaction of the FRW period $[\gamma_\g \vert \vphi_\h \ra$, we carry out the following steps:
\begin{enumerate}
    \item Draw all acyclic decorated graphs \(\f\) that fit between $\h$ and $\g$, in the sense that
    \[
    \h \longrightarrow \f \longrightarrow \g \, ,
    \]
    where an arrow represents the replacement of some (but possibly none) of the directed edges by pinched edges. (This give us the set $\pinch(\g,\h)$ from above.) 
    \item Delete all graphs $\f$ whose edges form an oriented cycle when all pinched edges are contracted.
    \item For each remaining $\f$, write down the expression  
    \[ C_{\f\f}^{-1} \
    [\gamma_\g|\varphi_\f\rangle \otimes [\gamma_\f|\varphi_\h\rangle\, .
    \]
    \item The coaction $ \Delta [\gamma_\g \vert \vphi_\h \ra$ is given by the sum of all these terms.
\end{enumerate}
\end{mybox}
\noindent It can be checked this reproduces the coaction of $[\gamma_\g \vert \vphi_\h \ra$, as computed using the more general formula from~\eqref{eq:coactionMaster}.

It is worth connecting the simplified formula~\eqref{eq:basis_simplifications_graphical_coaction} to the graphical notation introduced in~\eqref{eq:graphicalPeriods_2chain}. In that notation, the depicted acyclic minor specifies an FRW integrand, which just means it picks out a subset of the denominator factors that arise in the corresponding wavefunction coefficient. The cut tubing that we superimpose on the acyclic minor indicates the integration contour we pair this integrand with; in particular, each tube in this tubing indicates that we compute a residue with respect to a specific denominator factor. Clearly, if this contour computes a residue with respect to any denominator factor that is not present---which means there is a tube present that is not present in the cut tubing associated with the acyclic minor that is depicted---the integral will vanish. This is what allows us to restrict the sum in~\eqref{eq:basis_simplifications_graphical_coaction} just to pinches of $\h$. Moreover, we can think of the terms that do survive in the coaction sum as ranging over all further sets of residues that can be computed in the integrand, that are not computed via the original contour.

We can also compare this graphical coaction for wavefunction coefficients with the diagrammatic coaction developed for flat-space Feynman integrals~\cite{Abreu:2017enx,Abreu:2017mtm}. Both of these graphical operations encode the (co)homological properties of integrals---including the derivatives and discontinuities of these integrals---in terms of straightforward, combinatorial graph modifications. In both cases, the coaction also enables us to decompose an integral into simpler integrals (associated with modified graphs), in which edges are pinched or cut. However, while the diagrammatic coaction on Feynman integrals considers the integration over all components of the loop momenta, the graphical coaction we consider here just pays attention to the integrations over site energies (which encode the integrals over time). Even so, the FRW coaction relies on a richer graphical vocabulary, as it involves directed edges that encode the causal structure of the FRW wavefunction coefficients. Finally, while extending the strict diagrammatic coaction for flat-space Feynman integrals beyond one loop remains an active area of research~\cite{Abreu:2021vhb}, the full set of identities between cut Feynman integrals has not yet been characterized beyond one loop. In contrast, all of the identities between FRW integrals are generated by intersecting tubes via the relation~\eqref{eq:linRel}, which our graphical notation already accounts for; thus, the FRW graphical coaction~\eqref{eq:basis_simplifications_graphical_coaction} applies to graphs at arbitrary loop orders.


\section{Derivation of the twisted (co)homology from cuts and positive geometry}
\label{sec:cohomDets}

In this section, we provide more detail on the construction of our basis of the FRW (co)homology and its (Poincar\'e and de Rham) duals. We also justify the formulas we presented in the last section, which we made use of when constructing the coaction.

\subsection{A basis of homology from the positive geometry of physical cuts}
\label{sec:hom}

We begin by providing a short review of twisted homology, specialized to situations in which some divisors are untwisted (as is true here). 
Then, we use the geometry of the physical cuts from section \ref{sec:cutGeom} to build a basis of cycles for the physical subspace of FRW integrals. 
While the cycles corresponding to unphysical cuts can be built in the same manner, there is no (known) set of combinatorial objects (analogous to the decorated graphs $\g$) that characterize cycles outside the physical subspace. 

\paragraph{Partially twisted homology---generalities.}

A \textit{partially twisted chain} (or integration contour) is a chain (or contour) that avoids $\B$, but that can have boundaries on $\T$ as long as we provide an instruction for which branch of the twist $u_\G$ to use. 
This is usually represented as a tensor product $\gamma = \Gamma \otimes u_\gamma$.
Here, $\Gamma$ is a topological chain and $u_\gamma \in \mathcal{L}$ is a section of the line bundle $\mathcal{L} = \mathbb{C} u_\G$ (equivalently, a choice of branch for the multi-valued function $u_\G$).
In the literature, $\mathcal{L}$ is also called a \textit{local system}. 
It keeps track of the monodromies of $u$ (namely, the phases $e^{\pi i \alpha_\bullet}$ one picks up when encircling twisted branch points). 
While we have mostly suppressed the notation $\otimes u_\gamma$ in sections \ref{sec:FRWIntegrals}, \ref{sec:coaction} and \ref{sec:examples}, we will keep this factor explicit in the present section.

To define the twisted homology, we need to first explain how the twisted boundary operator $\partial_u$ is defined. 
This is simplest to describe when $\Gamma$ is a $m$-simplex, namely $\Gamma = \la p_1, \dots, p_m \ra$ where the $p_i$ are the vertices of the simplex. 
Then, 
\be \label{eq:twBd}
    \partial_u \gamma 
    = \partial_u (\Gamma \otimes u_\gamma)
    = \sum_{a} (-1)^a \la p_1, \dots, \hat{p}_a, \dots, p_m \ra 
    \otimes u_\gamma\vert_{\la p_1, \dots, \hat{p}_a, \dots, p_m \ra}
\ee
where $\la p_1, \dots, \hat{p}_a, \dots, p_m \ra$ indicates we should omit the $\hat{p}_a$ vertex to get the $a^\text{th}$ boundary component of $\Gamma$, and $u_\gamma\vert_{\la p_1, \dots, \hat{p}_a, \dots, p_m \ra}$ is just the twist $u_\gamma$ evaluated on the $a^\text{th}$ boundary component. 
If we set $u_\gamma = 1$, this reduces to the usual boundary operator. 

The $n$-th twisted homology is the subset of twisted $n$-chains that have no boundary and that are not boundaries themselves. 
Formally, it is given by 
\begin{align}
    H_n(M\setminus\B;{\mathcal{L}})
    := \frac{
        \ker \partial_u : C_n(M\setminus\B;{\mathcal{L}})
        \to C_{n-1}(M\setminus\B;\mathcal{L})
    }{
        \mathrm{im}\, \partial_u: C_{n+1}(M\setminus\B;{\mathcal{L}})
        \to C_n(M\setminus\B;{\mathcal{L}})
    }
    \,,
    \\
    = \frac{\{\text{$\partial_u$-closed twisted $n$-chains on $M\setminus\B$}\}}{\{\text{$\partial_u$-exact twisted $(n+1)$-chains on $M\setminus\B$}\}}
    \,,
\end{align}
where $C_p(M\setminus\B;{\mathcal{L}})$ is the set of twisted $p$-chains.
Note that, from \eqref{eq:twBd}, if each boundary component of $\Gamma$ lies inside $\T$, then the twisted chain $\gamma = \Gamma \otimes u_\gamma$ is automatically closed (namely, $\partial_u \gamma = 0$) since the twist vanishes on $\T$.
That is, the components of $\T$ behave like boundaries instead of divisors in the twisted homology.%
\footnote{%
    Chains that end on $\T$ are to be interpreted as locally finite or Borel-Moore chains. 
}
The only other kind of closed contours are residue contours that localize to components of $\B$. 
Indeed, all FRW cycles can be expressed in the form
\begin{align}
    \delta_\csf(\Delta_\csf) \otimes u_\G:= 
    \delta_{ \tau_{|\csf|} } \circ 
    \cdots \delta_{\tau_1}
    \circ \Delta_\csf 
    \otimes u_\G
\end{align}
where $\csf = (\tau_1, \cdots, \tau_{|\csf|})$ is a multi-index, $\delta_{\tau}$ is called the coboundary operator, and corresponds to a residue contour centered around the locus $V(B_\tau)$. Meanwhile, $\Delta_\csf$ is the chain (or integration contour) that has support on the cut space
\be\label{eq:defMJ}
    M_\csf := M \cap \B_\csf
    \,,
    \qquad
    \B_\csf = \bigcap_{\tau \in \csf} V(B_\tau)
    \,,
\ee
and that has boundaries on $\T\cap\B_\csf \supset \partial\Delta_\csf$
\be
    \int_{\delta_\csf(\Delta_\csf)} u_\G\;\vphi 
    = \int_{\Delta_\csf} \res_{B_{\tau_{|\csf|}}} \circ \cdots \circ \res_{B_{\tau_1}} [u_\G\;\vphi]
    \,.
\ee
For cuts $\csf$ associated to acyclic minors, the $\Delta_\csf$ become the $\Delta_\g$ in~\eqref{eq:Delta_g}.
Note that the physical contour is not associated to a cut, since no propagators are on shell ($\csf=\varnothing$). 
Explicitly, 
\be
    \gamma_\text{phys} := \delta_{\varnothing}(\Delta_\varnothing) \otimes u_\G
\ee
where $\Delta_\varnothing := \{\mbf{x} \in\mathbb{R}_+^{|\V_\g|}\}$ is the unbounded chamber that is the positive orthant.%

Remarkably, like in the fully twisted case, the partial twisted homology is (generically) middle dimensional; $H_n(M\setminus\B; \mathcal{L}) = 0$ if $n\neq |\V_\G|$.%
\footnote{%
    A proof and precise conditions for when middle dimensionality breaks in the partial twisted context will be published in a future article. 
    For now, the reader is directed to \cite{Caron-Huot:2021iev, Caron-Huot:2021xqj, matsumoto}. 
}
Moreover, for $n=|\V_\G|$, the twisted homology of $M\setminus\B$ can be computed by uplifting the homology of each cut $M_\csf$%
\footnote{%
    Formally, one should think of $\delta_\tau$ as the Leray coboundary map $\delta_\tau: H_{n-1}(\B_\tau \cap (M\setminus\B);{\mathcal{L}}\vert_{\B_\tau}) \to H_n(M\setminus\B;{\mathcal{L}})$ which uplifts a cycle from a lower dimensional subvariety by wrapping it with an infinitesimal tube.
    Leray's long exact sequence (in the twisted setting)
    facilitates the decomposition in \eqref{eq:homCutDecomp} \cite{pham,Hwa:1967csk}.
    Sequential coboundaries are generically anti-symmetric, and,
    when paired with a cocycle in an integral, the coboundary is equivalent to the residue operation. 
    They are also an important component to computing the homology intersection numbers.
}
\begin{align} \label{eq:homCutDecomp}
\begin{aligned}
    H_{|\V_\G|}(M\setminus\B;{\mathcal{L}}) 
    \subseteq \bigoplus_{p+q={|\V_\G|}} \left[
        \bigoplus_{\csf: |\csf| = q}
        \delta_{\csf} \Big( 
            H_p(
                M_{\csf}
                ;{\mathcal{L}}_{\csf} 
            ) 
        \Big)
    \right]
    \,.
\end{aligned}
\end{align}
It is worth stressing this last point. 
Equation \eqref{eq:homCutDecomp} tells us that the full twisted homology is spanned by uplifting the twisted homology of each cut $M_\csf$ into $M\setminus\B$.
Combining this statement with the discussion of section \ref{sec:cutGeom} suggests that the physical subspace of  the twisted homology $H^\text{phys}_{|\V_\G|} \subseteq H_{|\V_\G|}(M\setminus\B;{\mathcal{L}})$ is spanned by contours that localize to the physical cuts, whose remaining component $\Gamma$ is simply the unique bounded region on $M_\g$. 
Explicitly, 
\begin{align} 
\begin{aligned} \label{eq:physHom_sub}
    H_{|\V_\G|}^{\text{phys}} 
    \subseteq 
   \bigoplus_{\g}
    \left[
        \bigoplus_{\mathsf{c} \in \mathsf{C}_\mathfrak{g}}
        \delta_{\mathsf{c}} \Big( 
            H_{{|\V_\G|}-|\mathsf{c}|}(
                M_\g
                ; \mathcal{L}_\mathfrak{g}
            ) 
        \Big)
    \right]
    \,.
\end{aligned}
\end{align}
The above equation is not an equality due to the fact that the on-shell variety $\B$ is degenerate; the relations between residue contours give rise to relations between the coboundary operators $\delta_\csf$. 
We fix this degeneracy below, following the discussion in section \ref{sec:ResolveDegen}.

\paragraph{The physical subspace of partially twisted cycles from $\g$.}
\label{sec:physHom}

As suggested by the previous section, each twisted cycle in the physical subspace takes the form 
\begin{align}
    \gamma_\g 
    = \delta_{\Csf_\g} ( \Delta_\g )
    \otimes u_\G
    \,.
\end{align}
where $\Delta_{\g}$$:= {\bigtimes}_{\rsf \in \Rsf_\g} \Delta_\rsf$ is the unique bounded chamber on $M_\g$. 
Moreover, since the on-shell variety $\B$ is often degenerate, we only make use of the specific linear combinations of residue contours specified in \eqref{eq:degenRes}, namely
\begin{align} 
\label{eq:degenDelta}
    \delta_{\Csf_\g}
    &= \sum_{\csf \in \Csf_\g}
        \sgn_{\csf}
        \delta_{\csf}
        \,,
\end{align}
where the signs $\sgn_\csf$ are induced by the ordering chosen in box \ref{box:tubeOrd}; see \eqref{eq:sgn} for the explicit formula. 

\begin{mybox}[label={box:HomPhys}]{The physical twisted homology}
    With the sign in \eqref{eq:degenDelta} made explicit in \eqref{eq:sgn}, the precise linear combination of $\delta_\csf$ operators that correspond to physical cuts is determined, and we can fully specify $H_{|\V_\G|}^\text{phys}$:
    \begin{align} 
    \begin{aligned} \label{eq:physHom}
        H_{|\V_\G|}^\text{phys} := 
        \bigoplus_{\g}
        \; 
            \delta_{\Csf_\g} \Big( 
                H_{{|\V_\G|}-|\csf|}(
                    M_\g ; \mathcal{L}_\g
                ) 
            \Big)
        \subseteq  H_{|\V_\G|}(M\setminus\B;{\mathcal{L}}) 
        \,.
    \end{aligned}
    \end{align}
    Since each $M_\g$ is a positive geometry (with a single bounded chamber $\Delta_\g$), the twisted homology is built from the direct sum of twisted homologies of positive geometries. 
    Because there is only one bounded chamber on each $M_\g$, the basis of homology is essentially unique. 
\end{mybox}
\noindent

\paragraph{Example: the homology of the two-site chain.}
\label{sec:2chainHom}

\begin{figure}
\begin{center}
    \includegraphics[align=c,scale=1]{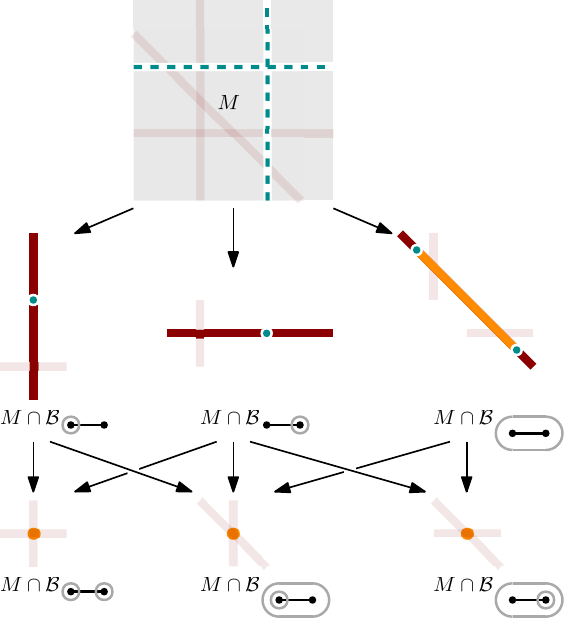}
    \caption{%
        Cut stratification of the pair $(M,\B)$ in the context of the two-site chain graph. 
        \label{eq:2ChainStrat}
    }
\end{center}
\end{figure}

Equation \eqref{eq:homCutDecomp} instructs us to look at the cut spaces that can be constructed using $B_{\includegraphics[scale=.3]{figs/2chainB1}}$, $B_{\includegraphics[scale=.3]{figs/2chainB2}}$, and $B_{\includegraphics[scale=.3]{figs/2chainB12}}$, and  compute the associated homology on each cut. 
From figure \ref{eq:2ChainStrat}, only the cuts associated to the acyclic minors in \eqref{eq:2chaing}
($M \cap \B_{\includegraphics[scale=.3,align=c]{figs/2ChainB12}}$, 
$M \cap \B_{\includegraphics[scale=.3,align=c]{figs/2ChainB1B2}}$, 
$M \cap \B_{\includegraphics[scale=.3,align=c]{figs/2ChainB12B1}}$ 
and 
$M \cap \B_{\includegraphics[scale=.3,align=c]{figs/2ChainB12B2}}$)
have bounded chambers (shown in {\color{Orange} orange}). 
Therefore, these are the only cuts that contribute to the homology
\be
    H_2( M \setminus \B; \mathcal{L})
    &= \delta_{\includegraphics[align=c, scale=.3]{figs/2ChainB12.pdf}}
    \left(
        H_1 \left(
            \includegraphics[align=c, scale=.5]{figs/2ChainM12.pdf}
            ; \mathcal{L}\vert_{\includegraphics[align=c, scale=.3]{figs/2ChainB12.pdf}}
        \right)
    \right) 
    \\&
    + \delta_{\includegraphics[align=c, scale=.3]{figs/2ChainB12B1.pdf}}
    \left(
        H_0 \left(
            \includegraphics[align=c, scale=.5]{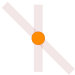}
        \right)
    \right) 
    + \delta_{\includegraphics[align=c, scale=.3]{figs/2ChainB12B2.pdf}}
    \left(
        H_0 \left(
            \includegraphics[align=c, scale=.5]{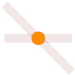}
        \right)
    \right) 
    + \delta_{\includegraphics[align=c, scale=.3]{figs/2ChainB1B2.pdf}}
    \left(
        H_0 \left(
            \includegraphics[align=c, scale=.5]{figs/2ChainM1_2.pdf}
        \right)
    \right) \, .
\ee
Notice that the homology group inside each $\delta_\bullet$ is generated by a single bounded chamber.

From section \ref{sec:cutGeom}, we have that each graph region corresponds to a region-simplex $\Delta_\rsf$, and that the Cartesian product of all region-simplices is the bounded chamber $\Delta_\g$ on $M_\g$. 
Moreover, each $\Delta_\g$ is a generator of $H_\bullet(M_\g,\mathcal{L}\vert_{\B_\g})$ and can be uplifted to a cycle on the larger space $M\setminus\B$ by acting with $\delta_{\Csf_\g}$. 
That is, we lift the chamber $\Delta_\g$ to a middle-dimensional contour in the space  $M\setminus\B$ by wrapping the lower-dimensional cycle $\Delta_\g$ with infinitesimal tubes:
\be
    \gamma_{\twoChaing}
    &= \delta_{\includegraphics[align=c, scale=.3]{figs/2ChainB12.pdf}}
    \left( \includegraphics[align=c,scale=.6]{figs/2ChainM12.pdf} \right) 
    \otimes u_{\includegraphics[align=b, scale=.7]{figs/2chainNoLabels.pdf}}
    \,,
    &\qquad\qquad
    \gamma_{\twoChaingg}
    &= \delta_{\includegraphics[align=c, scale=.3]{figs/2ChainB12B1.pdf}}
    \left( \includegraphics[align=c,scale=.6]{figs/2chainM12_1.pdf} \right) 
    \otimes u_{\includegraphics[align=b, scale=.7]{figs/2chainNoLabels.pdf}}
    \,,
    \\
    &= \includegraphics[align=c,scale=.6]{figs/2Chain_gamma4.pdf}
    \otimes u_{\includegraphics[align=b, scale=.7]{figs/2chainNoLabels.pdf}}
    \,,
    &\qquad
    &= \includegraphics[align=c,scale=.6]{figs/2Chain_gamma2.pdf}
    \otimes u_{\includegraphics[align=b, scale=.7]{figs/2chainNoLabels.pdf}}
    \,,
\ee
\be
    \gamma_{\twoChainggg}
    &= \delta_{\includegraphics[align=c, scale=.3]{figs/2ChainB12B2.pdf}}
    \left( \includegraphics[align=c,scale=.6]{figs/2ChainM12_2} \right) 
    \otimes u_{\includegraphics[align=b, scale=.7]{figs/2chainNoLabels.pdf}}
    \,,
    &\qquad\qquad
    \gamma_{\twoChaingggg}
    &= \delta_{\includegraphics[align=c, scale=.3]{figs/2ChainB1B2.pdf}}
    \left( \includegraphics[align=c,scale=.6]{figs/2ChainM1_2.pdf} \right) 
    \otimes u_{\includegraphics[align=b, scale=.7]{figs/2chainNoLabels.pdf}}
    \,,
    \\
    &= \includegraphics[align=c,scale=.6]{figs/2Chain_gamma3.pdf}
    \otimes u_{\includegraphics[align=b, scale=.7]{figs/2chainNoLabels.pdf}}
    \,,
    &\qquad\qquad
    &= \includegraphics[align=c,scale=.6]{figs/2Chain_gamma1.pdf}
    \otimes u_{\includegraphics[align=b, scale=.7]{figs/2chainNoLabels.pdf}}
    \,.
\ee
Indeed, we see that we can construct four independent integration contours in this way. One can verify that $|\chi\big(\mathbb{C}^2\setminus(\T\cup\B)\big)| = 4$, so we have indeed spanned the full space of twisted homology.

\subsection{A basis of the dual relative twisted homology}
\label{sec:dualhom}

Next, we introduce the (Poincar\'e) dual twisted homology $\check{H}_n^\text{phys} \subset H_n(M,\B; \check{\mathcal{L}})$, where $\check{\mathcal{L}} := \mathbb{C}\check{u}_\G$ is the local system of the inverse twist $\check{u}_\G=u^{-1}_\G$. 
There are two reasons for introducing the dual relative twisted homology:
\begin{itemize}
    \item[1)] We want to use positive geometry to build the space of differential forms on $M\setminus\B$. Since the canonical form is a map from the relative twisted homology of the pair $(M,\B)$ to the space of differential forms on $M\setminus\B$, we construct $\check{H}_n^\text{phys}$.
    \item[2)] It is also useful to make sure that the homology intersection matrix is diagonal. This ensures that the cohomology intersection matrix of canonical forms---which appears in the coaction---is diagonal.
\end{itemize}

\paragraph{The dual relative twisted homology---generalities.}

Let $C_p(M,\B;\check{\mathcal{L}})$ be the space of relative twisted chains%
\footnote{%
    More precisely, we interpret all chains as locally-finite/Borel-Moore.   
    Since the locally-finite twisted homology is isomorphic to the  compactly supported twisted homology we do not make the distinction between the two unless it is necessary.
    We only provide a working definition aimed at applications in cosmology. 
}
\be \label{eq:lfchains}
    C_p(M,\B;\check{\mathcal{L}})
    := \{ 
        \check{\gamma} = \check{\Gamma} \otimes \check{u}_\G
        : \partial \check{\Gamma} \subset \T\cup\B
    \}\,.
\ee
Here, $\check{\Gamma}$ is essentially the usual relative $p$-chain with boundaries on $\T \cup \B$. 
The dual twisted boundary operator $\partial_{\check{u}}$ works exactly like $\partial_u$. 
In particular, the boundaries of $\check{\Gamma}$ on any component of $\T$ are in the kernel of the twisted boundary operator.
The relative twisted homology
\begin{align}
    H_n(M,\B;\check{\mathcal{L}})
    := \frac{
        \ker \partial_{\check{u}} : C_n(M,\B;\check{\mathcal{L}})
        \to C_{n-1}(M,\B;\check{\mathcal{L}})
    }{
        \mathrm{im} \partial_{\check{u}}: C_{n+1}(M,\B;\check{\mathcal{L}})
        \to C_n(M,\B;\check{\mathcal{L}})
    }
\end{align}
is the space of relative twisted $n$-chains with boundaries on $\B$ that are not themselves a boundary of a relative twisted $(n+1)$-chain.
Like the partial twisted homology of section \ref{sec:hom}, the relative twisted homology is also middle dimensional; only $n={|\V_\G|}$ is non-trivial.

\paragraph{The physical subspace of relative twisted cycles from $\g$.}
The space of physical dual cycles is spanned by the set of relative twisted ${|\V_\G|}$-cycles whose maximal boundary corresponds to the bounded chamber $\Delta_\g$ on the cut space $M_\g$.
To build such dual cycles, we start with the $\Delta_\g$ and let it grow into the negative orthant until it touches the hyperplane at infinity. 
Explicitly, 
\be \label{eq:GammaCheck}
    \check{\Gamma}_\g := \{
        (x_1,\dots,x_n) \in \mathbb{C}^n
        : \substack{
            B_\tau \leq 0 \;\forall\; \tau\in\csf 
            \text{ and } \csf\in\Csf_\g, 
            \\
            x_v \leq0 \;\forall\; v\in\rsf \text{ and } \rsf\in\Rsf
            \text{ such that } |\V_\rsf|\geq2
        }
    \}
    \,.
\ee
This unbounded chamber has $\Delta_\g$ on its maximal boundary
\be
    \partial_{\Csf_\g} \check{\Gamma}_\g = |\Csf_\g| \Delta_\g
    \,,
\ee
where $\partial_{\Csf_{\g}} := \sum_{\csf\in\Csf_\g} \sgn_\csf \partial_\csf$, $\partial_\csf := \partial_{\tau_{|\csf|}} \circ \cdots \circ \partial_{\tau_1}$ with $\tau_1, \dots, \tau_{|\csf|} \in \csf$ and the $\tau_i$ are ordered as usual according to box \ref{box:tubeOrd}.
Here, $\partial_\tau \bullet := (\partial_{\check{u}} \bullet) \cap \B_\tau$ is the part of the twisted boundary that lies inside the boundary $\B_\tau$.

\paragraph{Example: the dual homology of the two-site chain.}
\label{sec:2chainHomDual}

Applying \eqref{eq:GammaCheck}, straightforwardly produces the relative twisted cycles $\check{\gamma}_\g = \check{\Gamma}_\g \otimes \check{u}_\G$ where 
\be
    \check{\Gamma}_{\twoChaing} 
    &= \includegraphics[align=c,scale=.7]{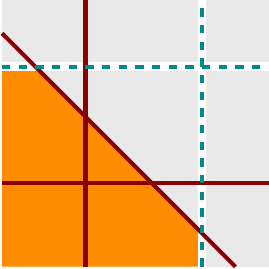}
    \,,
    &
    \check{\Gamma}_{\twoChaingg} 
    &= \includegraphics[align=c,scale=.7]{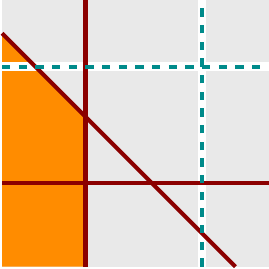}
    \,,
    \\
    \check{\Gamma}_{\twoChainggg} 
    &= \includegraphics[align=c,scale=.7]{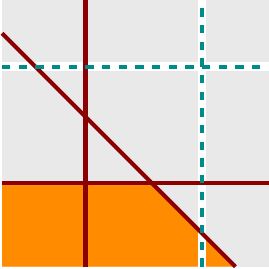}
    \,,
    &
    \check{\Gamma}_{\twoChaingggg} 
    &= \includegraphics[align=c,scale=.7]{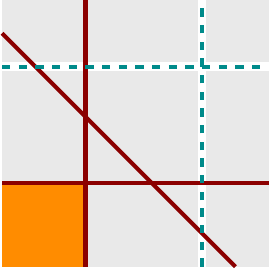}
    \,. 
\ee
These cycles are dual to the basis constructed in section \ref{sec:2chainHom} via the homology intersection number. 
Since the intersection number of twisted cycles is proportional to the usual topological intersection number, and, because the topological cycles $\delta_\g(\Delta_\g)$ and $\check{\Gamma}_{\g^\prime}$ share support (on $\Delta_\g)$ if and only if $\g=\g^\prime$, the resulting intersection matrix must be diagonal.

By ensuring that the dual contours have a diagonal intersection matrix with the physical FRW-cycles, the canonical forms of the dual cycles ($\Omega[\check{\Gamma}_\g]$) pair nicely with the physical FRW-cycles. 
To leading order in the parameters $\alpha_\bullet$, the resulting period matrix is diagonal.

\subsection{Positive geometry: from dual homology to FRW cohomology}
\label{sec:FRWCohom}

In this section, we provide an explicit representation for the canonical form map which generates our basis of FRW forms, and review some basis aspects of twisted cohomology.

\paragraph{Partially twisted cohomology---generalities.}
To work at the level of cohomology, we quotient by total derivatives since these integrals vanish.
Explicitly, for any $n$-form $\vphi$ on $M\setminus\B$, 
\begin{align}
    \int_{\gamma} \d(u_\G\;\vphi)
    = \int_{\gamma} u_\G\; (\nabla_\omega \vphi)
    = 0
    \quad\forall\quad
    \gamma \in H_n(M\setminus \B; \mathcal{L})
\end{align}
where
\begin{align}
    \nabla_{\omega} = \d+\omega\wedge
    \,,
    \qquad
    \omega := \dlog\; u_\G
    \,,
\end{align}
is the twisted differential. 
That is, the integral only depends on the equivalence class of $\vphi$: $\vphi \simeq \vphi + \nabla_\omega \psi$ for any $\vphi \in \Omega^n(M\setminus\B)$ and $\psi \in \Omega^{n-1}(M\setminus\B)$ where $\Omega^n(M\setminus\B)$ is the set of $n$-forms on $M\setminus\B$. 
The partial twisted cohomology is the set of equivalence classes 
\be
    H^p(M\setminus\B;\nabla_\omega)
    &= \frac{\ker\left[ \nabla_\omega: \Omega^p(M\setminus\B) \to \Omega^{p+1}(M\setminus\B) \right]}{\mathrm{im}\left[ \nabla_\omega: \Omega^{p-1}(M\setminus\B) \to \Omega^{p}(M\setminus\B) \right]}
    \,,
    \\
    &= \frac{
        \{ \vphi \in \Omega^p(M\setminus\B) : \nabla_\omega \vphi = 0\}
    }{
        \{ \nabla_\omega \psi : \psi \in \Omega^{p-1}(M\setminus\B) \}
    }
    \,.
\ee

\paragraph{The physical subspace of partially twisted cocycles from $\g$.}
In practice, because the underlying geometry (here, a hyperplane arrangement) is a positive geometry, the partial twisted cohomology can be simply generated by the \emph{canonical form} map \cite{Brown:2025jjg, Arkani-Hamed:2017tmz} 
\footnote{%
    Mathematically, there are still some unresolved subtleties with this map in the context of twisted (co)homology. 
    While the locally-finite twisted homology is isomorphic to the compactly-supported twisted homology, the space of locally-finite twisted chains is much larger than the space of compactly-supported twisted chains.
    In fact, the space of locally-finite twisted chains contains the space of compactly-supported twisted chains. 
    At the time of writing, the canonical form map is only well-defined for locally-finite twisted chains that are ``maximally'' locally-finite (see \cite{Brown:2019jng}). 
    In practice, we find that we span the space of the relative twisted homology with maximally locally-finite chains for which the canonical form map makes sense, so we ignore possible pathologies. 
}
\begin{align} \label{eq:canFormMap}
    {\Omega}: 
    H_{|\V_\G|}(M,\B;\check{\mathcal{L}}) 
    \to H^{|\V_\G|}(M \setminus \B;\nabla_\omega) 
    \,.
\end{align}
Restricting to the space of physical dual cycles $\check{\Gamma}_\g$ produces the physical subspace of differential forms $\vphi_\g := \Omega[\check{\Gamma}_\g]$ with logarithmic singularities on $\partial\check{\Gamma}_\g$. 
An explicit representation for $\vphi_\g$ was provided in equation \eqref{eq:Omega[CheckGamma]}.

\paragraph{Example: canonical forms of the two-site chain.}

The cocycles $\vphi_\g$ for the two-site chain were quoted without derivation in~\eqref{eq:vphi_2chain}. Here we recover them from the general construction \eqref{eq:Omega[CheckGamma]}, which expresses each $\vphi_\g$ as the canonical form of the dual unbounded chamber $\check\Gamma_\g$ built in section~\ref{sec:dualhom}.

Using equation \eqref{eq:Omega[CheckGamma]}, the canonical forms associated to our basis of dual cycles are
\be
    \Omega[\check{\Gamma}_{\twoChaing}]
    &= \dlog B_{\includegraphics[scale=.3,align=c]{figs/2chainB12}}
    \wedge \tilde{\Omega}[\Delta_{\twoChaing}]
    = \dlog B_{\includegraphics[scale=.3,align=c]{figs/2chainB12}}
    \wedge \dlog \frac{x_2}{x_1}
    \,,
    \\
    \Omega[\check{\Gamma}_{\twoChaingg}]
    &= - \dlog B_{\includegraphics[scale=.3,align=c]{figs/2chainB12}}
    \wedge \dlog B_{\includegraphics[scale=.3,align=c]{figs/2chainB1}}
    \,,
    \\
    \Omega[\check{\Gamma}_{\twoChainggg}]
    &= \dlog B_{\includegraphics[scale=.3,align=c]{figs/2chainB12}}
    \wedge \dlog B_{\includegraphics[scale=.3,align=c]{figs/2chainB2}}
    \,,
    \\
    \Omega[\check{\Gamma}_{\twoChaingggg}]
    &= \dlog B_{\includegraphics[scale=.3,align=c]{figs/2chainB1}}
    \wedge \dlog B_{\includegraphics[scale=.3,align=c]{figs/2chainB2}}
    \,.
\ee
The extra sign in $\Omega[\check{\Gamma}_{\twoChaing}]$ is crucial to make the combinatorics of the signs in the kinematic connection simple \cite{Glew:2025ypb}. 
It is easy to verify that these differential forms are indeed canonical forms of the bounded regions in \eqref{sec:2chainHomDual}. 
It is also easy to verify that the maximal cuts of these forms are the canonical form of $\Delta_\g$. 
For example, 
\be
    \res_{\includegraphics[align=c, scale=.3]{figs/2ChainB12.pdf}}\left[
    \Omega[\check{\Gamma}_{\twoChaing}]
    \right]
    &= \res_{\includegraphics[align=c, scale=.3]{figs/2ChainB12.pdf}}\left[
        \dlog B_{\includegraphics[scale=.3,align=c]{figs/2chainB12}}
        \wedge \dlog \frac{x_2}{x_1}
    \right]
    = \dlog\frac{
        x_2\vert_{B_{\includegraphics[scale=.2,align=c]{figs/2chainB12}}=0}
    }{
        x_1\vert_{B_{\includegraphics[scale=.2,align=c]{figs/2chainB12}}=0}
    }
    = \Omega[\Delta_{\twoChaing}]
    \,,
    \\
    \res_{\includegraphics[align=c, scale=.3]{figs/2ChainB1B2.pdf}}\left[
    \Omega[\check{\Gamma}_{\twoChaingggg}]
    \right]
    &= \res_{\includegraphics[align=c, scale=.3]{figs/2ChainB1B2.pdf}}\left[
        \dlog B_{\includegraphics[align=c, scale=.3]{figs/2ChainB1.pdf}} 
        \wedge 
        \dlog B_{\includegraphics[align=c, scale=.3]{figs/2ChainB2.pdf}}
    \right]
    = 1
    = \Omega[\Delta_{\twoChaingggg}]
    \,,
\ee
where $\Delta_{\twoChaing}$ is a line segment and $\Delta_{\twoChaingggg}$ is a point.

\subsection{Positive geometry: from the FRW-homology to dual cohomology}
\label{sec:dualCohom}

In section \ref{sec:dualhom}, we introduced the dual relative twisted homology and described a basis of physical cycles.
These cycles can be paired via integration with dual cocycles---elements of the dual relative twisted cohomology $\check{H}^n_\mathrm{phys}:= H^n(M,\B;\nabla_{-\omega})$---to form dual periods. 
Furthermore, the dual relative twisted cohomology can be paired with the FRW cohomology via intersection numbers to form an inner product on these vector spaces; we encountered these intersection numbers in~\eqref{eq:intersection_numbers}.
Like the FRW homology, the dual relative twisted cohomology can be geometrically decomposed into simple building blocks, each of which can be naturally associated with a physical cut. 

\paragraph{The dual relative twisted cohomology and the physical subspace.}
\label{sec:dualCohomGen}

More formally, the dual cohomology we are discussing is the Poincar\'e dual of the FRW-homology. There exists a \emph{dual canonical form}, which maps a cut contour to a dual form in the relative twisted cohomology
\begin{align} \label{eq:dualCanFormMap}
    {\check{\Omega}}: 
    H_{|\V_\G|}(M \setminus \B;\check{\mathcal{L}}) 
    \to H^{|\V_\G|}(M,\B;\nabla_{-\omega}) 
    \,,
    \quad\text{via}\quad
    \delta_{\Csf_\g} (\Gamma_\g) 
    \mapsto \delta^*_{\Csf_\g} (\Omega[\Gamma_\g]) 
    \,,
\end{align}
where
\begin{align} 
\label{eq:degenDelta*}
    \delta_{\Csf_\g}^*
    &= \sum_{\csf \in \Csf_\g}
        \sgn_{\csf}
        \delta_{\csf}
    \,,
    \qquad
    \delta_\csf^* = \delta_{\tau_{|\csf|}} 
    \circ \cdots \circ \delta_{\tau_1}
    \,,
    \qquad
    \csf = (\tau_1, \dots, \tau_{|\csf|})
    \,,
\end{align}
and $\delta^*_\tau$ is the dual coboundary symbol.
Similar to the coboundary map, the dual coboundary maps a form on a cut to a form in the larger ambient topological space in which the cut is embedded. 

Intuitively, one can think of $\delta_\tau^*(\check{\phi})$ as extending the support of the lower-degree form $\check{\phi}$ which is defined on the cut $M_\tau$ into an infinitesimal tubular neighborhood around the subvariety $\B_\tau$.%
\footnote{%
    More precisely, $\delta_\tau^*$ is a map 
    $\delta_\tau^*: 
    H^{n-1}( (M\setminus\B)\cap\B_\tau;\nabla_{-\omega}\vert_{\B_\tau}) 
    \to H^{n-1}(M\setminus\B;\nabla_{-\omega})
    $ and $\delta_\csf^*$ is the iteration of this map. 
    See \cite{pham2011singularities} for a construction of the coboundary and its dual in the untwisted context. 
    The generalization to the twisted context can be found in \cite{Caron-Huot:2021xqj} or \cite{Caron-Huot:2021iev}. 
}
While one can make $\delta_\csf^*$ explicit as a differential form using bump functions,%
\footnote{%
    More explicitly, the iterated dual coboundary can be represented as
    $
        \delta_\csf^* 
        := \frac{ \check{u} }{ \check{u}\vert_{\B_{\csf}} }
        \bigwedge_{\tau\in\csf} \d\theta_\tau
    $
    where $\theta_\tau = \theta(|B_\tau|>\epsilon)$ is the Heaviside step function (which can be replaced by a smooth bump function) that is unity in an infinitesimal neighborhood  $|B_\tau|<\epsilon \ll 1$. 
}
we will only need the following algebraic properties of the dual coboundary:
\begin{itemize}
    \item[1)] \emph{Multi-index notation}:
    \be
        \delta^*_{\csf'} \circ \delta^*_{\csf} 
        := \delta^*_{\csf, \csf'}
        \,.
    \ee
    
    \item[2)] \emph{Anti-symmetry}: whenever the multi-index $\Csf = (\csf_1, \dots, \csf_k)$ corresponds to a generic normal-crossing component $\B_\Csf$ in the on-shell variety, we have
    \be
        \delta^*_{\Csf} 
        := \sgn(\sigma) \delta^*_{\sigma(\Csf)}
    \ee
    where $\sigma$ is any permutation of $\Csf$. 
    
    \item[3)] \emph{Identities}: the coboundary operators satisfy the same identities as $\delta_\Csf$ and $\res_\Csf$ when the multi-index $\Csf$ corresponds to a component $\B_\Csf$ of the on-shell variety that is \emph{not} normal-crossing.

    \item[4)] \emph{Differentiation}: upon differentiation, the coboundary gives rise to boundary terms, namely
    \be \label{eq:ddeta*}
        \nabla_{-\omega} \delta^*_\csf(\phi) 
        := (-1)^{|\csf|} \left[
            \delta^*_\csf\Big( 
                (\nabla_{-\omega}\vert_{\B_\csf}) \phi
            \Big)
            - \sum_{\tau \notin \csf} \delta^*_{\csf\tau}(\phi\vert_{\B_{\csf\tau}})
        \right]\,.
    \ee
    These boundary terms are lower degree forms on cuts that can appear when using Stokes theorem (we make this explicit below). 
    
\end{itemize}
Like the partially twisted homology,  the relative twisted cohomology has a direct sum decomposition. 
While this is a property of the full relative twisted cohomology, we only exhibit this decomposition for the physical subspace: 
\begin{align} 
    \begin{aligned} \label{eq:physDualCohom}
        \check{H}^{|\V_\G|}_\text{phys} := 
        \bigoplus_\g
        \; 
            \delta_{\Csf_\g}^* \Big( 
                H^{{|\V_\G|}-|\csf|}(
                    M_\g ; \nabla_{-\omega}\vert_{\B_{\Csf_\g}}
                ) 
            \Big)
        \subseteq  H^{|\V_\G|}(M,\B; \nabla_{-\omega}) 
        \,,
    \end{aligned}
\end{align}
where $\delta^*_{\Csf_\g} := \sum_{\csf\in\Csf_\g} \sgn_\csf \delta_\csf^*$ like in \eqref{eq:degenDelta}. 
The relative twisted homology characterizes the ``interesting'' integrands on each cut $M_\g$---including those produced using Stokes theorem.  
Then, since each $M_\g$ is a positive geometry (with a single bounded chamber $\Delta_\g$), the twisted homology is built from the direct sum of twisted cohomologies of positive geometries.

Not surprisingly, we can pair dual cohomology classes with dual homology classes to construct dual periods
    \be \label{eq:delta*gamma}
        \int_{\check{\gamma}} \check{u}\; \delta^*_\csf(\vphi)
        = \int_{\partial_\csf \check{\gamma}} \check{u}\vert_{\B_\csf}\; \vphi
    \ee
where $\vphi \in H^{n-|\csf|}(M_\csf;\nabla_{-\omega}\vert_{\B_\csf})$, and $\partial_\csf \check{\gamma} \in H_{n-|\csf|}(M_\csf;\check{\mathcal{L}}\vert_{\B_\csf})$ are (co)cycles on the cut $M_\csf$. 
Inside an integral, the dual coboundary induces a boundary on the dual contour such that the integral pairs a contour and form on the same cut (topological space). 

Moreover, the boundary terms produced by differentiation of the dual coboundary ensure that exact dual forms vanish after integration: $\int_{\check{\gamma}} \check{u}\; \nabla_{-\omega} \delta^*_\csf(\phi) = 0$. 
Explicitly, 
\begin{align}
    \int_{\check{\gamma}} \check{u}\; 
    \nabla_{-\omega} \delta^*_\csf(\phi)
    &= \int_{\check{\gamma}} \check{u}\; (-1)^{|\csf|} \left[
        \delta^*_\csf\Big( 
           (\nabla_{-\omega}\vert_{\B_\csf}) \phi
        \Big)
        - \sum_{\tau \notin \csf} \delta^*_{\csf\tau}(\phi\vert_{\B_{\csf\tau}})
    \right]
    \,,
   \nn\\
    &= (-1)^{|\csf|} \left[
        \int_{\partial_{\csf}\check{\gamma}} (\check{u}\vert_{\B_\csf})\;
            (\nabla_{-\omega}\vert_{\B_\csf}) \phi
        - \sum_{\tau \notin \csf} 
            \int_{\partial_{\csf\tau}\check{\gamma}} (\check{u}\vert_{\B_{\csf\tau}})\;
            (\phi\vert_{\B_{\csf\tau}})
    \right]
    \,,
    \nn\\
    &= (-1)^{|\csf|} \left[
        \int_{\partial(\partial_{\csf}\check{\gamma})} (\check{u}\vert_{\B_\csf})\; \phi
        - \sum_{\tau \notin \csf} 
            \int_{\partial_{\csf\tau}\check{\gamma}} (\check{u}\vert_{\B_{\csf\tau}})\;
            (\phi\vert_{\B_{\csf\tau}})
    \right]
    \,,
    \nn\\
    &= (-1)^{|\csf|} \sum_{\tau \notin \csf} 
    \left[
        \int_{\partial_\tau\partial_{\csf}\check{\gamma}} (\check{u}\vert_{\B_\csf\tau})\; (\phi\vert_{\B_\tau})
        - \int_{\partial_{\csf\tau}\check{\gamma}} (\check{u}\vert_{\B_{\csf\tau}})\;
            (\phi\vert_{\B_{\csf\tau}})
    \right]
    = 0
    \,.
\end{align}
Here, we have used \eqref{eq:ddeta*} in the first line, and equation \eqref{eq:delta*gamma} in the second line. 
In the third line, we have decomposed the boundary operator into the parts that project onto one boundary component of $\B$: $\partial (\partial_\csf\check{\gamma}) = \sum_{\tau\notin\csf} \partial_\tau \partial\csf\check{\gamma}$. 
Finally, in the last line we have used the identification $\partial_\tau \circ \partial_\csf := \partial_{\csf\tau}$. 
Therefore, as usual, we can add the total derivative of any dual form without changing the value of the original integral. 

While we have only defined a subspace of the relative twisted cohomology indirectly through the association \eqref{eq:dualCanFormMap}, in general, it is the space of dual forms that are closed and not themselves total derivatives.



\paragraph{Example: the canonical dual forms of the two-site chain.}
Here, we build the dual cocycles $\check\vphi_\g$ as canonical forms of the FRW cycles $\gamma_\g$ from section~\ref{sec:2chainInto}. 
These supply one factor of the cohomology intersection pairing $\la\check\vphi_\g\vert\vphi_\h\ra$ used in the master coaction formula~\eqref{eq:coactionMaster}.

Relative cohomology tracks the boundary terms that arise from Stokes’ theorem when the integration contours end on the untwisted hyperplanes $\B_\tau$. For the two-site chain, the twisted variety $\B$ has three components, 
$$
    \B = \B_{\includegraphics[align=c, scale=.4]{figs/2ChainB1.pdf}} 
    \cup \B_{\includegraphics[align=c, scale=.4]{figs/2ChainB2.pdf}}
    \cup \B_{\includegraphics[align=c, scale=.4]{figs/2ChainB12.pdf}}
$$ 
where, as usual, $\B_\tau = V(B_\tau)$ and the $B_\tau$'s for the two-site chain graph are given in \eqref{eq:2chainBs}. 
The on-shell variety was depicted in equation \eqref{eq:2chainTvarBvar}, while the full hyperplane arrangement along with its cut spaces was depicted in figure \ref{eq:2ChainStrat}.
From \eqref{eq:2chainRg} we have 
\begin{align} \label{eq:2chainDualForms}
    \check{\vphi}_{\includegraphics[scale=.3]{figs/2chain_gs}}
    &= \delta^*_{\includegraphics[scale=.3]{figs/2chainCs.pdf}}
    \left(\Omega\left[
       \includegraphics[align=c,scale=.5]{figs/2chainM12.pdf}
    \right]\right)
    = \delta^*_{\includegraphics[scale=.3]{figs/2chainCs.pdf}}
    \left(
        \dlog\frac{x_2}{x_1}
        \Big\vert_{\includegraphics[scale=.3]{figs/2chainCs.pdf}}
    \right)
    = \delta^*_{\includegraphics[scale=.3]{figs/2ChainB12.pdf}}
    \left(
        \dlog\frac{
            x_2
        }{-x_2 - X_1-X_2}
    \right)
    \,,
    \nn\\[.2em]
    \check{\vphi}_{\includegraphics[scale=.3]{figs/2chain_gr}}
    &= \delta^*_{\includegraphics[scale=.3]{figs/2chainCr.pdf}}\left(\Omega\left[
        \includegraphics[scale=.5,align=c]{figs/2ChainM12_1.pdf}
    \right]\right)
    = \delta^*_{\includegraphics[scale=.3]{figs/2chainCr.pdf}}\left(1\right)
    = \delta^*_{\includegraphics[scale=.3]{figs/2chainB12.pdf},\includegraphics[scale=.3]{figs/2chainB1.pdf},}\left(1\right)
    \,,
    \nn\\[.2em]
    \check{\vphi}_{\includegraphics[scale=.3]{figs/2chain_gl}}
    &= \delta^*_{\includegraphics[scale=.3]{figs/2chainCr.pdf}}\left(\Omega\left[
        \includegraphics[scale=.5,align=c]{figs/2ChainM12_2.pdf}
    \right]\right)
    = \delta^*_{\includegraphics[scale=.3]{figs/2chainCl.pdf}}\left(1\right)
    = \delta^*_{\includegraphics[scale=.3]{figs/2chainB12.pdf},\includegraphics[scale=.3]{figs/2chainB2.pdf}}\left(1\right)
    \,,
    \nn\\[.2em]
    \check{\vphi}_{\includegraphics[scale=.3]{figs/2chain_gb}}
    &= \delta^*_{\includegraphics[scale=.3]{figs/2chainCr.pdf}}\left(\Omega\left[
        \includegraphics[scale=.5,align=c]{figs/2ChainM1_2.pdf}
    \right]\right)
    = \delta^*_{\includegraphics[scale=.3]{figs/2chainCb.pdf}}\left(1\right)
    = \delta^*_{\includegraphics[scale=.3]{figs/2chainB1.pdf},\includegraphics[scale=.3]{figs/2chainB2.pdf},}\left(1\right)
    \,,
\end{align}
where, in the first line above, we chose to eliminate $x_1$ when restricting to the loci $B_{\includegraphics[scale=.3]{figs/2ChainB12.pdf}}=0$.

Each form in \eqref{eq:2chainDualForms} generates the twisted cohomology of its corresponding cut and exemplifies the decomposition \eqref{eq:physDualCohom}:
\be 
    \check{H}^2_\mathrm{phys} 
    &= H^2\left(
        \mathbb{C}^2\setminus
        \includegraphics[align=c,scale=.3]{figs/2ChainVarT.pdf}, 
        \includegraphics[align=c,scale=.3]{figs/2ChainVarB.pdf}
        ; \nabla_{-\omega}
    \right)
    \\ &= \delta^*_{
         \includegraphics[scale=.3, align=c]{figs/2ChainB12}
    } \circ H^{1}\left(
        \includegraphics[scale=.5, align=c]{figs/2ChainM12}
        ; \nabla_{-\omega}\vert_{\B_{ \includegraphics[scale=.3, align=c]{figs/2ChainB12}}}
    \right)
    \\& \qquad
    \oplus \delta^*_{
        \includegraphics[scale=.3, align=c]{figs/2ChainCb}
    } \circ H^{0}\left(
        \includegraphics[scale=.5, align=c]{figs/2ChainM1_2}
    \right)
    \oplus \delta^*_{
         \includegraphics[scale=.3, align=c]{figs/2ChainCr}
    } \circ H^{0}\left(
        \includegraphics[scale=.5, align=c]{figs/2ChainM12_1}
    \right)
    \oplus \delta^*_{
         \includegraphics[scale=.3, align=c]{figs/2ChainCl}
    } \circ H^{0}\left(
        \includegraphics[scale=.5, align=c]{figs/2ChainM12_2}
    \right)
    \,.
\ee
Note that $\omega$ vanishes when restricted to a point. 
Therefore, the dual covariant derivative reduces to the usual exterior derivative on a point $\nabla_{-\omega}\vert_{\mathrm{point}} = \d$. 
The two-site chain is also a special case in which $H^2_\mathrm{phys}$ is equal to the full relative twisted cohomology.

\subsection{The intersection number as an inner product}
\label{sec:intersectionPairing}

In this section, we provide a lightning review of intersection theory and the importance of the (dual) canonical forms for formulating the coaction.
A pictorial summary of these building blocks and how they are related to each other was already presented in figure \ref{fig:pairings}. 

The (co)homology intersection number is a non-degenerate pairing between elements of the FRW (co)homology and dual (co)homology
\begin{align}
    \la\check{\vphi}\vert\vphi\ra:& 
    H^{|\V_\G|}(M,\B;\nabla_{-\omega}) 
    \times 
    H^{|\V_\G|}(M\setminus\B;\nabla_\omega) 
    \to \mathbb{C}[[\alpha_\bullet^\pm,X_\bullet^\pm,Y_\bullet^\pm]] \,,
    \\
    [\gamma\vert\check{\gamma}]: &
    H_{|\V_\G|}(M\setminus\B;\mathcal{L}) 
    \times 
    H_{|\V_\G|}(M,\B;\check{\mathcal{L}})
    \to \mathbb{C}[[e^{\pm i\pi \alpha_\bullet}, (1-e^{\pm i\pi \alpha_\bullet }) ]] \,.
\end{align}
Here, $\mathbb{C}[[x,y,\dots]]$ denotes the space of Laurent polynomials functions in the variables $\{x,y,\dots\}$ with $\mathbb{C}$-coefficients. 
These pairings can be used like an inner product to project any element of the (co)homology onto a chosen basis. 
Moreover, we have the following resolution of identity
\be
    \mathds{1} &= 
    \vert\vphi_a\ra C_{ab}^{-1} \la\check{\vphi}_b\vert
    \,,
\ee
where $\{\vphi_a\}$ and $\{\check{\vphi}_a\}$ 
constitute bases of the cohomology and dual cohomology,
and $C_{ab} = \la \check{\vphi}_a \vert \vphi_b \ra$ is 
the (co)homology intersection matrix. 

In the setting of partially twisted FRW hyperplane arrangements and logarithmic differential forms, the cohomology intersection number simplifies to a specific sequential residue built from the data of a dual form. 
For any cut $\csf$ (possibly not physical) in the on-shell variety and any bounded region $\Delta_\csf$ on $M_\csf$, one can define the dual form $\delta^*_\csf(\Omega[\Delta_\csf])$. 
In the intersection number, this dual-form induces the following action on any FRW-form $\vphi$
\be\label{eq:genCohomIntNum}
    \la 
        \delta_\csf^*(\Omega[\Delta_\csf])
    \vert
        \vphi 
    \ra
    = \sum_{\mbf{w} \in \V_{\Delta_\csf}} 
    \frac{1}{\prod_{i}^{|\mbf{w}|} \alpha_{w_i}}
    \res_{x_{w_1}\vert_{\B_\csf}=0} \circ \cdots \circ \res_{x_{w_{|\mathbf{w}|}}\vert_{\B_\csf}=0}
    \circ \res_{\csf}[\vphi]
    \,.
\ee
Here, $\V_{\Delta_\csf}$ is the set of all vertices of the bounded region $\Delta_\csf$; each element, $\mathbf{w}$, of this set indexes which twisted hyperplanes meet at a given vertex. 
Equation \eqref{eq:genCohomIntNum} is enough to prove all results following from intersection theory in this work; there are corrections when $\vphi$ or the argument of $\delta_\csf^*$ have higher order poles. 
In such situations, see \cite{Caron-Huot:2021xqj,Caron-Huot:2021iev, matsumoto} for more details. 
Using \eqref{eq:genCohomIntNum} and the basis constructed in sections \ref{sec:FRWCohom} and \ref{sec:dualCohom}, the intersection number between physical forms is 
\be \label{eq:Cggp}
    C_{\g\g'} 
    := \la \check{\vphi}_{\g_1}
    \vert \vphi_{\g_2} \ra 
    = \delta_{\g\g'}\;  
        |\Csf_\g| \; 
        \prod_{\rsf \in \Rsf_\g} 
        \frac{
            \sum_{v\in\V_\rsf} \alpha_v
        }{
            \prod_{v\in\V_\rsf} \alpha_v
        }
\ee
where $\delta_{\g\g'}$ is the Kronecker delta symbol and $|\Csf_\g|$ is the cardinality of the set $\Csf_\g$.

\subsection{Partial fractions, zonotopes and the flow of cuts/derivatives}
\label{sec:partialFracAndZono}

In this section, we review the partial fractions decomposition of the physical FRW-form $\vphi_\G$ into the cut basis of section \ref{sec:hom}. 
The partial-fractioned expression for $\vphi_\G$ manifests the claims about the independent physical residues in sections \ref{sec:degenArr} and \ref{sec:ResolveDegen}. 
It also leads to the identification of \emph{graphical zonotopes} in the on-shell variety. 
These zonotopes encode the combinatorics of the sequential residues and the differential equations of the physical subspace. 
These combinatorics will be used to simplify sums in the coaction. 

\paragraph{Partial fractions formula for the wavefunction.}

Let $\mathcal{A}_\E(\G)$ be the set of acyclic minors with fixed broken edge set $\E$ and $\mathcal{A}_\E^\varnothing(\G)$ be the subset of $\mathcal{A}_\E(\G)$ with no pinched edges
\begin{align}
    \mathcal{A}_{\E}(\G) = \{ 
        \g \in \mathcal{A}(\G) 
        : \E^{\!\brokenEdge{.8}}_\g = \E 
    \} 
    \supset 
    \mathcal{A}^{\varnothing}_{\E}(\G) := \{ 
        \g \in \mathcal{A}_\E(\G) 
        : \E^{\!\solidEdge{.8}}_\g = \varnothing 
    \}.
\end{align}
where $\E^{\!\brokenEdge{.8}}_\g$ and $\E^{\!\solidEdge{.8}}_\g$ denote the set of broken and pinched edges of the minor $\g$.
Then, the physical wavefunction has the following alternative expression that is fully partial fractioned in the $x_i$ variables \cite{Glew:2025ypb}:
\begin{align} \label{eq:partialFractionedPsi}
    \vphi_\G &= \sum_{\E \subset \E_\G} (-1)^{|\E|} \sum_{\g \in \mathcal{A}^{\varnothing}_\E(\G)} 
    \vphi_{\g}
    \,.
\end{align}

For example, consider the three-site chain. 
In the language of \eqref{eq:partialFractionedPsi}, the physical FRW-form is
\begin{align}\begin{aligned} \label{eq:3chainPsi}
    \vphi_{\includegraphics[align=c,scale=.2]{figs/3chain/3chain.pdf}} =&
    \left( 
        \vphi_{\includegraphics[align=c,scale=.2]{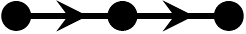}}
        + \vphi_{\includegraphics[align=c,scale=.2]{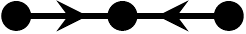}}
        + \vphi_{\includegraphics[align=c,scale=.2]{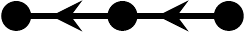}}
        + \vphi_{\includegraphics[align=c,scale=.2]{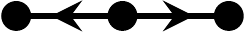}}
    \right)
    \\
    &-\left(
        \vphi_{\includegraphics[align=c,scale=.2]{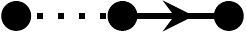}}
        + \vphi_{\includegraphics[align=c,scale=.2]{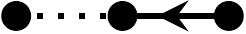}}
    \right)
    \\
    &-\left(
        \vphi_{\includegraphics[align=c,scale=.2]{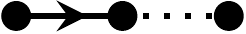}}
        + \vphi_{\includegraphics[align=c,scale=.2]{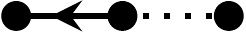}}
    \right)
    \\
    &+ \left(  
        \vphi_{\includegraphics[align=c,scale=.2]{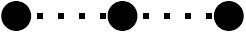}}
    \right)
    \,.
\end{aligned}\end{align}
With the exception of one acyclic minor, each $\g$ corresponds to a unique set of cut-tubings $\Csf_\g$. 
Thus, most $\dlog\Csf_\g$ are a single term:
\begin{align}\begin{aligned}
    \vphi_{\includegraphics[align=c,scale=.2]{figs/3chain/3chain_rr.pdf}} &= \dlog B_{\includegraphics[align=c,scale=.5]{figs/3chainT1.pdf}}
    \wedge
    \dlog B_{\includegraphics[align=c,scale=.5]{figs/3chainT12.pdf}}
    \wedge
    \dlog B_{\includegraphics[align=c,scale=.5]{figs/3chainT123.pdf}}
    \,,
    \\
    \vphi_{\includegraphics[align=c,scale=.2]{figs/3chain/3chain_bl.pdf}} &= \dlog B_{\includegraphics[align=c,scale=.5]{figs/3chainT1.pdf}}
    \wedge
    \dlog B_{\includegraphics[align=c,scale=.5]{figs/3chainT23.pdf}}
    \wedge
    \dlog B_{\includegraphics[align=c,scale=.5]{figs/3chainT3.pdf}}
    \,,
    \\
    \vphi_{\includegraphics[align=c,scale=.2]{figs/3chain/3chain_bb.pdf}} &= \dlog B_{\includegraphics[align=c,scale=.5]{figs/3chainT1.pdf}}
    \wedge
    \dlog B_{\includegraphics[align=c,scale=.5]{figs/3chainT2.pdf}}
    \wedge
    \dlog B_{\includegraphics[align=c,scale=.5]{figs/3chainT3.pdf}}
    \,.
\end{aligned}\end{align}
The $\dlog$'s that correspond to a degenerate cut are captured by a single linear combination:
\begin{align}\begin{aligned}\label{eq:phi3chian_lr}
    \vphi_{\includegraphics[align=c,scale=.2]{figs/3chain/3chain_lr.pdf}} 
    &= \dlog B_{\includegraphics[align=c,scale=.5]{figs/3chainT12.pdf}}
        \wedge
        \dlog B_{\includegraphics[align=c,scale=.5]{figs/3chainT2.pdf}}
        \wedge
        \dlog B_{\includegraphics[align=c,scale=.5]{figs/3chainT123.pdf}}
    \\&\quad
    + \dlog B_{\includegraphics[align=c,scale=.5]{figs/3chainT123.pdf}}
        \wedge
        \dlog B_{\includegraphics[align=c,scale=.5]{figs/3chainT2.pdf}}
        \wedge
        \dlog B_{\includegraphics[align=c,scale=.5]{figs/3chainT23.pdf}}
        \,.
\end{aligned}\end{align}
There is a strong sense in which  \eqref{eq:partialFractionedPsi} better represents the structure of FRW cosmological integrals than \eqref{eq:Psi}. 
It manifests the fact that each cut $\Csf_\g$ contributes a specific combination of $\dlog$'s as claimed in sections \ref{sec:degenArr} and \ref{sec:ResolveDegen}. 

\begin{figure}
    \centering
    \includegraphics[width=.55\textwidth,align=c]{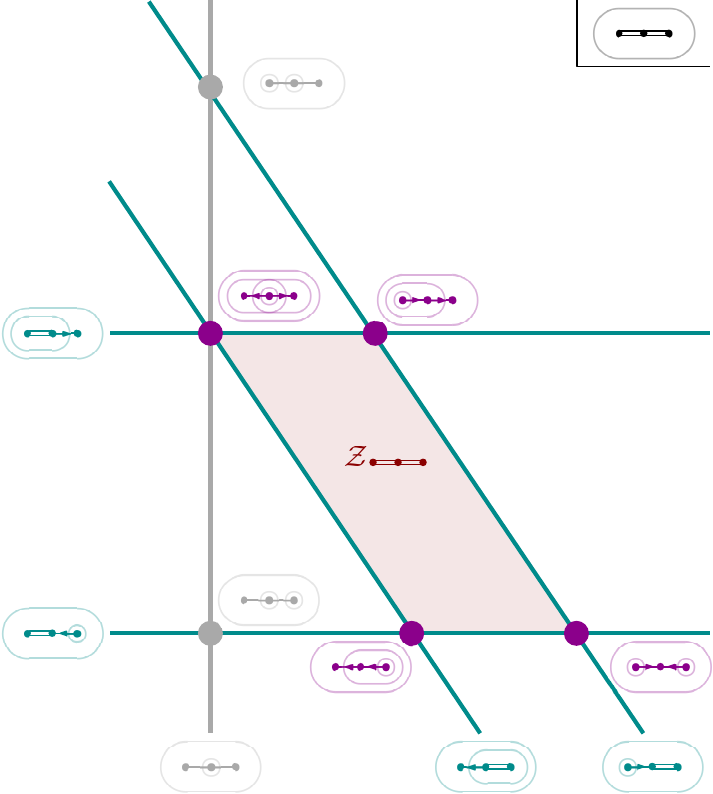} 
    \hspace{2em}
    \includegraphics[width=.3\textwidth,align=c]{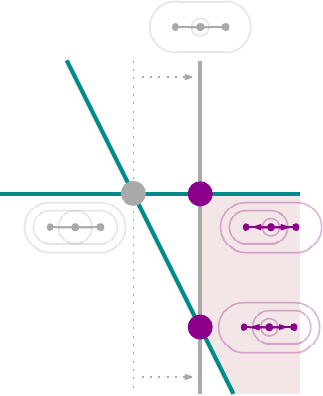} 
    \caption{%
    The on-shell variety restricted to the total energy plane of the three-site chain graph     
    ($\B \cap \B_{\includegraphics[scale=.5,align=c]{figs/3chainT123}}$, not the cut space 
    $
        M_{\includegraphics[scale=.2,align=c]{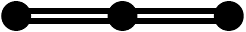}}
        = \T \cap \B_{\includegraphics[scale=.5,align=c]{figs/3chainT123}}
    $). 
    The {\color{darkCyan}cyan} and {\color{gray}gray} lines give the intersection of the total energy plane and one of the other on-shell planes $\B_\tau$.
    The points in {\color{darkMagenta} magenta} correspond to the maximal cuts associated to the acyclic minors 
    \includegraphics[align=c,scale=.2]{figs/3chain/3chain_rr.pdf},
    \includegraphics[align=c,scale=.2]{figs/3chain/3chain_rl.pdf},
    \includegraphics[align=c,scale=.2]{figs/3chain/3chain_ll.pdf},
    and,
    \includegraphics[align=c,scale=.2]{figs/3chain/3chain_lr.pdf}; 
    all of these appear in \eqref{eq:3chainSquareNNMaxCuts}.
    Note that \includegraphics[align=c,scale=.2]{figs/3chain/3chain_lr.pdf} corresponds to the intersection of four propagators and is a degenerate point of the on-shell variety. 
    One way to resolve this degeneracy is to move the {\color{gray}gray} line in the direction normal to itself (see right above). 
    After deforming the {\color{gray}gray} line, the canonical form of the {\color{BrickRed}red} region gets contributions from five vertices. 
    In fact,  $\vphi_{\includegraphics[align=c,scale=.2]{figs/3chain/3chain_lr.pdf}}$ is the sum of two terms corresponding to the two vertices after resolving the degeneracy as depicted in the right figure above. 
    }
    \label{fig:3chainTotEnergyZono}
\end{figure}

\paragraph{Zonotopes in the on-shell variety.}
Here, we show that the residues of the physical FRW-form $\vphi_\G$ are canonical forms of graphical zonotopes. 
A zonotope is a convex polytope that is the Minkowski sum of finitely many line segments. 
Some familiar three-dimensional examples are the triangular prism, cube, and rhombic dodecahedron. 
The above three-dimensional examples are also examples of graphical zonotopes. 
For graphical zonotopes, the line segments in the Minkowski sum correspond to the edges of $G$, encapsulating both combinatorial and geometric data.

To see how graphical zonotopes arise in our context, consider taking the $B_{\includegraphics[scale=.2]{figs/3chain/3chain_ss.pdf}}=0$ residue of the form $\vphi_{\includegraphics[align=c,scale=.2]{figs/3chain/3chain.pdf}}$:
\be \label{eq:3chainSquareNNMaxCuts}
    \res_{\includegraphics[scale=.2]{figs/3chain/3chain_ss.pdf}}[
        \vphi_{\includegraphics[align=c,scale=.2]{figs/3chain/3chain.pdf}}
    ]
    &= \res_{\includegraphics[scale=.2]{figs/3chain/3chain_ss.pdf}}[
        \vphi_{\includegraphics[align=c,scale=.2]{figs/3chain/3chain_rr.pdf}}
        + \vphi_{\includegraphics[align=c,scale=.2]{figs/3chain/3chain_rl.pdf}}
        + \vphi_{\includegraphics[align=c,scale=.2]{figs/3chain/3chain_ll.pdf}}
        + \vphi_{\includegraphics[align=c,scale=.2]{figs/3chain/3chain_lr.pdf}}
    ]
    \,.
\ee
The $B_{\includegraphics[scale=.2]{figs/3chain/3chain_ss.pdf}}=0$ residue annihilates all $\vphi_\g$ for $\g \notin \mathcal{A}^{\varnothing}_{\includegraphics[scale=.2]{figs/3chain/3chain_ss.pdf}}(\includegraphics[align=c,scale=.2]{figs/3chain/3chain.pdf})$. 
Examining the surviving terms reveals that 
$\res_{\includegraphics[scale=.2]{figs/3chain/3chain_ss.pdf}}[
        \vphi_{\includegraphics[align=c,scale=.2]{figs/3chain/3chain.pdf}}
]$ is a two-form that has singularities on the boundary of the {\color{BrickRed}red} shaded rectangle (zonotope) labeled by $\mathcal{Z}_{\includegraphics[scale=.2]{figs/3chain/3chain_ss.pdf}}$ in figure \ref{fig:3chainTotEnergyZono}. 
In fact, on the total energy cut, the physical form $\vphi_{\includegraphics[scale=0.2]{figs/3chain/3chain.pdf}}$ is proportional to the canonical form of the zonotope $\mathcal{Z}_{\includegraphics[scale=.2]{figs/3chain/3chain_ss.pdf}}$ 
$
    \res_{\includegraphics[scale=.2]{figs/3chain/3chain_ss.pdf}}[
        \vphi_{\includegraphics[align=c,scale=.2]{figs/3chain/3chain.pdf}}
    ]
    \propto \Omega[\mathcal{Z}_{\includegraphics[scale=.2]{figs/3chain/3chain_ss.pdf}}]
$. 

\begin{figure}
    \centering
    \includegraphics[width=.55\textwidth,align=c]{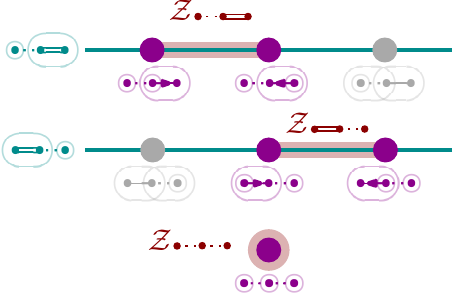}  
    \caption{%
        The {\color{darkCyan} cyan}
        lines correspond to the intersection of two propagators. 
        On these lines, there are {\color{darkMagenta} magenta} and {\color{gray} gray} points where three propagators intersect; only the {\color{darkMagenta} magenta} points are physical.
    }
    \label{fig:3chainZonoOther}
\end{figure}

To verify this, we can compute all other residues $B_\tau = 0$ for $\tau \in \{ \includegraphics[scale=.5]{figs/3chainT1}, \includegraphics[scale=.5]{figs/3chainT3}, \includegraphics[scale=.5]{figs/3chainT12}, \includegraphics[scale=.5]{figs/3chainT23} \}$. This yields the canonical forms of the boundaries of $\mathcal{Z}_{\includegraphics[scale=.2]{figs/3chain/3chain_ss.pdf}}$:
\be \label{eq:3chainSquareNMaxCuts}
    \res_{\includegraphics[scale=.2]{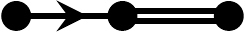}}[
        \vphi_{\includegraphics[align=c,scale=.2]{figs/3chain/3chain.pdf}}
    ] &= \res_{\includegraphics[scale=.2]{figs/3chain/3chain_rs.pdf}}[
        \vphi_{\includegraphics[align=c,scale=.2]{figs/3chain/3chain_rr.pdf}}
        + \vphi_{\includegraphics[align=c,scale=.2]{figs/3chain/3chain_rl.pdf}}
    ]
    \propto \Omega\left[\includegraphics[align=c,scale=.5]{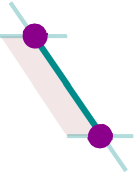}\right]
    =: \Omega\left[\mathcal{Z}_{\includegraphics[scale=.2]{figs/3chain/3chain_rs.pdf}}\right]
    \,,
    \\
    \res_{\includegraphics[scale=.2]{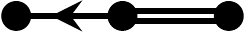}}[
        \vphi_{\includegraphics[align=c,scale=.2]{figs/3chain/3chain.pdf}}
    ] &= \res_{\includegraphics[scale=.2]{figs/3chain/3chain_ls.pdf}}[
        \vphi_{\includegraphics[align=c,scale=.2]{figs/3chain/3chain_ll.pdf}}
        + \vphi_{\includegraphics[align=c,scale=.2]{figs/3chain/3chain_lr.pdf}}
    ]
    \propto \Omega\left[\includegraphics[align=c,scale=.5]{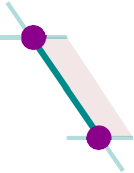}\right]
    =: \Omega\left[\mathcal{Z}_{\includegraphics[scale=.2]{figs/3chain/3chain_ls.pdf}}\right]
    \,,
    \\
    \res_{\includegraphics[scale=.2]{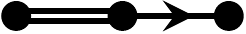}}[
        \vphi_{\includegraphics[align=c,scale=.2]{figs/3chain/3chain.pdf}}
    ] &= \res_{\includegraphics[scale=.2]{figs/3chain/3chain_sr.pdf}}[
        \vphi_{\includegraphics[align=c,scale=.2]{figs/3chain/3chain_rr.pdf}}
        + \vphi_{\includegraphics[align=c,scale=.2]{figs/3chain/3chain_lr.pdf}}
    ]
    \propto \Omega\left[\includegraphics[align=c,scale=.5]{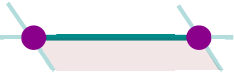}\right]
    =: \Omega\left[\mathcal{Z}_{\includegraphics[scale=.2]{figs/3chain/3chain_sr.pdf}}\right]
    \,,
    \\
    \res_{\includegraphics[scale=.2]{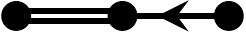}}[
        \vphi_{\includegraphics[align=c,scale=.2]{figs/3chain/3chain.pdf}}
    ] &= \res_{\includegraphics[scale=.2]{figs/3chain/3chain_sl.pdf}}[
        \vphi_{\includegraphics[align=c,scale=.2]{figs/3chain/3chain_rl.pdf}}
        + \vphi_{\includegraphics[align=c,scale=.2]{figs/3chain/3chain_ll.pdf}}
    ]
    \propto \Omega\left[\includegraphics[align=c,scale=.5]{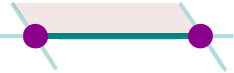}\right]
    =: \Omega\left[\mathcal{Z}_{\includegraphics[scale=.2]{figs/3chain/3chain_sl.pdf}}\right]
    \,.
\ee
Each of the above forms is a one-form on the {\color{darkCyan}cyan} lines in figure \ref{fig:3chainZonoOther}, and, is singular on the {\color{darkMagenta} magenta} points that lie on said line. 
Finally, we take the maximal cuts to find that 
\be \label{eq:3chainSquareMaxCuts}
    \res_{\includegraphics[scale=.2]{figs/3chain/3chain_rr.pdf}}[
        \vphi_{\includegraphics[align=c,scale=.2]{figs/3chain/3chain.pdf}}
    ] &= \res_{\includegraphics[scale=.2]{figs/3chain/3chain_rr.pdf}}[
        \vphi_{\includegraphics[align=c,scale=.2]{figs/3chain/3chain_rr.pdf}}
    ] \propto \Omega\left[\includegraphics[align=c,scale=.5]{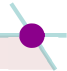}\right]
    =: \Omega\left[\mathcal{Z}_{\includegraphics[scale=.2]{figs/3chain/3chain_ll.pdf}}\right]
    \,,
    \\
    \res_{\includegraphics[scale=.2]{figs/3chain/3chain_ll.pdf}}[
        \vphi_{\includegraphics[align=c,scale=.2]{figs/3chain/3chain.pdf}}
    ] &= \res_{\includegraphics[scale=.2]{figs/3chain/3chain_ll.pdf}}[
        \vphi_{\includegraphics[align=c,scale=.2]{figs/3chain/3chain_ll.pdf}}
    ] \propto \Omega\left[\includegraphics[align=c,scale=.5]{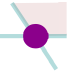}\right]
    =: \Omega\left[\mathcal{Z}_{\includegraphics[scale=.2]{figs/3chain/3chain_rr.pdf}}\right]
    \,,
    \\
    \res_{\includegraphics[scale=.2]{figs/3chain/3chain_lr.pdf}}[
        \vphi_{\includegraphics[align=c,scale=.2]{figs/3chain/3chain.pdf}}
    ] &= \res_{\includegraphics[scale=.2]{figs/3chain/3chain_lr.pdf}}[
        \vphi_{\includegraphics[align=c,scale=.2]{figs/3chain/3chain_lr.pdf}}
    ] \propto \Omega\left[\includegraphics[align=c,scale=.5]{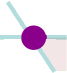}\right]
    =: \Omega\left[\mathcal{Z}_{\includegraphics[scale=.2]{figs/3chain/3chain_lr.pdf}}\right]
    \,,
    \\
    \res_{\includegraphics[scale=.2]{figs/3chain/3chain_rl.pdf}}[
        \vphi_{\includegraphics[align=c,scale=.2]{figs/3chain/3chain.pdf}}
    ] &= \res_{\includegraphics[scale=.2]{figs/3chain/3chain_rl.pdf}}[
        \vphi_{\includegraphics[align=c,scale=.2]{figs/3chain/3chain_rl.pdf}}
    ] \propto \Omega\left[\includegraphics[align=c,scale=.5]{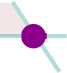}\right]
    =: \Omega\left[\mathcal{Z}_{\includegraphics[scale=.2]{figs/3chain/3chain_rl.pdf}}\right]
    \,.
\ee
Since the residues in \eqref{eq:3chainSquareNMaxCuts} and \eqref{eq:3chainSquareMaxCuts} are canonical forms of the   codimension-one and -two boundaries of $\mathcal{Z}_{\includegraphics[scale=.2]{figs/3chain/3chain_ss.pdf}}$, 
$ 
    \res_{\includegraphics[scale=.2]{figs/3chain/3chain_ss.pdf}}[
        \vphi_{\includegraphics[align=c,scale=.2]{figs/3chain/3chain.pdf}}
    ]
$
is indeed the canonical form of $\mathcal{Z}_{\includegraphics[scale=.2]{figs/3chain/3chain_ss.pdf}}$. 

An analogous story holds for each line of \eqref{eq:3chainPsi}.
The forms 
\be
    \res_{\includegraphics[scale=.2]{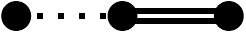}}[
        \vphi_{\includegraphics[align=c,scale=.2]{figs/3chain/3chain.pdf}}
    ]&= -
    \res_{\includegraphics[scale=.2]{figs/3chain/3chain_bs.pdf}}[
        \vphi_{\includegraphics[align=c,scale=.2]{figs/3chain/3chain_br.pdf}}
        + \vphi_{\includegraphics[align=c,scale=.2]{figs/3chain/3chain_bl.pdf}}
    ] 
    \propto \Omega[\mathcal{Z}_{\includegraphics[scale=.2]{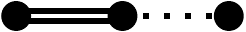}}]
    \,,
    \\
    \res_{\includegraphics[scale=.2]{figs/3chain/3chain_sb.pdf}}[
        \vphi_{\includegraphics[align=c,scale=.2]{figs/3chain/3chain.pdf}}
    ]&= -
    \res_{\includegraphics[scale=.2]{figs/3chain/3chain_sb.pdf}}[
        \vphi_{\includegraphics[align=c,scale=.2]{figs/3chain/3chain_rb.pdf}}
        + \vphi_{\includegraphics[align=c,scale=.2]{figs/3chain/3chain_lb.pdf}}
    ]
    \propto \Omega[\mathcal{Z}_{\includegraphics[scale=.2]{figs/3chain/3chain_sb.pdf}}]
    \,,
    \\
    \res_{\includegraphics[scale=.2]{figs/3chain/3chain_bb.pdf}}[
        \vphi_{\includegraphics[align=c,scale=.2]{figs/3chain/3chain.pdf}}
    ]&=
    \res_{\includegraphics[scale=.2]{figs/3chain/3chain_bb.pdf}}[
        \vphi_{\includegraphics[align=c,scale=.2]{figs/3chain/3chain_bb.pdf}}
    ]
    \propto \Omega[\mathcal{Z}_{\includegraphics[scale=.2]{figs/3chain/3chain_bb.pdf}}]
    \,,
\ee
are proportional to the canonical forms of the line segments highlighted in {\color{BrickRed}red} in figure \ref{fig:3chainZonoOther}.
Note that 
$\vphi_{\includegraphics[scale=0.2]{figs/3chain/3chain.pdf}}$ 
has no residue on the {\color{gray}gray} line in figure \ref{fig:3chainTotEnergyZono} as well as the {\color{gray}gray} points in both figures \ref{fig:3chainTotEnergyZono} and \ref{fig:3chainZonoOther}. 
The {\color{gray}gray} line and points do not correspond to tubings associated to acyclic minors; this verifies that 
$\vphi_{\includegraphics[scale=0.2]{figs/3chain/3chain.pdf}}$ 
is only singular on the cuts indexed by acyclic minors. 

\begin{mybox}[label={box:zonotopes}]{Graphical zonotopes}
    
To each decorated graph $\mathfrak{z}$ with only broken (\!\!\brokenEdge{1}\!\!) and pinched (\!\!\solidEdge{1}\!\!) edges we associate a \emph{graphical zonotope} $\mathcal{Z}_\mathfrak{z}$, defined as the unique polytope satisfying
    \[
        \res_{\mathfrak{z}}[\vphi_\G] \;=\; \Omega[\mathcal{Z}_\mathfrak{z}]
        \,,
    \]
whose canonical form is the corresponding residue of $\vphi_\G$.
    
\paragraph{Faces from anti-pinches.} 
Lower-dimensional faces of $\mathcal{Z}_\mathfrak{z}$ are themselves graphical zonotopes $\mathcal{Z}_{\tilde{\mathfrak{z}}}$, where $\tilde{\mathfrak{z}}$ is an \emph{anti-pinch} of $\mathfrak{z}$: an acyclic minor obtained by turning some pinched edges into oriented ones, $\solidEdge{1} \to \directedEdge{1}$ or $\solidEdge{1} \to \ddirectedEdge{1}$. We write $\antipinch(\mathfrak{z})$ for the set of all anti-pinches of $\mathfrak{z}$ (including $\mathfrak{z}$ itself), and $\antipinch_n(\mathfrak{z}) := \{\mathfrak{f} \in \antipinch(\mathfrak{z}) : |\Rsf_\mathfrak{f}| - |\Rsf_\mathfrak{z}| = n\}$. The facets of $\mathcal{Z}_\mathfrak{z}$ are labeled by $\antipinch_1(\mathfrak{z})$.

\end{mybox}


The zonotope $\mathcal{Z}_{\mathfrak{z}}$ embedded into the on-shell variety is obtained by taking the convex hull of its vertices. 
The vertices of $\mathcal{Z}_{\mathfrak{z}}$ are labeled by the acyclic minors that descend from $\mathfrak{z}$ and have no pinched edges (namely, all pinched edges of $\mathfrak{z}$ are turned into oriented edges).
Let $\antipinch_\mathrm{max}(\mathfrak{z})$ 
be the set of acyclic minors where all pinched edges have been replaced by all possible oriented edges that do not produce a cyclic graph; then 
\be 
    \mathcal{Z}_{\mathfrak{z}}
    = \mathrm{ConvHull}\Big(\{
        \B_{
            \mathfrak{u} \in \antipinch_\mathrm{max}(\mathfrak{z})
        }
    \}\Big)
    \,,
\ee
where we recall that $\B_{\mathfrak{u}} = \bigcap_{\tau \in \csf} V(S_\tau)$ for any $\csf \in \Csf_{\mathfrak{u}}$. 
The convex hull definition is valid for any lower-dimension faces. 
The only difference is that $\mathfrak{z}$ starts with some oriented edges that are shared with all of its descendants $\tilde{\mathfrak{z}}$.

\paragraph{The flow of cuts and derivatives from zonotopes.}
It is well known that the discontinuity of an integral can be computed as a linear combination of sequential residue contours (see \cite{Cutkosky, Hwa:1967csk, pham2011singularities, Hannesdottir:2022xki, Vergu:2023rqz} for a very incomplete survey).
A number of sequential discontinuities are known to annihilate the physical subspace, due to the Steinmann relations~\cite{Steinmann,Steinmann2,Caron-Huot:2016owq,Caron-Huot:2019bsq,Benincasa:2020aoj, Bourjaily:2020wvq,Hannesdottir:2025bss} and the hierarchical principle~\cite{boyling1968homological,Landshoff1966,Hannesdottir:2022xki,Hannesdottir:2024cnn}. 
These relations constrain the kind of functions FRW-integrals evaluate to.
In this work we have classified the set of physical sequential residues that do not annihilate the physical subspace. 
This classification is of course consistent with the Steinmann relations and hierarchical principle, but is even more complete. 

Following \cite{Glew:2025arc}, we can also formalize the combinatorics of the physical sequential residues in what is called the \emph{flow of cuts}. 
Through \eqref{eq:partialFractionedPsi}, the combinatorics of the physical sequential residues is inherited from the collection of zonotopes $\{\mathcal{Z}_\mathfrak{z}\}$, 
since the combinatorics of the residues of $\Omega[\mathcal{Z}_\mathfrak{z}]$ is inherited from the combinatorial structure of the zonotope $\mathcal{Z}_\mathfrak{z}$:
\begin{mybox}[label={box:flowOfCuts}]{The flow of physical cuts}
    The cut tubing $\Csf_\g$ and its associated residue operator $\res_{\Csf_{\g}}$ projects $H^{|\V_\G|}_\text{phys}$ onto the subspace of physical forms that only have on-shell singularities compatible with the zonotope $\mathcal{Z}_{\g}$. 
    In order to not leave the space $H^{|\V_\G|}_\text{phys}$, further compatible residues must localize to lower-dimensional faces of $\mathcal{Z}_{\g}$:
    \be
        \res_{\Csf_\g} \to \res_{\Csf_{\g'}} 
        \quad\text{if}\quad
        \g \in \antipinch_1(\g')
        \,.
    \ee
\end{mybox}
\noindent

\noindent As a consequence of the combinatorics of physical cuts, and because the cut-basis of FRW-forms is tailored to these combinatorics, the resulting differential equations have a block triangular form in which each block corresponds to a top-dimensional zonotope $\mathcal{Z}_\mathfrak{z}$ where $\mathfrak{z}$ has no oriented edges.
The kinematic derivative of $\vphi_\g$ couples to itself and all $\vphi_{\g'}$ such that $\g'\in \antipinch_1(\g)$. 
Note that the flow (arrows between $\g$ and $\g'$) is reversed compared to the flow of residues; if $\mathfrak{z}\in\antipinch(\g)$ corresponds to an $m$-dimensional face of $\mathcal{Z}_\mathfrak{z}$, then $\mathfrak{z}\in\antipinch(\g')$ corresponds to a compatible $(m{+}1)$-dimensional face of $\mathcal{Z}_\mathfrak{z}$. 
This is a natural consequence of the duality between discontinuities and derivatives.

\section{Examples and applications}
\label{sec:examples}

In this section, we work through a pair of examples that go beyond the two-site chain---first the three-site chain, and then (going beyond tree level) the box and kite graphs. We then discuss the connection between our coaction and the kinematic flow of the wavefunction of the universe, and (separately) the monodromy group that governs these periods. Finally, we highlight simplifications that occur in the coaction when considering in-in correlators, rather than the wavefunction of the universe.

Whenever possible, we have compared our coaction formula with the coaction of hypergeometric functions in \cite{Brown:2019jng,Abreu:2019wzk}, and found perfect agreement. Additionally, we have checked the compatibility of our coaction with the coaction on MPLs by comparing the $\alpha_\bullet\sim0$ expansion of the coaction against the coaction of the polylogarithmic coefficients of the $\alpha_\bullet\sim0$ expansion, to several orders in $\alpha_\bullet$.

\subsection{The three-site chain graph}
\label{sec:3chain}

As detailed in section \ref{sec:partialFracAndZono}, the space of acyclic minors label graphical zonotopes, each of which represents a disconnected group of periods.
Because of this, the period matrix is block diagonal; in the case of the three-site chain,
\be
    \mat{P} = \mat{P}[\mathcal{Z}_{\includegraphics[scale=.2]{figs/3chain/3chain_bb.pdf}}]
    \oplus
    \mat{P}[\mathcal{Z}_{\includegraphics[scale=.2]{figs/3chain/3chain_sb.pdf}}]
    \oplus
    \mat{P}[\mathcal{Z}_{\includegraphics[scale=.2]{figs/3chain/3chain_bs.pdf}}]
    \oplus
    \mat{P}[\mathcal{Z}_{\includegraphics[scale=.2]{figs/3chain/3chain_ss.pdf}}]
\ee 
As can be checked, the periods associated to $\mathcal{Z}_{\includegraphics[scale=.2]{figs/3chain/3chain_bb.pdf}}$, $\mathcal{Z}_{\includegraphics[scale=.2]{figs/3chain/3chain_sb.pdf}}$ and $\mathcal{Z}_{\includegraphics[scale=.2]{figs/3chain/3chain_bs.pdf}}$ evaluate to two-site periods with shifted kinematics, up to multiplication by a power function:
\begin{align}
    \mathcal{Z}_{\includegraphics[scale=.2]{figs/3chain/3chain_bb.pdf}}: 
    &\quad
    \begin{matrix}
        \g_1 = \includegraphics[scale=.3,align=c]{figs/3chain/3chain_bb.pdf}
    \end{matrix}
    \quad&
    \mat{P}[\mathcal{Z}_{\includegraphics[scale=.2]{figs/3chain/3chain_bb.pdf}}]
    &:= \begin{pmatrix} f^{\alpha_1}_{\includegraphics[scale=.2]{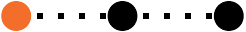}}
    f^{\alpha_2}_{\includegraphics[scale=.2]{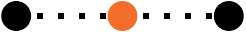}}
    f^{\alpha_3}_{\includegraphics[scale=.2]{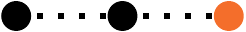}}
    \end{pmatrix}
    \,,
    \\
    \mathcal{Z}_{\includegraphics[scale=.2]{figs/3chain/3chain_sb.pdf}}:
    &\quad
    \begin{matrix}
        \g_2 = \includegraphics[scale=.3,align=c]{figs/3chain/3chain_sb.pdf}
        \\
        \g_3 = \includegraphics[scale=.3,align=c]{figs/3chain/3chain_rb.pdf}
        \\
        \g_4 = \includegraphics[scale=.3,align=c]{figs/3chain/3chain_lb.pdf}
    \end{matrix}
    &
    \mat{P}[\mathcal{Z}_{\includegraphics[scale=.2]{figs/3chain/3chain_sb.pdf}}]
    &:= \mat{P}[\mathcal{Z}_{\includegraphics[scale=.4]{figs/2chain_gs}}]\;
    f^{\alpha_3}_{\includegraphics[scale=.2]{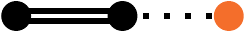}}
    \,,
    \\
    \mathcal{Z}_{\includegraphics[scale=.2]{figs/3chain/3chain_bs.pdf}}:
    &\quad 
    \begin{matrix}
        \g_5 = \includegraphics[scale=.3,align=c]{figs/3chain/3chain_bs.pdf}
        \\
        \g_6 = \includegraphics[scale=.3,align=c]{figs/3chain/3chain_br.pdf}
        \\
        \g_7 = \includegraphics[scale=.3,align=c]{figs/3chain/3chain_bl.pdf}
    \end{matrix}
    &
    \mat{P}[\mathcal{Z}_{\includegraphics[scale=.2]{figs/3chain/3chain_bs.pdf}}]
    &:= f^{\alpha_1}_{\includegraphics[scale=.2]{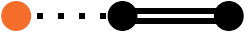}}\;
    \left(\mat{P}[\mathcal{Z}_{\includegraphics[scale=.4]{figs/2chain_gs}}]
    \vert_{
        \alpha_i \to \alpha_{i+1}, 
        X_i \to X_{i+1}, 
        Y_{12} \to Y_{23}
    }\right)
    \,,
\end{align}
where we recall that each of the letters $f_\rsf$ is a polynomial, as given in~\eqref{eq:partialRg}. Thus, the new periods we encounter are associated only with $\mathcal{Z}_{\includegraphics[scale=.2]{figs/3chain/3chain_ss.pdf}}$:
\be
    \g_{8} &= \includegraphics[scale=.3,align=c]{figs/3chain/3chain_ss.pdf}
    \,,
    &
    \g_{9} &= \includegraphics[scale=.3,align=c]{figs/3chain/3chain_ls.pdf}
    \,,
    &
    \g_{10} &= \includegraphics[scale=.3,align=c]{figs/3chain/3chain_rs.pdf}
    \,,
    &
    \g_{11} &= \includegraphics[scale=.3,align=c]{figs/3chain/3chain_sl.pdf}
    \,,
    &
    \g_{12} &= \includegraphics[scale=.3,align=c]{figs/3chain/3chain_sr.pdf}
    \,,
    \\
    \g_{13} &= \includegraphics[scale=.3,align=c]{figs/3chain/3chain_ll.pdf}
    \,,
    &
    \g_{14} &= \includegraphics[scale=.3,align=c]{figs/3chain/3chain_lr.pdf}
    \,,
    &
    \g_{15} &= \includegraphics[scale=.3,align=c]{figs/3chain/3chain_rl.pdf}
    \,,
    &
    \g_{16} &= \includegraphics[scale=.3,align=c]{figs/3chain/3chain_rr.pdf}
    \,.
\ee
As always, the diagonal of the period matrix are power functions 
\begin{align}
    &\mathrm{diag}\mat{P}[\mathcal{Z}_{\includegraphics[scale=.2]{figs/3chain/3chain_ss.pdf}}]
    :=\bigg(
        \frac{\Gamma \left(\alpha_1\right) \Gamma \left(\alpha_2\right) \Gamma \left(\alpha_3\right)}{\Gamma \left(\alpha_1{+}\alpha_2{+}\alpha_3\right)}
        f_{\includegraphics[scale=.2]{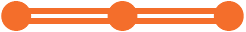}}^{\alpha_1{+}\alpha_2{+}\alpha_3}
        ,\,
        \frac{
            \Gamma(\alpha_2) 
            \Gamma(\alpha_3)
        }{  
            \Gamma(\alpha_2{+}\alpha_3)
        }
        f_{\includegraphics[scale=.2]{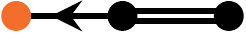}}^{\alpha_1}
        f_{\includegraphics[scale=.2]{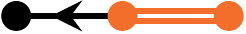}}^{\alpha_2{+}\alpha_3}
        ,\,
        \nn\\&\qquad
        \frac{
            \Gamma(\alpha_2) 
            \Gamma(\alpha_3)
        }{  
            \Gamma(\alpha_2{+}\alpha_3)
        }
        f_{\includegraphics[scale=.2]{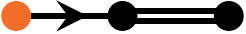}}^{\alpha_1}
        f_{\includegraphics[scale=.2]{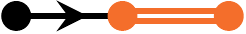}}^{\alpha_2{+}\alpha_3}
        ,\,
        \frac{
            \Gamma(\alpha_1) 
            \Gamma(\alpha_2)
        }{  
            \Gamma(\alpha_1{+}\alpha_2)
        }
        f_{\includegraphics[scale=.2]{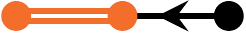}}^{\alpha_1{+}\alpha_2}
        f_{\includegraphics[scale=.2]{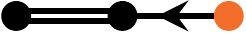}}^{\alpha_3}
        ,\,
        \frac{
            \Gamma(\alpha_1) 
            \Gamma(\alpha_2)
        }{  
            \Gamma(\alpha_1{+}\alpha_2)
        }
        f_{\includegraphics[scale=.2]{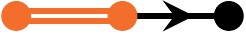}}^{\alpha_1{+}\alpha_2}
        f_{\includegraphics[scale=.2]{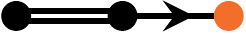}}^{\alpha_3}
        ,\, \label{eq:diagP3Z4}
        \\&\qquad
        f_{\includegraphics[scale=.2]{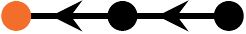}}^{\alpha_1}
        f_{\includegraphics[scale=.2]{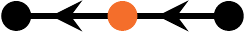}}^{\alpha_2}
        f_{\includegraphics[scale=.2]{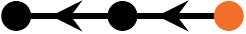}}^{\alpha_3}
        ,\,
        2
        f_{\includegraphics[scale=.2]{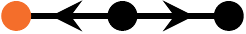}}^{\alpha_1}
        f_{\includegraphics[scale=.2]{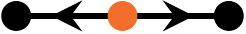}}^{\alpha_2}
        f_{\includegraphics[scale=.2]{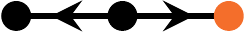}}^{\alpha_3}
        ,\,
        f_{\includegraphics[scale=.2]{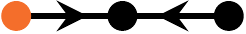}}^{\alpha_1}
        f_{\includegraphics[scale=.2]{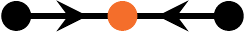}}^{\alpha_2}
        f_{\includegraphics[scale=.2]{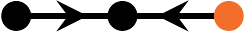}}^{\alpha_3}
        ,\,
        f_{\includegraphics[scale=.2]{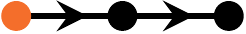}}^{\alpha_1}
        f_{\includegraphics[scale=.2]{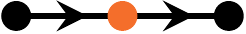}}^{\alpha_2}
        f_{\includegraphics[scale=.2]{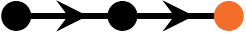}}^{\alpha_3}
    \bigg)
    \,.
    \nn
\end{align}
The most interesting periods are the $P_{\includegraphics[scale=.2]{figs/3chain/3chain_ss.pdf},\h}$-entries. 
There are four $P_{\includegraphics[scale=.2]{figs/3chain/3chain_ss.pdf},\h}$ that are analogous to the off-diagonals of the two-site chain:
\begin{align}
    \frac{
        P_{
            \includegraphics[scale=.2]{figs/3chain/3chain_ss.pdf},
            \includegraphics[scale=.2]{figs/3chain/3chain_ls.pdf}
        }
    }{
        P_{\includegraphics[scale=.2]{figs/3chain/3chain_ss.pdf},\includegraphics[scale=.2]{figs/3chain/3chain_ss.pdf}}
    } 
    &= \alpha_1(\alpha_2{+}\alpha_3) 
        \frac{\Gamma(\alpha _1{+}\alpha_2{+}\alpha_3)}{\Gamma(\alpha _1{+}\alpha_2{+}\alpha_3{+}2)}
        \;{}_2\tilde{F}_1\left(
            1,
            \alpha _1{+}1;
            \alpha _1{+}\alpha_2{+}\alpha_3{+}2;
            \frac{
                f_{\includegraphics[scale=.2]{figs/3chain/3chain_ss_R1.pdf}}
            }{
                f_{\includegraphics[scale=.2]{figs/3chain/3chain_ls_R1.pdf}}
            }
        \right)
    \,,
    \nn\\
    &= -\alpha_1(\alpha_2{+}\alpha_3) 
        \frac{\Gamma(\alpha _1{+}\alpha_2{+}\alpha_3)}{\Gamma(\alpha_1{+}\alpha_2{+}\alpha_3{+}2)}
        \;{}_2\tilde{F}_1\left(
            1,
            \alpha_2{+}\alpha_3{+}1;
            \alpha_1{+}\alpha_2{+}\alpha_3{+}2;
            \frac{
                f_{\includegraphics[scale=.2]{figs/3chain/3chain_ss_R1.pdf}}
            }{
                f_{\includegraphics[scale=.2]{figs/3chain/3chain_ls_R2.pdf}}
            }
        \right)
    \,,
    \nn\\
    P_{
        \includegraphics[scale=.2]{figs/3chain/3chain_ss.pdf},
        \includegraphics[scale=.2]{figs/3chain/3chain_rs.pdf}
    }
    &= - P_{
            \includegraphics[scale=.2]{figs/3chain/3chain_ss.pdf},
            \includegraphics[scale=.2]{figs/3chain/3chain_ls.pdf}
        }\vert_{Y_{12} \to -Y_{12}}
    \,,
    \nn\\
    P_{
        \includegraphics[scale=.2]{figs/3chain/3chain_ss.pdf},
        \includegraphics[scale=.2]{figs/3chain/3chain_sl.pdf}
    }
    &= - P_{
            \includegraphics[scale=.2]{figs/3chain/3chain_ss.pdf},
            \includegraphics[scale=.2]{figs/3chain/3chain_ls.pdf}
        }\vert_{
            \alpha_{1} \leftrightarrow \alpha_3,
            X_{1} \leftrightarrow X_3,
            Y_{12} \to -Y_{23}
        }
    \,,
    \nn\\
    P_{
        \includegraphics[scale=.2]{figs/3chain/3chain_ss.pdf},
        \includegraphics[scale=.2]{figs/3chain/3chain_sr.pdf}
    }
    &= P_{
            \includegraphics[scale=.2]{figs/3chain/3chain_ss.pdf},
            \includegraphics[scale=.2]{figs/3chain/3chain_ls.pdf}
        }\vert_{
            \alpha_{1} \leftrightarrow \alpha_3,
            X_{1} \leftrightarrow X_3,
            Y_{12} \to Y_{23}
        }
    \,.
\end{align}
The most complex periods in this row are Appell $F_3$'s (a specific example of Lauricella's $F_B$ functions). 
These are the hypergeometric functions that go beyond those we saw in the two-site chain:\footnote{Because $\vphi_{\includegraphics[align=c,scale=.2]{figs/3chain/3chain_lr.pdf}}$ is a sum of two $\dlog$-forms (recall \eqref{eq:phi3chian_lr}), there is also the two term representation:
\be
    \frac{
        P_{
            \includegraphics[scale=.2]{figs/3chain/3chain_ss.pdf},
            \includegraphics[scale=.2]{figs/3chain/3chain_lr.pdf}
        }
    }{
        P_{\includegraphics[scale=.2]{figs/3chain/3chain_ss.pdf},\includegraphics[scale=.2]{figs/3chain/3chain_ss.pdf}}
    }&=-\frac{
            \alpha_1\alpha_2\alpha_3
            \Gamma(\alpha _1{+}\alpha_2{+}\alpha_3) 
        }{
            \Gamma(3+\alpha _1{+}\alpha_2{+}\alpha_3) 
        }
        \bigg[
        \tilde{F}_B\left(
            1,
            1;
            \alpha_1{+}1,
            \alpha_2{+}1;
            \alpha_1{+}\alpha_2{+}\alpha_3{+}3;
            \frac{
                f_{\includegraphics[scale=.2]{figs/3chain/3chain_ss_R1.pdf}}
            }{
                f_{\includegraphics[scale=.2]{figs/3chain/3chain_ls_R1.pdf}}
            },
            \frac{
                f_{\includegraphics[scale=.2]{figs/3chain/3chain_ss_R1.pdf}}
            }{
                f_{\includegraphics[scale=.2]{figs/3chain/3chain_lr_R2.pdf}}
            }
        \right)
    \nn\\&\qquad 
    + \tilde{F}_B\left(
            1,
            1;
            \alpha_2{+}1,
            \alpha_3{+}1;
            \alpha_1{+}\alpha_2{+}\alpha_3{+}3;
            \frac{
                f_{\includegraphics[scale=.2]{figs/3chain/3chain_ss_R1.pdf}}
            }{
                f_{\includegraphics[scale=.2]{figs/3chain/3chain_lr_R2.pdf}}
            },
            \frac{
                f_{\includegraphics[scale=.2]{figs/3chain/3chain_ss_R1.pdf}}
            }{
                f_{\includegraphics[scale=.2]{figs/3chain/3chain_sr_R2.pdf}}
            }
        \right)
    \bigg]
    \,.
    \nn
\ee
}
\begin{align}
    \frac{
        P_{
            \includegraphics[scale=.2]{figs/3chain/3chain_ss.pdf},
            \includegraphics[scale=.2]{figs/3chain/3chain_ll.pdf}
        }
    }{
        P_{\includegraphics[scale=.2]{figs/3chain/3chain_ss.pdf},\includegraphics[scale=.2]{figs/3chain/3chain_ss.pdf}}
    }
    &= \frac{
            \alpha_1\alpha_2\alpha_3
            \Gamma(\alpha _1{+}\alpha_2{+}\alpha_3) 
        }{
            \Gamma(3+\alpha _1{+}\alpha_2{+}\alpha_3) 
        }
        \tilde{F}_B\left(
            1,
            1;
            \alpha_1{+}1,
            \alpha_3{+}1;
            \alpha_1{+}\alpha_2{+}\alpha_3{+}3;
            \frac{
                f_{\includegraphics[scale=.2]{figs/3chain/3chain_ss_R1.pdf}}
            }{
                f_{\includegraphics[scale=.2]{figs/3chain/3chain_ls_R1.pdf}}
            },
            \frac{
                f_{\includegraphics[scale=.2]{figs/3chain/3chain_ss_R1.pdf}}
            }{
                f_{\includegraphics[scale=.2]{figs/3chain/3chain_sl_R2.pdf}}
            }
        \right)
    \,,
    \nn\\
    P_{
        \includegraphics[scale=.2]{figs/3chain/3chain_ss.pdf},
        \includegraphics[scale=.2]{figs/3chain/3chain_lr.pdf}
    }
    &= P_{
        \includegraphics[scale=.2]{figs/3chain/3chain_ss.pdf},
        \includegraphics[scale=.2]{figs/3chain/3chain_ll.pdf}
    }\vert_{Y_{23}\to-Y_{23}}
    \,,
    \nn\\
    P_{
        \includegraphics[scale=.2]{figs/3chain/3chain_ss.pdf},
        \includegraphics[scale=.2]{figs/3chain/3chain_rl.pdf}
    }    
    &= P_{
        \includegraphics[scale=.2]{figs/3chain/3chain_ss.pdf},
        \includegraphics[scale=.2]{figs/3chain/3chain_ll.pdf}
    }\vert_{
        Y_{12} \to -Y_{12}
    }
    \,,
    \nn\\
    P_{
        \includegraphics[scale=.2]{figs/3chain/3chain_ss.pdf},
        \includegraphics[scale=.2]{figs/3chain/3chain_rr.pdf}
    }    
    &= P_{
        \includegraphics[scale=.2]{figs/3chain/3chain_ss.pdf},
        \includegraphics[scale=.2]{figs/3chain/3chain_ll.pdf}
    }\vert_{
        Y_{12} \to -Y_{12},
        Y_{23} \to -Y_{23}
    }
\end{align}
where 
\be
    \tilde{F}_B(\bs{a};\bs{b};c;\bs{z}) = 
    F_B(\bs{a};\bs{b};c;\bs{z})\; \prod_{k=1}^m z_k
\ee
is a rescaled Lauricella $F_B$ function.

The rest of the entries in this block are simpler. For the rows $P_{\includegraphics[scale=0.2]{figs/3chain/3chain_ll.pdf},\h}$, $P_{\includegraphics[scale=0.2]{figs/3chain/3chain_lr.pdf},\h}$, $P_{\includegraphics[scale=0.2]{figs/3chain/3chain_rl.pdf},\h}$, and $P_{\includegraphics[scale=0.2]{figs/3chain/3chain_rr.pdf},\h}$, the only nonzero terms are the diagonal entries, which are given in (\ref{eq:diagP3Z4}). 
In the remaining four rows, corresponding to $P_{\includegraphics[scale=0.2]{figs/3chain/3chain_ls.pdf},\h}$, $P_{\includegraphics[scale=0.2]{figs/3chain/3chain_rs.pdf},\h}$, $P_{\includegraphics[scale=0.2]{figs/3chain/3chain_sl.pdf},\h}$, and $P_{\includegraphics[scale=0.2]{figs/3chain/3chain_sr.pdf},\h}$, the only nonzero off-diagonal entries are 
\be 
    P_{
        \includegraphics[scale=.2]{figs/3chain/3chain_ls.pdf},
        \includegraphics[scale=.2]{figs/3chain/3chain_ll.pdf}
    }
    &= f^{\alpha_1}_{\includegraphics[scale=.2]{figs/3chain/3chain_ls_R1.pdf}}\;
    \left(
    P_{\includegraphics[scale=.4]{figs/2chain_gs},\includegraphics[scale=.4]{figs/2chain_gl}}
    \vert_{
        \alpha_i \to \alpha_{i+1}, 
        X_1 \to X_{2}+Y_{12}, 
        X_2 \to X_3,
        Y_{12} \to Y_{23}
    }
    \right)
    \,,
    \\
    P_{
        \includegraphics[scale=.2]{figs/3chain/3chain_ls.pdf},
        \includegraphics[scale=.2]{figs/3chain/3chain_lr.pdf}
    }
    &= f^{\alpha_1}_{\includegraphics[scale=.2]{figs/3chain/3chain_ls_R1.pdf}}\;
    \left(
    P_{\includegraphics[scale=.4]{figs/2chain_gs},\includegraphics[scale=.4]{figs/2chain_gr}}
    \vert_{
        \alpha_i \to \alpha_{i+1}, 
        X_1 \to X_{2}+Y_{12}, 
        X_2 \to X_3,
        Y_{12} \to Y_{23}
    }
    \right)
    \,,
    \\
    P_{
        \includegraphics[scale=.2]{figs/3chain/3chain_rs.pdf},
        \includegraphics[scale=.2]{figs/3chain/3chain_rl.pdf}
    }
    &= f^{\alpha_1}_{\includegraphics[scale=.2]{figs/3chain/3chain_rs_R1.pdf}}\;
    \left(
    P_{\includegraphics[scale=.4]{figs/2chain_gs},\includegraphics[scale=.4]{figs/2chain_gl}}
    \vert_{
        \alpha_i \to \alpha_{i+1}, 
        X_1 \to X_{2}-Y_{12}, 
        X_2 \to X_3,
        Y_{12} \to Y_{23}
    }
    \right)
    \,,
    \\
    P_{
        \includegraphics[scale=.2]{figs/3chain/3chain_rs.pdf},
        \includegraphics[scale=.2]{figs/3chain/3chain_rr.pdf}
    }
    &= f^{\alpha_1}_{\includegraphics[scale=.2]{figs/3chain/3chain_rs_R1.pdf}}\;
    \left(
    P_{\includegraphics[scale=.4]{figs/2chain_gs},\includegraphics[scale=.4]{figs/2chain_gr}}
    \vert_{
        \alpha_i \to \alpha_{i+1}, 
        X_1 \to X_{2}-Y_{12}, 
        X_2 \to X_3,
        Y_{12} \to Y_{23}
    }
    \right)
    \,,
    \\
    P_{
        \includegraphics[scale=.2]{figs/3chain/3chain_sl.pdf},
        \includegraphics[scale=.2]{figs/3chain/3chain_ll.pdf}
    }
    &= f^{\alpha_3}_{\includegraphics[scale=.2]{figs/3chain/3chain_sl_R2.pdf}}\;
    \left(
    P_{\includegraphics[scale=.4]{figs/2chain_gs},\includegraphics[scale=.4]{figs/2chain_gl}}
    \vert_{
        X_2 \to X_2-Y_{23}
    }
    \right)
    \,,
    \\
    P_{
        \includegraphics[scale=.2]{figs/3chain/3chain_sl.pdf},
        \includegraphics[scale=.2]{figs/3chain/3chain_rl.pdf}
    }
    &= f^{\alpha_3}_{\includegraphics[scale=.2]{figs/3chain/3chain_sl_R2.pdf}}\;
    \left(
    P_{\includegraphics[scale=.4]{figs/2chain_gs},\includegraphics[scale=.4]{figs/2chain_gr}}
    \vert_{
        X_2 \to X_2-Y_{23}
    }
    \right)
    \,,
    \\
    P_{
        \includegraphics[scale=.2]{figs/3chain/3chain_sr.pdf},
        \includegraphics[scale=.2]{figs/3chain/3chain_lr.pdf}
    }
    &= f^{\alpha_3}_{\includegraphics[scale=.2]{figs/3chain/3chain_sr_R2.pdf}}\;
    \left(
    P_{\includegraphics[scale=.4]{figs/2chain_gs},\includegraphics[scale=.4]{figs/2chain_gl}}
    \vert_{
        X_2 \to X_2+Y_{23}
    }
    \right)
    \,,
    \\
    P_{
        \includegraphics[scale=.2]{figs/3chain/3chain_sr.pdf},
        \includegraphics[scale=.2]{figs/3chain/3chain_rr.pdf}
    }
    &= f^{\alpha_3}_{\includegraphics[scale=.2]{figs/3chain/3chain_sr_R2.pdf}}\;
    \left(
    P_{\includegraphics[scale=.4]{figs/2chain_gs},\includegraphics[scale=.4]{figs/2chain_gr}}
    \vert_{
        X_2 \to X_2+Y_{23}
    }
    \right)
    \,,
\ee
which correspond to off-diagonal entries that appeared for the two-site chain but with shifted kinematics, and multiplied by power functions.

To illustrate the coaction in this context, consider the coaction of the period 
$P_{\includegraphics[scale=.2]{figs/3chain/3chain_ss.pdf},\includegraphics[scale=.2]{figs/3chain/3chain_ll.pdf}}$
\begin{align}
    \Delta \left[\includegraphics[scale=.8,align=c]{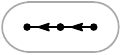}\right]
    &= \frac{\alpha_1\alpha_2\alpha_3}{\alpha_1+\alpha_2+\alpha_3}
    \includegraphics[scale=.8,align=c]{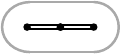} 
    \otimes \includegraphics[scale=.8,align=c]{figs/P_ss_ll.pdf} 
    + \frac{\alpha_2\alpha_3}{\alpha_2+\alpha_3}
    \includegraphics[scale=.8,align=c]{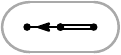} 
    \otimes \includegraphics[scale=.8,align=c]{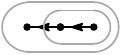} 
    \nn\\&
    + \frac{\alpha_1\alpha_2}{\alpha_1+\alpha_2}
    \includegraphics[scale=.8,align=c]{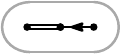} 
    \otimes \includegraphics[scale=.8,align=c]{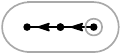} 
    + \includegraphics[scale=.8,align=c]{figs/P_ss_ll.pdf} 
    \otimes \includegraphics[scale=.8,align=c]{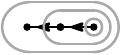} 
    \,.
\end{align}
The reader can check that this formula agrees with the with the coaction of the Appell $F_3$ in \cite{Brown:2019jng,Abreu:2019wzk}. 

For periods whose contour corresponds to a degenerate cut tubing, we write out the linear combination of periods dictated by the linear combination of cut contours \eqref{eq:degenRes} or \eqref{eq:degenDelta}. 
For example, we write
\begin{align}
    P_{
        \includegraphics[scale=.6,align=b]{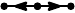},
        \includegraphics[scale=.6,align=b]{figs/g_lr.pdf}
    }
    =\includegraphics[scale=.8,align=c]{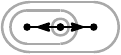}
    - \includegraphics[scale=.8,align=c]{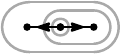}
    \,.
\end{align}
The coaction works as before; the only thing that changes is that the factor of $|\Csf_\g|$ in the intersection number is no longer one. 
This is how degenerate contours are represented both in the \texttt{Mathematica} notebook \github\;, and the \texttt{webapp} \webappAlt\!.

\subsection{Beyond tree level}
\label{sec:4gon}

The one-loop four-gon graph has a basis of 226 elements, making it too large to showcase in full detail here. 
Instead, we focus on a particular integral in the most complicated sector, which corresponds to the zonotope $\smash{\mathcal{Z}_{\includegraphics[align=b,scale=.3]{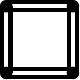}}}$. 
Even this zonotope contains 51 acyclic minors (14 dimension-0 facets, 24 dimension-1 facets, 12 dimension-2 facets, and 1 dimension-3 facet). 
Furthermore, we focus on the coaction of $\smash{\includegraphics[align=c, scale=.3]{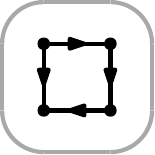}}$, which only mixes with 16 periods from this sector.
Explicitly, 
\begin{align}
    \Delta 
    \;\includegraphics[align=c, scale=.5]{figs/C1234Box.pdf}\;
    &{=} 
    \;\includegraphics[align=c, scale=.5]{figs/C1234Box.pdf}\;
    {\otimes}
    \;\includegraphics[align=c, scale=.5]{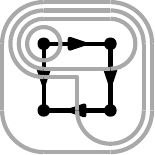}\;
    {+} 
    \frac{\alpha_1\alpha_2}{\alpha_1{+}\alpha_2} 
    \;\includegraphics[align=c, scale=.5]{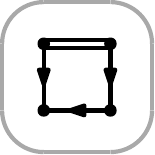}\;
    {\otimes}
    \;\includegraphics[align=c, scale=.5]{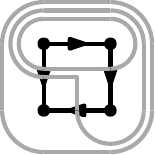}\;
    \nn
    \\& 
    {+} 
    \frac{\alpha_2\alpha_3}{\alpha_2{+}\alpha_3} 
    \;\includegraphics[align=c, scale=.5]{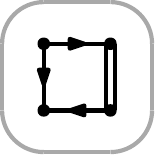}\;
    {\otimes}
    \;\includegraphics[align=c, scale=.5]{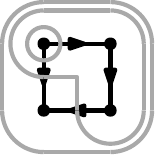}\;
    {+} 
    \frac{\alpha_3\alpha_4}{\alpha_3{+}\alpha_4} 
    \;\includegraphics[align=c, scale=.5]{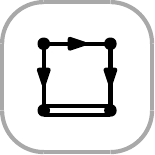}\;
    {\otimes}
    \;\includegraphics[align=c, scale=.5]{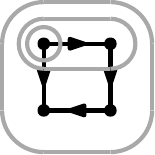}\;
    \\& 
    {+} 
    \frac{\alpha_1\alpha_2 \alpha_3}{\alpha_1{+}\alpha_2{+}\alpha_3} 
    \;\includegraphics[align=c, scale=.5]{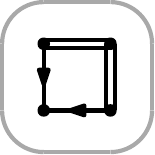}\;
    {\otimes}
    \;\includegraphics[align=c, scale=.5]{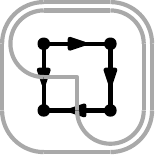}\;
    {+} 
    \frac{\alpha_2\alpha_3\alpha_4}{\alpha_2{+}\alpha_3{+}\alpha_4} 
    \;\includegraphics[align=c, scale=.5]{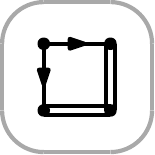}\;
    {\otimes}
    \;\includegraphics[align=c, scale=.5]{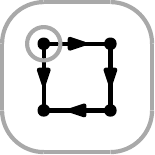}\;
    \nn
    \\& 
    {+} 
    \frac{\alpha_1\alpha_2\alpha_3\alpha_4}{(\alpha_1{+}\alpha_2)(\alpha_3{+}\alpha_4)} 
    \;\includegraphics[align=c, scale=.5]{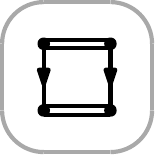}\;
    {\otimes}
    \;\includegraphics[align=c, scale=.5]{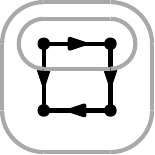}\;
    {+} 
    \frac{\alpha_1\alpha_2\alpha_3\alpha_4}{\alpha_1{+}\alpha_2{+}\alpha_3{+}\alpha_4} 
    \;\includegraphics[align=c, scale=.5]{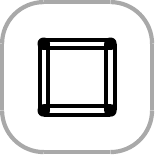}\;
    {\otimes}
    \;\includegraphics[align=c, scale=.5]{figs/C1234Box.pdf}\;
    \,,
    \nn
\end{align}
where for the $\alpha_i$ labels we adopt the convention that the vertices are ordered clockwise, starting from the top left. 
Note that for loop graphs, the acyclic condition starts to matter. In this example,
the decorated minors 
\begin{align}
    \left\{
    \;\includegraphics[align=c, scale=.5]{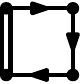}\;, 
    \;\includegraphics[align=c, scale=.5]{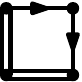}\;,
    \;\includegraphics[align=c, scale=.5]{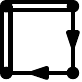}\;,
    \;\includegraphics[align=c, scale=.5]{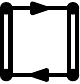}\;,
    \;\includegraphics[align=c, scale=.5]{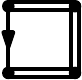}\;,
    \;\includegraphics[align=c, scale=.5]{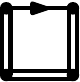}\;,
    \;\includegraphics[align=c, scale=.5]{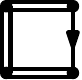}\;,
    \;\includegraphics[align=c, scale=.5]{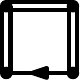}\;
    \right\}
\end{align}
do not appear because they contain a cycle of directed edges, once the pinched edges are contracted.

As a two-loop example, let us now consider the kite graph. 
The physical subspace for this graph is spanned by 614 acyclic minors, which are organized into 32 zonotopes. The largest zonotope, $\mathcal{Z}_{\includegraphics[align=c, scale=.2]{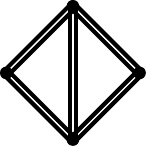}}$, contains 59 periods (one for each facet: 18 dimension-0 facets, 28 dimension-1 facets, 8 dimension-2 facets, 4 dimension-3 facets and 1 dimension-4 facet). 
Like for the box, we will only exhibit the coaction for one of the maximally-complex periods that appears:
\begin{align}
    &\Delta 
    \;\includegraphics[align=c, scale=.5]{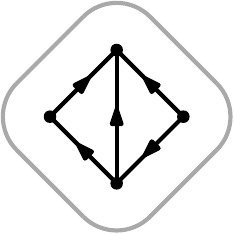}\;
    {=} 
    \;\includegraphics[align=c, scale=.5]{figs/P81.pdf}\;
    {\otimes}
    \;\includegraphics[align=c, scale=.5]{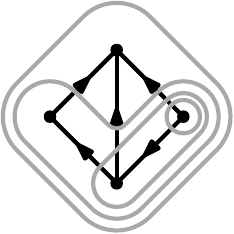}\;
    {+} 
    \frac{\alpha_4\alpha_1}{\alpha_4{+}\alpha_1} 
    \;\includegraphics[align=c, scale=.5]{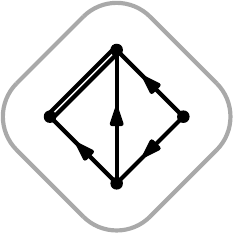}\;
    {\otimes}
    \;\includegraphics[align=c, scale=.5]{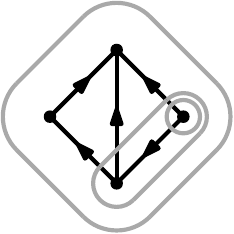}\;
    \nn\\
    &{+}
    \frac{\alpha_3\alpha_4}{\alpha_3{+}\alpha_4} 
    \;\includegraphics[align=c, scale=.5]{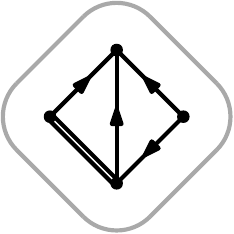}\;
    {\otimes}
    \;\includegraphics[align=c, scale=.5]{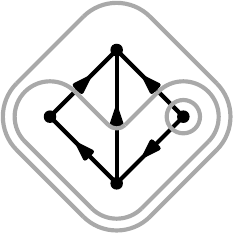}\;
    {+} 
    \frac{\alpha_3\alpha_4\alpha_1}{\alpha_3{+}\alpha_4{+}\alpha_1} 
    \;\includegraphics[align=c, scale=.5]{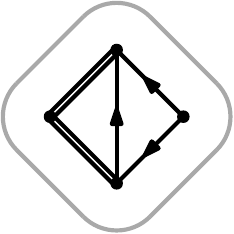}\;
    {\otimes}
    \;\includegraphics[align=c, scale=.5]{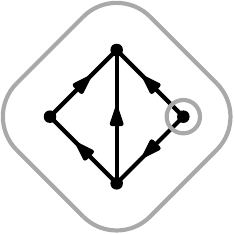}\;
    \\
    &{+}
    \frac{\alpha_2\alpha_3}{\alpha_2{+}\alpha_3} 
    \;\includegraphics[align=c, scale=.5]{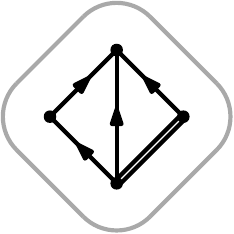}\;
    {\otimes}
    \;\includegraphics[align=c, scale=.5]{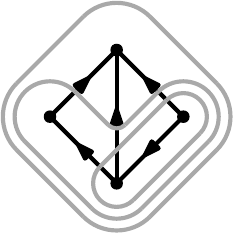}\;
    {+} 
    \frac{\alpha_4\alpha_1\alpha_2\alpha_3}{(\alpha_4{+}\alpha_1)(\alpha_2{+}\alpha_3)} 
    \;\includegraphics[align=c, scale=.5]{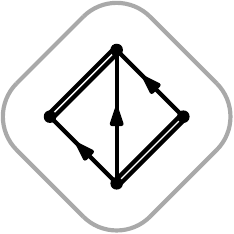}\;
    {\otimes}
    \;\includegraphics[align=c, scale=.5]{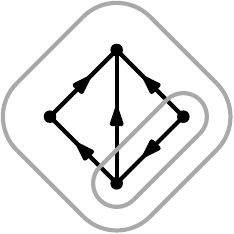}\;
    \nn\\
    &{+}
    \frac{\alpha_2\alpha_3\alpha_4}{\alpha_2{+}\alpha_3{+}\alpha_4}
    \;\includegraphics[align=c, scale=.5]{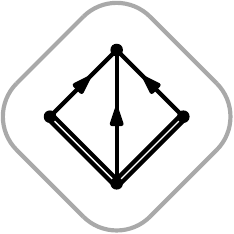}\;
    {\otimes}
    \;\includegraphics[align=c, scale=.5]{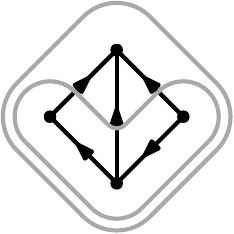}\;
    {+} 
    \frac{\alpha_1\alpha_2\alpha_3\alpha_4}{\alpha_1{+}\alpha_2{+}\alpha_3{+}\alpha_4}
    \;\includegraphics[align=c, scale=.5]{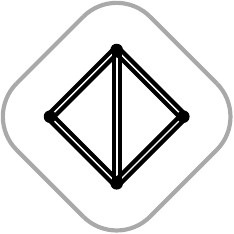}\;
    {\otimes}
    \;\includegraphics[align=c, scale=.5]{figs/P81.pdf}\;
    \,,
    \nn
\end{align}
where we have again labeled the vertices clockwise, this time starting from the top. 

Although the 4-gon has $226$ basis elements in total, the integral we have chosen above only mixes with $16$ of them; this is representative of the more general behavior that, even at one loop, the block-diagonal structure of the period matrix dictated by zonotopes (see section~\ref{sec:partialFracAndZono}) keeps the coaction of any particular period manageable. 

\subsection{Kinematic flow from the coaction}

The coaction fully encodes the system of differential equations that govern the space of periods that we associate with each FRW graph. In particular, in the graphical language of acyclic minors, we have the following differential equation:
\begin{align} \label{eq:DEQFromCoaction}
    \d [ \gamma \vert \vphi_{\g} \ra
    = \sum_{\mathfrak{h} \in \pinch(\g)}
    \frac{[ \gamma \vert \vphi_{\mathfrak{h}} \ra}{\la \check{\vphi}_\mathfrak{h} \vert \vphi_\mathfrak{h} \ra }
    \;
    \d W_1^\mathrm{MPL,\alpha}\Big[
    [ \gamma_\mathfrak{h} \vert \vphi_{\mathfrak{g}} \ra
    \Big]
    \implies
    A_{\g\mathfrak{h}} 
    = \frac{
        \d W_1^\mathrm{MPL,\alpha}\Big[
            [ \gamma_\mathfrak{h} \vert \vphi_{\mathfrak{g}} \ra
        \Big]
    }{
        \la \check{\vphi}_\mathfrak{h} \vert \vphi_\mathfrak{h} \ra 
    } 
    \,.
\end{align}
In order to compute these differential equations, we see that we must compute the weight-one part of each of the periods $[ \gamma_\mathfrak{h} \vert \vphi_{\mathfrak{g}} \ra$ in our basis; an explicit formula for this weight-one part is given in~\eqref{eq:weight1_hg_app} of appendix \ref{app:weight1}. There, we see that there are two situations in which the period $[ \gamma_\mathfrak{h} \vert \vphi_{\mathfrak{g}} \ra$ can have a non-trivial weight-one part. 
The first is when $\h=\g$; then, we have
\be \label{eq:weight1_gg}
    W_1^\mathrm{MPL,\alpha}\Big[
        [ \gamma_\mathfrak{g} \vert \vphi_{\mathfrak{g}} \ra
    \Big]
    =\la \check{\vphi}_\mathfrak{g} \vert \vphi_\mathfrak{g} \ra
    \left( \sum_{\rsf \in \Rsf_\g} \beta_\rsf \ \log f_\rsf \right)
    \,,
    \qquad \h=\g 
    \,.
\ee
The second case is when a region $\rsf^*$ of $\h$ splits into two regions of $\g$. 
That is, $\rsf^* \in \Rsf_\h$ such that $\Rsf_\h \setminus (\Rsf_\g \cap \Rsf_\h) = \{\rsf^*\} $ and $\Rsf_\g \setminus (\Rsf_\h \cap \Rsf_\g) = \{\rsf^*_\uparrow, \rsf^*_\downarrow\}$.%
\footnote{Recall that, by definition, the regions $\rsf^*_\uparrow$ and $\rsf^*_\downarrow$ are both connected subgraphs with pinched edges.}
Pictorially, we represent this as 
\be
    \mathfrak{h} &= \includegraphics[align=c,scale=.6]{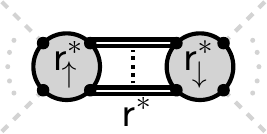} 
    \,,
    &\mathrm{and}&\quad
    \g &= \includegraphics[align=c,scale=.6]{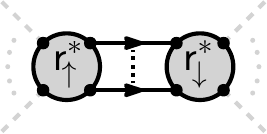}
    \,,
    \\
    &
    &\mathrm{or}&
    & 
    \\
    \mathfrak{h} &= \includegraphics[align=c,scale=.6]{figs/gsg.pdf}
    \,,
    &\mathrm{and}&\quad
    \g &= \includegraphics[align=c,scale=.6]{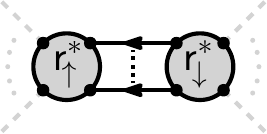}
    \,.
\ee 
Here, the gray dashed lines represent possible connections to other regions via broken edges. 
Then, the weight-one part only depends on how $\rsf^*$ breaks into $\rsf^*_\uparrow$ and $ \rsf^*_\downarrow$:
\be\label{eq:weight1_hg}
    W_1^\mathrm{MPL,\alpha}\Big[
        [ \gamma_\mathfrak{h} \vert \vphi_{\mathfrak{g}} \ra
    \Big]
    &= \begin{cases}
        C_{\g\g} 
        \log \frac{
            f_{\includegraphics[align=c, scale=.35]{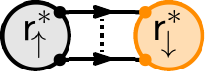}}
        }{
            f_{\includegraphics[align=c, scale=.35]{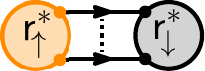}}
        }
        & \mathfrak{h} = \includegraphics[align=c,scale=.4]{figs/gsg.pdf} 
            \quad\mathrm{and}\quad
            \g = \includegraphics[align=c,scale=.4]{figs/grg.pdf}
        \,,
        \\[1em]
        C_{\g\g} 
        \log \frac{
            f_{\includegraphics[align=c, scale=.35]{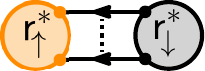}}
        }{
            f_{\includegraphics[align=c, scale=.35]{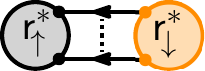}}
        }
        & \mathfrak{h} = \includegraphics[align=c,scale=.4]{figs/gsg.pdf}
            \quad\mathrm{and}\quad
            \g = \includegraphics[align=c,scale=.4]{figs/glg.pdf}
        \,
    \end{cases}
\ee
(see appendix \ref{app:weight1} for details). Putting \eqref{eq:DEQFromCoaction}, \eqref{eq:weight1_gg} and \eqref{eq:weight1_hg} together, we reproduce kinematic flow, as derived in \cite{Glew:2025ypb}:
\be
    A_{\g\mathfrak{h}} 
    &= \begin{cases}
        \sum_{\rsf \in \Rsf_\g} \beta_\rsf \ \dlog f_\rsf 
        & \h=\g \,,
        \\[1em]
        \frac{ C_{\g\g} }{ C_{\h\h} }
        \dlog \frac{
            f_{\includegraphics[align=c, scale=.35]{figs/f_u_right_dc.pdf}}
        }{
            f_{\includegraphics[align=c, scale=.35]{figs/f_uc_right_d.pdf}}
        }
        & \mathfrak{h} = \includegraphics[align=c,scale=.4]{figs/gsg.pdf} 
            \quad\mathrm{and}\quad
            \g = \includegraphics[align=c,scale=.4]{figs/grg.pdf}
        \,,
        \\[1em]
        \frac{ C_{\g\g} }{ C_{\h\h} }
        \dlog \frac{
            f_{\includegraphics[align=c, scale=.35]{figs/f_uc_left_d.pdf}}
        }{
            f_{\includegraphics[align=c, scale=.35]{figs/f_u_left_dc.pdf}}
        }
        & \mathfrak{h} = \includegraphics[align=c,scale=.4]{figs/gsg.pdf}
            \quad\mathrm{and}\quad
            \g = \includegraphics[align=c,scale=.4]{figs/glg.pdf}
        \,.
    \end{cases}
\ee

\subsection{Monodromy from the coaction}

Recall from section~\ref{sec:coaction} that, just as derivatives only act on the last entry of the coaction, discontinuities only act on the first entry. Thus, given the coaction
\begin{align}
    \Delta [ \gamma_\g \vert \vphi \ra
    = \sum_{\mathfrak{h} :  \g \in \pinch(\h)}
    \frac{
        [\gamma_\g \vert \vphi_\h \ra 
    }{
        \la \check{\vphi}_\h \vert \vphi_\h \ra
    }
    \otimes 
    [\gamma_\h \vert \vphi \ra 
    \,, 
\end{align} 
we have that
\begin{align}
    \disc_{\bullet} [ \gamma_\g \vert \vphi \ra
    = \sum_{\mathfrak{h} : \g \in \pinch(\h)}
    \frac{
        \disc_{\bullet} W_1^\mathrm{MPL,\alpha}\Big[
            [\gamma_\g \vert \vphi_\h \ra 
        \Big]
    }{
        \la \check{\vphi}_\h \vert \vphi_\h \ra
    }
    [\gamma_\h \vert \vphi \ra 
    := \sum_{\h} M_{\g\h} [\gamma_\h \vert \vphi \ra 
    \,,
\end{align} 
where the last equality defines the \emph{monodromy matrix}  $M_{\g\h}$, 
and $\disc_{\bullet}$ denotes a discontinuity with respect to a given kinematic locus.
In appendix \ref{app:weight1}, we show that the entries of this matrix are only nonzero when one of the graphical conditions (on $\g$ and $\h$) are satisfied:
\be
    M_{\g\h}
    = \begin{cases}
        \sum_{\rsf \in \Rsf_\g} \beta_\rsf \ \disc_{\bullet}\left[ \log f_\rsf \right]
        & \h=\g \,,
        \\[1em]
        \disc_{\bullet}\left[ 
        \log \frac{
            f_{\includegraphics[align=c, scale=.35]{figs/f_u_right_dc.pdf}}
        }{
            f_{\includegraphics[align=c, scale=.35]{figs/f_uc_right_d.pdf}}
        }
        \right]
        & \mathfrak{g} = \includegraphics[align=c,scale=.4]{figs/gsg.pdf} 
            \quad\mathrm{and}\quad
            \h = \includegraphics[align=c,scale=.4]{figs/grg.pdf}
        \,,
        \\[1em]
        \disc_{\bullet}\left[ \log \frac{
            f_{\includegraphics[align=c, scale=.35]{figs/f_uc_left_d.pdf}}
        }{
            f_{\includegraphics[align=c, scale=.35]{figs/f_u_left_dc.pdf}}
        }
        \right]
        & \mathfrak{g} = \includegraphics[align=c,scale=.4]{figs/gsg.pdf}
            \quad\mathrm{and}\quad
            \h = \includegraphics[align=c,scale=.4]{figs/glg.pdf}
        \,.
    \end{cases}
\ee
Note that the discontinuity operator will only return a nonzero result if we compute a discontinuity with respect to one of the branch cuts that starts where the arguments of these logarithms either vanish or become infinite.

At any order in the $\alpha_\bullet$-expansion, the monodromy matrix $M_{\g\h}$ tells us how to build the symbol starting from the first entry (just as the connection matrix $A_{\g\h}$ instructs us how to build the symbol starting from the last entry). As an example, consider the two-site chain graph. 
It has the monodromy matrix 
\be \label{eq:two_site_monodromy_matrix}
    \mat{M}[\mathcal{Z}_{\twoChaing}] {=} \begin{pmatrix}
        (\alpha_1{+}\alpha_2) \disc_{\bullet}[
            \log f_{\includegraphics[scale=.4]{figs/2chain_Rs1}}
        ]
        &
            \disc_{\bullet}\left[
                \log \frac{
                    f_{\includegraphics[scale=.4]{figs/2chain_Rr2}}
                }{
                    f_{\includegraphics[scale=.4]{figs/2chain_Rr1}}
                }
            \right]
        & 
            \disc_{\bullet}\left[
                \log \frac{
                    f_{\includegraphics[scale=.4]{figs/2chain_Rl1}}
                }{
                    f_{\includegraphics[scale=.4]{figs/2chain_Rl2}}
                }
            \right]
        \\
        & \Large \substack{
            \alpha_1 \disc_{\bullet}[
                \log f_{\includegraphics[scale=.4]{figs/2chain_Rr2}}
            ]
            \\ \qquad
            {+} \alpha_2 \disc_{\bullet}[
                \log f_{\includegraphics[scale=.4]{figs/2chain_Rr1}}
            ]
        }
        &
        \\
        &
        & \Large \substack{
            \alpha_1 \disc_{\bullet}[
                \log f_{\includegraphics[scale=.4]{figs/2chain_Rl2}}
            ]
            \\ \qquad 
            {+} \alpha_2 \disc_{\bullet}[
                \log f_{\includegraphics[scale=.4]{figs/2chain_Rl1}}
            ]
        }
    \end{pmatrix}
\ee
In fact, we can make the discontinuity structure of the $\mathcal{Z}_{\twoChaing}$ sector more transparent by writing out the monodromy matrix associated with each logarithmic branch point  $f_\rsf$:
\be
    \frac{
        \mat{M}_{\includegraphics[scale=.4]{figs/2chain_Rs1}}[\mathcal{Z}_{\twoChaing}]
    }{2 \pi i} 
    &= 
    \begin{pmatrix}
        \alpha_1{+}\alpha_2
        & 0
        & 0
        \\
        & 0
        &
        \\
        &
        & 0
    \end{pmatrix}
    \,,
    \\
    \frac{
        \mat{M}_{\includegraphics[scale=.4]{figs/2chain_Rr1}}[\mathcal{Z}_{\twoChaing}]
    }{2 \pi i} 
    &=
    \begin{pmatrix}
        0
        & -1
        & 0
        \\
        & \alpha_2
        &
        \\
        &
        & 0
    \end{pmatrix}
    \,,
    &
    \frac{
        \mat{M}_{\includegraphics[scale=.4]{figs/2chain_Rr2}}[\mathcal{Z}_{\twoChaing}]
    }{2 \pi i} 
    &= 
    \begin{pmatrix}
        0
        & 1
        & 0
        \\
        & \alpha_1
        &
        \\
        &
        & 0
    \end{pmatrix}
    \,,
    \\
    \frac{
        \mat{M}_{\includegraphics[scale=.4]{figs/2chain_Rl1}}[\mathcal{Z}_{\twoChaing}]
    }{2 \pi i} 
    &= 
    \begin{pmatrix}
        0
        & 1
        & 0
        \\
        & \alpha_2
        &
        \\
        &
        & 0
    \end{pmatrix}
    \,,
    &
    \frac{
        \mat{M}_{\includegraphics[scale=.4]{figs/2chain_Rl2}}[\mathcal{Z}_{\twoChaing}]
    }{2 \pi i} 
    &= 
    \begin{pmatrix}
        0
        & -1
        & 0
        \\
        & \alpha_1
        &
        \\
        &
        & 0
    \end{pmatrix}
    \,.
\ee
These monodromy matrices predict the non-trivial sequential discontinuities that can appear in the FRW periods contained in $\mathcal{Z}_{\twoChaing}$. 
For example, 
\begin{align}
    \frac{
        \disc_{f_{\includegraphics[scale=.4]{figs/2chain_Rs1}}=0}
    }{
        2 \pi i
    }
    \begin{pmatrix}
        [\gamma_{\twoChaing} \vert \vphi_{\twoChaingg} \ra
        \\
        [\gamma_{\twoChaingg} \vert \vphi_{\twoChaingg} \ra
        \\
        \cancel{[\gamma_{\twoChainggg} \vert \vphi_{\twoChaingg} \ra}
    \end{pmatrix}
    = 
    \begin{pmatrix}
        \alpha_1{+}\alpha_2
        & 0
        & 0
        \\
        & 0
        & 
        \\
        &
        & 0
    \end{pmatrix}
    \begin{pmatrix}
        [\gamma_{\twoChaing} \vert \vphi_{\twoChaingg} \ra
        \\
        [\gamma_{\twoChaingg} \vert \vphi_{\twoChaingg} \ra
        \\
        0
    \end{pmatrix}
    \,,
    \\
    \frac{
        \disc_{f_{\includegraphics[scale=.4]{figs/2chain_Rr1}}=0}
    }{
        2 \pi i
    }
    \begin{pmatrix}
        [\gamma_{\twoChaing} \vert \vphi_{\twoChaingg} \ra
        \\
        [\gamma_{\twoChaingg} \vert \vphi_{\twoChaingg} \ra
        \\
        \cancel{[\gamma_{\twoChainggg} \vert \vphi_{\twoChaingg} \ra}
    \end{pmatrix}
    = 
    \begin{pmatrix}
        0
        & -1
        & 0
        \\
        & \alpha_1
        &
        \\
        &
        & 0
    \end{pmatrix}
    \begin{pmatrix}
        [\gamma_{\twoChaing} \vert \vphi_{\twoChaingg} \ra
        \\
        [\gamma_{\twoChaingg} \vert \vphi_{\twoChaingg} \ra
        \\
        0
    \end{pmatrix}
    \,,
    \\
    \frac{
        \disc_{f_{\includegraphics[scale=.4]{figs/2chain_Rr2}}=0}
    }{
        2 \pi i
    }
    \begin{pmatrix}
        [\gamma_{\twoChaing} \vert \vphi_{\twoChaingg} \ra
        \\
        [\gamma_{\twoChaingg} \vert \vphi_{\twoChaingg} \ra
        \\
        \cancel{[\gamma_{\twoChainggg} \vert \vphi_{\twoChaingg} \ra}
    \end{pmatrix}
    = 
    \begin{pmatrix}
        0
        & 1
        & 0
        \\
        & \alpha_2
        &
        \\
        &
        & 0
    \end{pmatrix}
    \begin{pmatrix}
        [\gamma_{\twoChaing} \vert \vphi_{\twoChaingg} \ra
        \\
        [\gamma_{\twoChaingg} \vert \vphi_{\twoChaingg} \ra
        \\
        0
    \end{pmatrix}
    \,,
\end{align}
where we have drawn a slash through periods that identically vanish. All discontinuities with repspect to other branch points also vanish. 
Since the FRW integrals we are considering do not give rise to algebraic discontinuities, each nonzero sequence of discontinuities yields a distinct word in the symbol.\footnote{For similar examples of this monodromy matrix construction in the context of flat-space Feynman integrals, see for instance~\cite{Bourjaily:2020wvq}.}

The monodromy matrix of FRW integrals can also be seen to encode nontrivial types of analytic structure. For instance, the sequences of nonzero discontinuities that appear in the integral ${[\gamma_{\twoChaing} \vert \vphi_{\twoChaingg} \ra}$ can be diagrammatically summarized as follows:
\be\begin{tikzcd}
    {[\gamma_{\twoChaing} \vert \vphi_{\twoChaingg} \ra} 
    \arrow[loop left, thick, "\substack{ \disc_{f_{\includegraphics[scale=.4]{figs/2chain_Rs1}}=0} \\ \disc_{f_{\includegraphics[scale=.4]{figs/2chain_Rr1}}=0} \\ \disc_{f_{\includegraphics[scale=.4]{figs/2chain_Rr2}}=0}}"]
    \arrow[rr,thick, "\substack{ \disc_{f_{\includegraphics[scale=.4]{figs/2chain_Rr1}}=0} \\ \disc_{f_{\includegraphics[scale=.4]{figs/2chain_Rr2}}=0}}"]
    &
    {}
    &
    {[\gamma_{\twoChaingg} \vert \vphi_{\twoChaingg} \ra} 
    \arrow[loop right,thick, "\substack{ \disc_{f_{\includegraphics[scale=.4]{figs/2chain_Rr1}}=0} \\ \disc_{f_{\includegraphics[scale=.4]{figs/2chain_Rr2}}=0}}"]
\end{tikzcd}
\,,
\ee
where each arrow represents a nonzero entry in the matrix~\eqref{eq:two_site_monodromy_matrix}. Namely, the symbol of the $\alpha_\bullet\sim0$ expansion of ${[\gamma_{\twoChaing} \vert \vphi_{\twoChaingg} \ra}$ can be read off of this diagram by considering all directed paths that begin at the ${[\gamma_{\twoChaing} \vert \vphi_{\twoChaingg} \ra}$ node; each path represents a sequence of nonzero discontinuities, corresponding to the sequence of arrows one has followed. Even though this integral has an especially simple structure, we can still read nontrivial information off of this diagram. In particular, we see that by computing discontinuities with respect to $f_{\includegraphics[scale=.4]{figs/2chain_Rr1}}$ or $f_{\includegraphics[scale=.4]{figs/2chain_Rr2}}$ we can move from the ${[\gamma_{\twoChaing} \vert \vphi_{\twoChaingg} \ra}$ to the $[\gamma_{\twoChaingg} \vert \vphi_{\twoChaingg} \ra$ node in the diagram; conversely, there are no discontinuities that take us back in the other direction. This phenomenon---that some discontinuities no longer exist once others have been computed---also appears in flat-space Feynman integrals, where it goes under the name of the \emph{hierarchical principle}~\cite{boyling1968homological,Landshoff1966,pham,Hannesdottir:2022xki,Hannesdottir:2024cnn}. In this case, the implications of the hierarchical principle are not drastic, since the same pair of discontinuities that take us between nodes in this diagram also appear via arrows that send each node back to itself; however, in more complicated examples, these hierarchical constraints hold important implications for the structure of the symbol.

\subsection{In-in correlators}
\label{sec:inin}

Let us finish this section by considering interesting simplifications that occur for in-in correlators. Following \cite{Arkani-Hamed:2025mce}, we have that the only difference between wavefunction coefficients and in-in correlators (in the class of theories we are considering) is the removal of the alternating sign in equation \eqref{eq:partialFractionedPsi}, namely
\be
    \psi_\G &= \int u_\G\; \vphi_\G 
    \,,
    &
    \vphi_\G &= \sum_{\E \subset \E_\G} (-1)^{|\E|} \sum_{\g \in \mathcal{A}^{\varnothing}_\E(\G)} 
    \vphi_{\g}
    \,,
    \\ & \qquad\downarrow &&\qquad\downarrow \\
    \la \G \ra_{\mathrm{in-in}} &= \int u_\G\; \vphi_{\la \G \ra}
    \,,
    &
    \vphi_{\la \G \ra} &= \sum_{\E \subset \E_\G}\sum_{\g \in \mathcal{A}^{\varnothing}_\E(\G)} 
    \vphi_{\g}
    \,.
\ee
At tree-level, the two formulas are of the same relative complexity. 
However, at loop-level, cancellations occur such that only the fully-oriented and fully-disconnected acyclic minors contribute:
\be \label{eq:phi_inin}
    \vphi_{\la \G \ra} &= \sum_{\g \in \mathrm{Vert}(\mathcal{Z}_{\G_{\includegraphics[scale=.3,align=c]{figs/2chain_gs.pdf}}})}
    \vphi_{\g}
    + \sum_{\g \in \mathrm{Vert}(\mathcal{Z}_{\G_{\includegraphics[scale=.3,align=c]{figs/2chain_gb.pdf}}})}
    \vphi_{\g}
    & 
    \mathrm{if}\; \mathrm{loops}(\G) > 0
    \,, 
\ee
where $\G_{ \includegraphics[scale=.5,align=c]{figs/2chain_gs.pdf} }$ and $\G_{ \includegraphics[scale=.5,align=c]{figs/2chain_gb.pdf} }$ denote the zonotope where all edges of $\G$ are either replaced by pinched edges \includegraphics[scale=.5,align=c]{figs/2chain_gs.pdf} or broken edges \includegraphics[scale=.5,align=c]{figs/2chain_gb.pdf}, and $\mathrm{Vert}(\mathcal{Z}_{\G_{\bullet}})$ denotes the acyclic minors that correspond to the vertices of the zonotope. 
For $\mathcal{Z}_{\G_{\includegraphics[scale=.3,align=c]{figs/2chain_gb.pdf}}}$, there is only a single vertex labeled by $\g = \G_{\includegraphics[scale=.3,align=c]{figs/2chain_gb.pdf}}$. 
On the other hand, the vertices of  $\mathcal{Z}_{\G_{\includegraphics[scale=.3,align=c]{figs/2chain_gs.pdf}}}$ correspond to the acyclic minors of $\G$ that involve only oriented edges. 

\begin{figure}
    \centering
    \includegraphics[scale=.8,align=c]{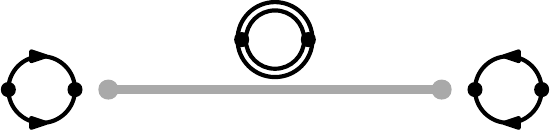}
    \hspace{4em}
    \includegraphics[scale=.8,align=c]{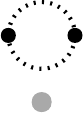}
    \caption{%
        The relevant zonotopes for the in-in correlator for the one-loop two-gon: 
        $\mathcal{Z}_{ \includegraphics[scale=.3,align=c]{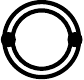} }$ (right) and 
        $\mathcal{Z}_{ \includegraphics[scale=.3,align=c]{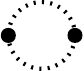} }$ (left). 
    }
    \label{fig:bubZono}
\end{figure}

Recently, it was observed that the cancellations at the integrand level \eqref{eq:phi_inin}, also lead to simplifications after loop integration \cite{Chowdhury:2023arc}; namely, there is an effective weight drop when passing from the de Sitter wavefunction coefficient to the in-in correlator.%
\footnote{
    Technically, both the wavefunction coefficient and the in-in correlator have the same weight. 
    However, in the case of the correlator, part of this weight is accounted for by trivial factors of $\pi$. 
    Therefore, the functions appearing in the correlator are genuinely simpler. 
}
While we have not given any attention to the integrals over spatial loop momentum in this work, our coaction applies straightforwardly to the loop integrand, since the function space of in-in correlator (loop) integrands is a subspace of the function space of wavefunction coefficient (loop) integrands. 

For example, consider the loop-level integrand of the one-loop bubble in-in correlator:
\be
    \left\la \includegraphics[scale=.5,align=c]{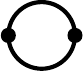} \right\ra
    &= \int_0^\infty u_{\includegraphics[scale=.3,align=c]{figs/bub.pdf}}\; \vphi_{\left\la \includegraphics[scale=.3,align=c]{figs/bub.pdf} \right\ra}
    \,,
    &
    \qquad
    \vphi_{\left\la \includegraphics[scale=.3,align=c]{figs/bub.pdf} \right\ra}
    &= \vphi_{\includegraphics[scale=.3,align=c]{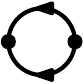}}
    + \vphi_{\includegraphics[scale=.3,align=c]{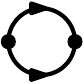}}
    + \vphi_{\includegraphics[scale=.3,align=c]{figs/bub_bb.pdf}}
    \,,
\ee
where we have summed over the vertices of the zonotopes depicted in figure \ref{fig:bubZono}.
In terms of our basis integrals, its coaction is simply
\be
    \Delta \left\la \includegraphics[scale=.5,align=c]{figs/bub.pdf} \right\ra
    &= \frac{\alpha_1\alpha_2}{\alpha_1+\alpha_2}
    \;\includegraphics[scale=.5,align=c]{figs/bub_pp.pdf}\;
    \otimes \left(  
        \includegraphics[scale=.5,align=c]{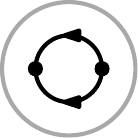}
        + 
        \includegraphics[scale=.5,align=c]{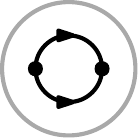}
    \right)
    \\&
    + \;\includegraphics[scale=.5,align=c]{figs/bub_ll.pdf}\; 
    \otimes 
    \;\includegraphics[scale=.5,align=c]{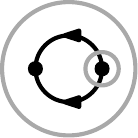}\; 
    + \;\includegraphics[scale=.5,align=c]{figs/bub_rr.pdf}\; 
    \otimes 
    \;\includegraphics[scale=.5,align=c]{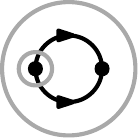}\; 
    + \;\includegraphics[scale=.5,align=c]{figs/bub_bb.pdf}\; 
    \otimes 
    \;\includegraphics[scale=.5,align=c]{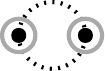}\; 
    . 
\ee
Thus, we see that all the same technology carries over. 

Although we will not explore the analytic structure of in-in correlators further in this work, is interesting to note that most terms in the discontinuity of such correlators are proportional to integrals that involve contours that enforce total energy conservation.
This means that for most terms in the discontinuity, we end up integrating over the (integrand of the) flat-space amplitude $\res_{B_\mathrm{tot}=0} [u_\G\; \vphi_\G] = (u_\G\vert_{B_\mathrm{tot}=0}) \; \mathcal{A}_\G(\mbf{X}+\mbf{x},\mbf{Y})\vert_{B_\mathrm{tot}=0}\;  \d^{|\V_\G|-1}\mbf{x}$.

\section{Conclusion}
\label{sec:conclusion}

In this work, we formulated a graphical coaction for the wavefunction of the universe, which applies to all FRW integrals that contribute to theories of conformally-coupled scalars in power-law FRW cosmologies. 
By leveraging the twisted (co)homology of the associated integral families, the coaction decomposes wavefunction coefficients in these theories into tensor products of simpler FRW periods, each of which can be represented by a decoration of the original Feynman diagram. 
This yields a unified, graphical language that encapsulates both the differential equations and the sequential discontinuities of cosmological correlators. 
Our construction is analogous to (and may provide hints for studying) the diagrammatic coaction for flat-space Feynman integrals, but involves a richer graphical language that incorporates the direction of time. 
 
As part of our analysis, we demonstrated that the coaction formula considered in \cite{Abreu:2019wzk, Brown:2019jng, Britto:2021prf} straightforwardly generalizes to the setting in which not all divisors are twisted (which is to say, to partial/relative twisted (co)homology). 
This is expected, because hypergeometric functions whose arguments take integer values are generally smooth, as has been seen in the context of flat-space Feynman integrals~\cite{Abreu:2014cla, Abreu:2017enx, Abreu:2021vhb}. 
However, even though hypergeometric functions themselves often have smooth limits, the relevant (co)homology changes discontinuously. 
Thus, our analysis in this paper has been restricted to the (co)homology that actually appears for FRW integrals, rather than more general (co)homologies that reduce to the FRW (co)homology in special limits. 

We have also developed a user-friendly web application that computes the graphical coaction of any graph: \webappAlt\!.  
A \texttt{Mathematica} notebook with the same functionality is also hosted on the repository: \github.

Several natural extensions of our investigation suggest themselves. 
It would be interesting to connect this construction more explicitly to the diagrammatic coaction for flat-space Feynman integrals~\cite{Abreu:2017enx,Abreu:2017mtm,Abreu:2021vhb} in the $\vep\to-1$ limit, and to explore whether the coaction structure persists for theories with spinning fields, non-power-law scale factors, or non-conformally coupled scalars. 
It would also be interesting to see how this can be applied to the recent understanding of in-in correlators as dispersive integrals \cite{Chowdhury:2026upp}. 
In particular, most of the structure of in-in correlators at loop-level should be determined by flat-space-like physics, as discussed at the end of section \ref{sec:inin}. 
Lastly, it would be worthwhile to connect our formalism to that of the cosmological cutting rules and cosmological optical theorem \cite{Melville:2021lst, Goodhew:2021oqg}. We leave these investigations to future work.

\acknowledgments
The authors would like to thank D.~Baumann, R.~Britto, C.~Dupont, E.~Gardi, C.~Larkin, A.~Lipstein, and O.~Schlotterer for stimulating discussions and comments on the draft. 
We also thank C.~Duhr, whose original suggestion led to this work.
AJM is supported by the Royal Society grant URF{\textbackslash}R1{\textbackslash}221233, and additionally acknowledges support from the European Research Council (ERC) under the European Union’s Horizon Europe research and innovation program grant agreement 101163627 (ERC Starting Grant
“AmpBoot”).
AP is supported by the European Union (ERC, UNIVERSE PLUS, 101118787). 
LR is supported by the Royal Society via a Newton International Fellowship.

\appendix

\section{Weight-one part of the basis integrals}
\label{app:weight1}

Diagonal pairings $[\g\vert\g\ra$ are proportional to (products of) power functions: $[\g\vert\g\ra = c_\g(\alpha_\bullet) \prod_{\rsf \in \Rsf_\g} f_{\rsf}^{\beta_\rsf}$ where $\beta_\rsf = \sum_{v\in\V_\rsf} \alpha_v$ are linear in the $\alpha_\bullet$.
The small $\alpha_i$ ($\beta_\rsf$) expansion of a power function is 
\be
    f_{\rsf}^{\beta_\rsf}
    = 1 + \beta_\rsf \log f_\rsf + \mathcal{O}(\beta_\rsf)
    \,.
\ee
We also know that the leading term of 
$[ \gamma_\mathfrak{h} \vert \vphi_{\mathfrak{g}} \ra$
in the small $\alpha_i$ limit is the self-intersection number 
$\la \check{\vphi}_\mathfrak{g} \vert \vphi_\mathfrak{g} \ra$: 
\begin{align} 
    [ \gamma_\mathfrak{g} \vert \vphi_{\mathfrak{g}} \ra = \la \check{\vphi}_\mathfrak{g} \vert \vphi_\mathfrak{g} \ra 
    \Big( 
        1 + \mathcal{O}(\alpha_\bullet) 
    \Big) 
    \,, 
\end{align}
and hence 
\be
   W_1^\mathrm{MPL,\alpha}\Big[[ \gamma_\mathfrak{g} \vert \vphi_{\mathfrak{g}} \ra \Big]
    &= \la \check{\vphi}_\mathfrak{g} \vert \vphi_\mathfrak{g} \ra
    \left( 
        \sum_{\rsf \in \Rsf_\g} \beta_\rsf \ \log f_\rsf 
    \right)
    \,.
\ee
These statements can be proved by repeated use of the basic identity 
\be \label{eq:powerFnId}
    \int_\mathbb{R} 
    \theta(-x) \theta(x+c) 
    x^{\alpha} (-x-c)^{\beta}
    \dlog\frac{x+c}{x}
    = \frac{
        \theta(c) (-c)^{\alpha+\beta}
        \Gamma(\alpha)
        \Gamma(\beta)
    }{
        \Gamma(\alpha+\beta)
    }
    \,,
\ee
whenever a diagonal period is not fully localized by the contour.

The off-diagonal pairings $[\g\vert\mathfrak{h}\ra$ are genuine hypergeometric functions. 
They contain a weight-one MPL term if 
and only if a region $\rsf^*$ of $\h$ splits to make $\g$. 
That is, $\rsf^* = \Rsf_\h \setminus (\Rsf_\g \cap \Rsf_\h)$ and $\Rsf_\g \setminus (\Rsf_\h \cap \Rsf_\g) = \{\rsf^*_\uparrow, \rsf^*_\downarrow\}$.%
\footnote{%
    If $\h$ is a $p$-pinch of $\g$, we can perform $|\V_\G|-p$ of the integrations via \eqref{eq:powerFnId} in $[\g\vert\mathfrak{h}\ra$ where each results in a power function. 
    The remaining $p$-integrations have $p$-many poles outside of the twisted loci. 
    To get the integral as a series in $\alpha_\bullet$, we expand the twist inside the integral and note that, after integration,  the lowest weight MPL has weight $p$.
}
Pictorially, we represent this as 
\be
    \mathfrak{h} &= \includegraphics[align=c,scale=.6]{figs/gsg.pdf} 
    \,,
    &\mathrm{and}&\quad
    \g &= \includegraphics[align=c,scale=.6]{figs/grg.pdf}
    \,,
    \\
    &
    &\mathrm{or}&
    & 
    \\
    \mathfrak{h} &= \includegraphics[align=c,scale=.6]{figs/gsg.pdf}
    \,,
    &\mathrm{and}&\quad
    \g &= \includegraphics[align=c,scale=.6]{figs/glg.pdf}
    \,.
\ee 
Here, the gray dashed lines represent possible connections to other regions via broken or oriented edges. 
Since all other regions are the same, we can focus solely on how $\rsf^*$ is broken into $\rsf^*_\uparrow$ and $ \rsf^*_\downarrow$.
Next, note that up to an overall sign, that we will reinstate later, 
\be
    {[} \gamma_{\h} \vert \vphi_{\g} \ra
    &\propto |\Csf_\g|
    \left(    
        \prod_{\rsf \in \Rsf_\h \setminus \rsf^*}
        \int_{\Delta_\rsf} 
        \left(\prod_{v\in\V_\rsf} x_v^{\alpha_v}\right)
        \Omega[\Delta_\rsf]
    \right) I^*
    \\
    I^*&= \pm
    \int_{\Delta_{\rsf^*}} 
    \res_{B_\mathrm{out}=0}
    \left[
        \left(\prod_{v\in\V_{\rsf^*}} x_v^{\alpha_v}\right)
        \dlog B_\mathrm{out}
        \wedge \dlog B_\mathrm{in} 
        \wedge \tilde{\Omega}[\Delta_{\rsf^*_\uparrow}]
        \wedge \tilde{\Omega}[\Delta_{\rsf^*_\downarrow}]
    \right]
    \\
    B_\mathrm{out} &= \includegraphics[align=c,scale=.5]{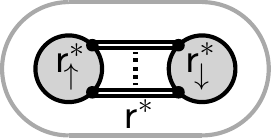}
    = \sum_{v\in \V_\uparrow} x_v + \sum_{v\in \V_\downarrow} x_v + S_\mathrm{out}
    \\
    B_\mathrm{in} &= \includegraphics[align=c,scale=.5]{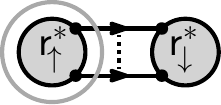}
    = \sum_{v\in \V_\uparrow} x_v + f_{\rsf^*_\uparrow}
    \qquad \mathrm{or} \qquad
    B_\mathrm{in} = \includegraphics[align=c,scale=.5]{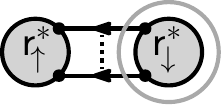}
    = \sum_{v\in \V_\downarrow} x_v + f_{\rsf^*_\downarrow}
    \,,
\ee
where $\V_\uparrow = \V_{\rsf^*_\uparrow}$ and $\V_\downarrow = \V_{\rsf^*_\downarrow}$.
To evaluate $I^*$, choose a $v' \in \V_\uparrow$ and a $v'' \in \V_\downarrow$. 
Then, set 
\be
    x_{v'} &= {-} \!\!\!\! \sum_{v \in \V_\uparrow \setminus v'} \!\! x_v {-} f_{\rsf^*_\uparrow} {+} z
    \,,
    \quad
    x_{v''} &= {-} \!\!\!\! \sum_{v \in \V_\downarrow \setminus v''} \!\! x_v 
    {-} f_{\rsf^*}
    {+} f_{\rsf^*_\uparrow}
    {+} y
    {-} z
    \quad\text{if}\quad
    \g = \includegraphics[align=c,scale=.5]{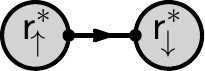}
    \\
    x_{v'} &= {-} \!\!\!\! \sum_{v \in \V_\downarrow \setminus v'} \!\! x_v {-} f_{\rsf^*_\downarrow} {+} z
    \,,
    \quad
    x_{v''} &= {-} \!\!\!\! \sum_{v \in \V_\uparrow \setminus v''} \!\! x_v 
    {-} f_{\rsf^*}
    {+} f_{\rsf^*_\downarrow}
    {+} y
    {-} z
    \quad\text{if}\quad
    \g = \includegraphics[align=c,scale=.5]{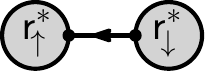}
\ee
After this change of variables, $B_\mathrm{out} = y$, $B_\mathrm{in}=z$ and 
\be
    I^* &\propto
    \int_{\Delta_{\rsf^*}} 
    \res_{y=0}
    \left[
        \left(\prod_{v\in\V_{\rsf^*}} x_v^{\alpha_v}\right)
        \dlog\, y
        \wedge \dlog\, z 
        \wedge \tilde{\Omega}[\Delta_{\rsf^*_\uparrow}]
        \wedge \tilde{\Omega}[\Delta_{\rsf^*_\downarrow}]
    \right]
\ee
Specializing to the case where $B_\mathrm{in} = \includegraphics[align=c,scale=.5]{figs/Bup.pdf}$, 
\be
    I^*
    &\propto \int_{-\infty}^\infty \dlog\, z\;  I^*_\uparrow I^*_\downarrow
\ee
where
\begin{align}
    I^*_\uparrow
    &\propto \int_{-\infty}^\infty 
    \d^{|\V_\uparrow|-1}x_\uparrow\;\;
    \theta\left(
        \sum_{v \in \V_\uparrow \setminus v'} \!\! x_v {+} f_{\rsf^*_\uparrow} {-} z
    \right)
    \left(
        {-} \!\!\!\! \sum_{v \in \V_\uparrow \setminus v'} \!\! x_v {-} f_{\rsf^*_\uparrow} {+} z
    \right)^{\alpha_{v'}-1}
    \left(
        \prod_{v\in \V_\uparrow \setminus v'}
        \theta(-x_v) x_v^{\alpha_v-1}
    \right) 
    \,,
    \nn\\
    &\propto \theta\big(
        f_{\rsf^*_\uparrow} 
        {-} z
    \big) 
    \big(
        {-} f_{\rsf^*_\uparrow} 
        {+} z
    \big)^{\alpha_\uparrow}
    \frac{
        \prod_{v\in\V_\uparrow}
        \Gamma(\alpha_v)
    }{
       \Gamma(\alpha_\uparrow)
    }
    \,,
\end{align}
and
\begin{align}
    I^*_\downarrow
    &\propto \int_{-\infty}^\infty 
    \d^{|\V_\downarrow|-1}x_\downarrow\;\;
    \theta\left(
        \sum_{v \in \V_\downarrow \setminus v''} \!\! x_v 
        {+} f_{\rsf^*}
        {-} f_{\rsf^*_\uparrow}
        {+} z
    \right)
    \nn\\&\hspace{4em}\times
    \left(
        {-} \!\!\!\! \sum_{v \in \V_\downarrow \setminus v''} \!\! x_v 
        {-} f_{\rsf^*}
        {+} f_{\rsf^*_\uparrow}
        {-} z
    \right)^{\alpha_{v''}-1}
    \left(
        \prod_{v\in \V_\downarrow \setminus v''}
        \!\!\!\!
        \theta(-x_v) x_v^{\alpha_v-1}
    \right) 
    \,,
    \nn\\
    &\propto \theta\big(
        f_{\rsf^*}
        {-} f_{\rsf^*_\uparrow}
        {+} z
    \big) 
    \big(
        {-} f_{\rsf^*}
        {+} f_{\rsf^*_\uparrow}
        {-} z
    \big)^{\alpha_\downarrow}
    \frac{
        \prod_{v\in\V_\downarrow}
        \Gamma(\alpha_v)
    }{
       \Gamma(\alpha_\downarrow)
    }
    \,,
\end{align}
where $\alpha_{\uparrow (\downarrow)} = \sum_{v\in\V_{\uparrow(\downarrow)}} \alpha_v$. 
Putting things together, 
\be
    W_1^{\mathrm{MPL,\alpha}}\Big[ {[} \gamma_{\h} \vert \vphi_{\g} \ra \Big]
    &\propto |\Csf_\g|
    \left(
    \prod_{\rsf \in \Rsf_\h \setminus \rsf^*}
    \left(    
        f_\rsf^{\alpha_\rsf}
        \frac{
            \prod_{v\in\V_\rsf}
            \Gamma(\alpha_v)
        }{
           \Gamma(\alpha_\rsf)
        }
    \right) 
    \right)
    \frac{
        \prod_{v\in\V_\uparrow}
        \Gamma(\alpha_v)
    }{
       \Gamma(\alpha_\uparrow)
    }
    \frac{
        \prod_{v\in\V_\downarrow}
        \Gamma(\alpha_v)
    }{
       \Gamma(\alpha_\downarrow)
    }
    \\&\times
    \int_{-\infty}^\infty \dlog\, z\;  
    \theta\big(
        f_{\rsf^*_\uparrow} 
        {-} z
    \big) 
    \big(
        {-} f_{\rsf^*_\uparrow} 
        {+} z
    \big)^{\alpha_\uparrow}
    \theta\big(
        f_{\rsf^*}
        {-} f_{\rsf^*_\uparrow}
        {+} z
    \big) 
    \big(
        {-} f_{\rsf^*}
        {+} f_{\rsf^*_\uparrow}
        {-} z
    \big)^{\alpha_\downarrow}
\ee
Keeping only the leading term in the $\alpha_\bullet$, yields
\be
    W_1^{\mathrm{MPL,\alpha}}\Big[ {[} \gamma_{\h} \vert \vphi_{\g} \ra \Big]
    &\propto C_{\g\g}
    \int_{-\infty}^\infty \dlog\, z\;  
    \theta\big(
        f_{\rsf^*_\uparrow} 
        {-} z
    \big) 
    \theta\big(
        f_{\rsf^*}
        {-} f_{\rsf^*_\uparrow}
        {+} z
    \big) 
    \propto C_{\g\g} \log \frac{
        f_{\rsf^*_\uparrow}
    }{
        f_{\rsf^*_\downarrow}
    }
\ee
where 
\be
    |\Csf_\g|
    \prod_{\rsf \in \Rsf_\g}
    \left(    
        \frac{
            \prod_{v\in\V_\rsf}
            \Gamma(\alpha_v)
        }{
           \Gamma(\alpha_\rsf)
        }
    \right) 
    = 
    C_{\g\g} \left( 1 + \mathcal{O}\left(\alpha_\bullet\right) \right)
    \,.
\ee
Performing the same computation for $B_\mathrm{in} = \includegraphics[align=c,scale=.5]{figs/Bdown.pdf}$ and reinserting the overall sign by carefully keeping track of the wedge product manipulations yields 
\be\label{eq:weight1_hg_app}
    W_1^\mathrm{MPL,\alpha}\Big[
        [ \gamma_\mathfrak{h} \vert \vphi_{\mathfrak{g}} \ra
    \Big]
    &= \begin{cases}
        C_{\g\g}
        \log \frac{
            f_{\includegraphics[align=c, scale=.35]{figs/f_u_right_dc.pdf}}
        }{
            f_{\includegraphics[align=c, scale=.35]{figs/f_uc_right_d.pdf}}
        }
        & \mathfrak{h} = \includegraphics[align=c,scale=.4]{figs/gsg.pdf} 
            \quad\mathrm{and}\quad
            \g = \includegraphics[align=c,scale=.4]{figs/grg.pdf}
        \,,
        \\[1em]
        C_{\g\g} 
        \log \frac{
            f_{\includegraphics[align=c, scale=.35]{figs/f_uc_left_d.pdf}}
        }{
            f_{\includegraphics[align=c, scale=.35]{figs/f_u_left_dc.pdf}}
        }
        & \mathfrak{h} = \includegraphics[align=c,scale=.4]{figs/gsg.pdf}
            \quad\mathrm{and}\quad
            \g = \includegraphics[align=c,scale=.4]{figs/glg.pdf}
        \,.
    \end{cases}
\ee

\bibliographystyle{JHEP}
\bibliography{reference}

@article{Chowdhury:2023arc,
    author = "Chowdhury, Chandramouli and Lipstein, Arthur and Mei, Jiajie and Sachs, Ivo and Vanhove, Pierre",
    title = "{The subtle simplicity of cosmological correlators}",
    eprint = "2312.13803",
    archivePrefix = "arXiv",
    primaryClass = "hep-th",
    reportNumber = "LMU-ASC 37/23, IPhT-t23/119",
    doi = "10.1007/JHEP03(2025)007",
    journal = "JHEP",
    volume = "03",
    pages = "007",
    year = "2025"
}

@article{brown2011multiple,
  title={Multiple elliptic polylogarithms},
  author={Brown, Francis and Levin, Andrey},
  journal={arXiv preprint arXiv:1110.6917},
  year={2011}
}

@article{Goodhew:2021oqg,
    author = "Goodhew, Harry and Jazayeri, Sadra and Lee, Mang Hei Gordon and Pajer, Enrico",
    title = "{Cutting cosmological correlators}",
    eprint = "2104.06587",
    archivePrefix = "arXiv",
    primaryClass = "hep-th",
    doi = "10.1088/1475-7516/2021/08/003",
    journal = "JCAP",
    volume = "08",
    pages = "003",
    year = "2021"
}

@article{Landshoff1966,
    Author = {Landshoff, P. V. and Olive, D. I. and Polkinghorne, J. C.},
    Da = {1966/05/01},
    Doi = {10.1007/BF02752870},
    Id = {Landshoff1966},
    Isbn = {1826-9869},
    Journal = {Il Nuovo Cimento A (1971-1996)},
    Number = {2},
    Pages = {444--453},
    Title = {The hierarchical principle in perturbation theory},
    Ty = {JOUR},
    Url = {https://doi.org/10.1007/BF02752870},
    Volume = {43},
    Year = {1966}}

@article{Caron-Huot:2016owq,
    author = "Caron-Huot, Simon and Dixon, Lance J. and McLeod, Andrew and von Hippel, Matt",
    title = "{Bootstrapping a Five-Loop Amplitude Using Steinmann Relations}",
    eprint = "1609.00669",
    archivePrefix = "arXiv",
    primaryClass = "hep-th",
    reportNumber = "SLAC-PUB-16811",
    doi = "10.1103/PhysRevLett.117.241601",
    journal = "Phys. Rev. Lett.",
    volume = "117",
    number = "24",
    pages = "241601",
    year = "2016"
}

@article{boyling1968homological,
  title={{A homological approach to parametric Feynman integrals}},
  author={Boyling, JB},
  journal={Il Nuovo Cimento A (1965-1970)},
  volume={53},
  number={2},
  pages={351--375},
  year={1968},
  publisher={Springer}
}

@article{Melville:2021lst,
    author = "Melville, Scott and Pajer, Enrico",
    title = "{Cosmological Cutting Rules}",
    eprint = "2103.09832",
    archivePrefix = "arXiv",
    primaryClass = "hep-th",
    doi = "10.1007/JHEP05(2021)249",
    journal = "JHEP",
    volume = "05",
    pages = "249",
    year = "2021"
}

@article{Chowdhury:2026upp,
    author = "Chowdhury, Chandramouli and Jazayeri, Sadra and Lipstein, Arthur and Marshall, Joe and Mei, Jiajie and Sachs, Ivo",
    title = "{Cosmological Correlator Discontinuities from Scattering Amplitudes}",
    eprint = "2602.03841",
    archivePrefix = "arXiv",
    primaryClass = "hep-th",
    month = "2",
    year = "2026"
}

@article{Arkani-Hamed:2025mce,
    author = "Arkani-Hamed, Nima and Glew, Ross and Vaz{\~a}o, Francisco",
    title = "{Correlators are simpler than wavefunctions}",
    eprint = "2512.23795",
    archivePrefix = "arXiv",
    primaryClass = "hep-th",
    month = "12",
    year = "2025"
}

@article{Caron-Huot:2019bsq,
    author = "Caron-Huot, Simon and Dixon, Lance J. and Dulat, Falko and Von Hippel, Matt and McLeod, Andrew J. and Papathanasiou, Georgios",
    title = "{The Cosmic Galois Group and Extended Steinmann Relations for Planar $\mathcal{N} = 4$ SYM Amplitudes}",
    eprint = "1906.07116",
    archivePrefix = "arXiv",
    primaryClass = "hep-th",
    reportNumber = "DESY 19-062, DESY-19-062, HU-EP-19/05, SLAC--PUB--17414",
    doi = "10.1007/JHEP09(2019)061",
    journal = "JHEP",
    volume = "09",
    pages = "061",
    year = "2019"
}

@article{Hannesdottir:2025bss,
    author = "Hannesdottir, Holmfridur S. and Lippstreu, Luke and McLeod, Andrew J. and Polackova, Maria",
    title = "{Steinmann Violation and Minimal Cuts}",
    eprint = "2512.24010",
    archivePrefix = "arXiv",
    primaryClass = "hep-th",
    month = "12",
    year = "2025"
}

@article{Vergu:2023rqz,
    author = "Vergu, C.",
    title = "{Cutkosky representation and direct integration}",
    eprint = "2311.16069",
    archivePrefix = "arXiv",
    primaryClass = "hep-th",
    doi = "10.1007/JHEP05(2024)302",
    journal = "JHEP",
    volume = "05",
    pages = "302",
    year = "2024"
}

@article{Cutkosky,
    author = {Cutkosky, R. E.},
    title = {Singularities and Discontinuities of Feynman Amplitudes},
    journal = {Journal of Mathematical Physics},
    volume = {1},
    number = {5},
    pages = {429-433},
    year = {1960},
    month = {09},
    issn = {0022-2488},
    doi = {10.1063/1.1703676},
    url = {https://doi.org/10.1063/1.1703676}
}

@article{Benincasa:2020aoj,
    author = "Benincasa, Paolo and McLeod, Andrew J. and Vergu, Cristian",
    title = "{Steinmann Relations and the Wavefunction of the Universe}",
    eprint = "2009.03047",
    archivePrefix = "arXiv",
    primaryClass = "hep-th",
    doi = "10.1103/PhysRevD.102.125004",
    journal = "Phys. Rev. D",
    volume = "102",
    pages = "125004",
    year = "2020"
}

@article{Hannesdottir:2022xki,
    author = "Hannesdottir, Holmfridur S. and McLeod, Andrew J. and Schwartz, Matthew D. and Vergu, Cristian",
    title = "{Constraints on sequential discontinuities from the geometry of on-shell spaces}",
    eprint = "2211.07633",
    archivePrefix = "arXiv",
    primaryClass = "hep-th",
    reportNumber = "CERN-TH-2022-189",
    doi = "10.1007/JHEP07(2023)236",
    journal = "JHEP",
    volume = "07",
    pages = "236",
    year = "2023"
}

@article{Bourjaily:2020wvq,
    author = "Bourjaily, Jacob L. and Hannesdottir, Holmfridur and McLeod, Andrew J. and Schwartz, Matthew D. and Vergu, Cristian",
    title = "{Sequential Discontinuities of Feynman Integrals and the Monodromy Group}",
    eprint = "2007.13747",
    archivePrefix = "arXiv",
    primaryClass = "hep-th",
    doi = "10.1007/JHEP01(2021)205",
    journal = "JHEP",
    volume = "01",
    pages = "205",
    year = "2021"
}

@article{Arkani-Hamed:2017tmz,
    author = "Arkani-Hamed, Nima and Bai, Yuntao and Lam, Thomas",
    title = "{Positive Geometries and Canonical Forms}",
    eprint = "1703.04541",
    archivePrefix = "arXiv",
    primaryClass = "hep-th",
    doi = "10.1007/JHEP11(2017)039",
    journal = "JHEP",
    volume = "11",
    pages = "039",
    year = "2017"
}

@article{Caron-Huot:2020bkp,
    author = {Caron-Huot, Simon and Dixon, Lance J. and Drummond, James M. and Dulat, Falko and Foster, Jack and G{\"u}rdo{\u{g}}an, {\"O}mer and von Hippel, Matt and McLeod, Andrew J. and Papathanasiou, Georgios},
    title = "{The Steinmann Cluster Bootstrap for $N$ = 4 Super Yang-Mills Amplitudes}",
    eprint = "2005.06735",
    archivePrefix = "arXiv",
    primaryClass = "hep-th",
    reportNumber = "DESY-20-087",
    doi = "10.22323/1.376.0003",
    journal = "PoS",
    volume = "CORFU2019",
    pages = "003",
    year = "2020"
}

@article{matsumoto,
      title={Relative twisted homology and cohomology groups associated with Lauricella's $F_D$}, 
      author={Keiji Matsumoto},
      year={2019},
      eprint={1804.00366},
      archivePrefix={arXiv},
      primaryClass={math.AG},
      url={https://arxiv.org/abs/1804.00366}, 
}

@article{Britto:2021prf,
    author = "Britto, Ruth and Mizera, Sebastian and Rodriguez, Carlos and Schlotterer, Oliver",
    title = "{Coaction and double-copy properties of configuration-space integrals at genus zero}",
    eprint = "2102.06206",
    archivePrefix = "arXiv",
    primaryClass = "hep-th",
    reportNumber = "TCDMATH 21-06, IPhT-t21/030, UUITP-08/21",
    doi = "10.1007/JHEP05(2021)053",
    journal = "JHEP",
    volume = "05",
    pages = "053",
    year = "2021"
}

@article{Glew:2025arc,
    author = "Glew, Ross",
    title = "{Correlators from Amplitubes}",
    eprint = "2507.07199",
    archivePrefix = "arXiv",
    primaryClass = "hep-th",
    month = "7",
    year = "2025"
}

@article{Baumann:2025qjx,
    author = "Baumann, Daniel and Goodhew, Harry and Joyce, Austin and Lee, Hayden and Pimentel, Guilherme L. and Westerdijk, Tom",
    title = "{Geometry of Kinematic Flow}",
    eprint = "2504.14890",
    archivePrefix = "arXiv",
    primaryClass = "hep-th",
    month = "4",
    year = "2025"
}

@inproceedings{Duhr:2014woa,
    author = "Duhr, Claude",
    title = "{Mathematical aspects of scattering amplitudes}",
    booktitle = "{Theoretical Advanced Study Institute in Elementary Particle Physics}: {Journeys Through the Precision Frontier: Amplitudes for Colliders}",
    eprint = "1411.7538",
    archivePrefix = "arXiv",
    primaryClass = "hep-ph",
    reportNumber = "CP3-14-70",
    doi = "10.1142/9789814678766_0010",
    pages = "419--476",
    year = "2015"
}

@article{Caron-Huot:2021xqj,
    author = "Caron-Huot, Simon and Pokraka, Andrzej",
    title = "{Duals of Feynman integrals. Part I. Differential equations}",
    eprint = "2104.06898",
    archivePrefix = "arXiv",
    primaryClass = "hep-th",
    doi = "10.1007/JHEP12(2021)045",
    journal = "JHEP",
    volume = "12",
    pages = "045",
    year = "2021"
}

@article{Caron-Huot:2021iev,
    author = "Caron-Huot, Simon and Pokraka, Andrzej",
    title = "{Duals of Feynman Integrals. Part II. Generalized unitarity}",
    eprint = "2112.00055",
    archivePrefix = "arXiv",
    primaryClass = "hep-th",
    doi = "10.1007/JHEP04(2022)078",
    journal = "JHEP",
    volume = "04",
    pages = "078",
    year = "2022"
}

@book{pham2011singularities,
  title={Singularities of integrals: Homology, hyperfunctions and microlocal analysis},
  author={Pham, Fr{\'e}d{\'e}ric},
  year={2011},
  publisher={Springer Science \& Business Media}
}

@article{Glew:2025ypb,
    author = "Glew, Ross and Pokraka, Andrzej",
    title = "{Kinematic flow from the flow of cuts}",
    eprint = "2508.11568",
    archivePrefix = "arXiv",
    primaryClass = "hep-th",
    month = "8",
    year = "2025"
}

@article{Hwa:1967csk,
    author = "Hwa, R. C. and Teplitz, V. L.",
    title = "{Homology and Feynman integrals}",
    doi = "10.1016/0375-9474(67)90109-1",
    journal = "Nucl. Phys. A",
    volume = "98",
    number = "3",
    year = "1967"
}

@article{Brown:2025jjg,
    author = "Brown, Francis and Dupont, Cl{\'e}ment",
    title = "{Positive Geometries and Canonical Forms via Mixed Hodge Theory}",
    eprint = "2501.03202",
    archivePrefix = "arXiv",
    primaryClass = "math.AG",
    doi = "10.1007/s00220-025-05399-y",
    journal = "Commun. Math. Phys.",
    volume = "406",
    number = "11",
    pages = "267",
    year = "2025"
}

@article{Gardi:2022wro,
    author = "Gardi, Einan and Abreu, Samuel and Britto, Ruth and Duhr, Claude and Matthew, James",
    title = "{The diagrammatic coaction}",
    eprint = "2207.07843",
    archivePrefix = "arXiv",
    primaryClass = "hep-th",
    reportNumber = "CERN-TH-2022-122, BONN-TH-2022-18",
    doi = "10.22323/1.416.0015",
    journal = "PoS",
    volume = "LL2022",
    pages = "015",
    year = "2022"
}

@article{FBThesis,
      author         = "Brown, Francis C.S.",
      title          = "{Multiple zeta values and periods of moduli spaces $\overline{\mathfrak{M}}_{0,n}(\mathbb{R})$}",
      journal        = "Annales Sci.Ecole Norm.Sup.",
      volume         = "42",
      pages          = "371",
      year           = "2009",
      eprint         = "math/0606419",
      archivePrefix  = "arXiv",
      primaryClass   = "math.AG",
      SLACcitation   = "%%CITATION = ARXIV:MATH/0606419;%%",
}

@article{Gonch3,
      author         = "Goncharov, A. B.",
      title          = "{Multiple polylogarithms and mixed Tate motives}",
      year           = "2001",
      eprint         = "math/0103059",
      archivePrefix  = "arXiv",
      primaryClass   = "math.AG",
      SLACcitation   = "%%CITATION = MATH/0103059;%%",
}

@ARTICLE{2015arXiv151206410B,
   author = "Brown, Francis",
    title = "{Notes on Motivic Periods}",
archivePrefix = "arXiv",
   eprint = {1512.06410},
 primaryClass = "math.NT",
     year = 2015,
    month = dec,
   adsurl = {http://adsabs.harvard.edu/abs/2015arXiv151206410B}
}

@ARTICLE{Brown1102.1312,
  author = {Brown, Francis},
  title = {Mixed {T}ate motives over {$\mathbb{Z}$}},
  journal = {Ann. of Math. (2)},
  year = {2012},
  volume = {175},
  pages = {949--976},
  number = {2},
  coden = {ANMAAH},
  doi = {10.4007/annals.2012.175.2.10},
  fjournal = {Annals of Mathematics. Second Series},
  issn = {0003-486X},
  mrclass = {19Exx (11M32 14Fxx)},
  mrnumber = {2993755},
  eprint         = "1102.1312",
  archivePrefix  = "arXiv",
      primaryClass   = "math.AG",
}

@article{Abreu:2021vhb,
    author = "Abreu, Samuel and Britto, Ruth and Duhr, Claude and Gardi, Einan and Matthew, James",
    title = "{The diagrammatic coaction beyond one loop}",
    eprint = "2106.01280",
    archivePrefix = "arXiv",
    primaryClass = "hep-th",
    doi = "10.1007/JHEP10(2021)131",
    journal = "JHEP",
    volume = "10",
    pages = "131",
    year = "2021"
}

@article{Abreu:2019xep,
    author = "Abreu, Samuel and Britto, Ruth and Duhr, Claude and Gardi, Einan and Matthew, James",
    title = "{Generalized hypergeometric functions and intersection theory for Feynman integrals}",
    eprint = "1912.03205",
    archivePrefix = "arXiv",
    primaryClass = "hep-th",
    reportNumber = "CERN-TH-2019-219, CP3-19-58",
    doi = "10.22323/1.375.0067",
    journal = "PoS",
    number = "RACOR2019",
    pages = "067",
    year = "2019"
}

@book{ELOP,
	Author = {R. J. Eden and P. V. Landshoff and D. I. Olive and J. C. Polkinghorne},
	Publisher = {Cambridge University Press},
	Title = {The Analytic S-Matrix},
	Year = {1966}}

@article{Goncharov:1998kja,
      author         = "Goncharov, Alexander B.",
      title          = "{Multiple polylogarithms, cyclotomy and modular
                        complexes}",
      journal        = "Math. Res. Lett.",
      volume         = "5",
      year           = "1998",
      pages          = "497-516",
      doi            = "10.4310/MRL.1998.v5.n4.a7",
      eprint         = "1105.2076",
      archivePrefix  = "arXiv",
      primaryClass   = "math.AG",
      SLACcitation   = "%%CITATION = ARXIV:1105.2076;%%"
}

@article{G91b,
title = "Geometry of Configurations, Polylogarithms, and Motivic Cohomology ",
journal = "Adv. Math.",
volume = "114",
number = "2",
pages = "197--318",
year = "1995",
note = "",
issn = "0001-8708",
doi = "10.1006/aima.1995.1045",
url = "http://www.sciencedirect.com/science/article/pii/S0001870885710456",
author = "Goncharov, A. B."
}

@article{Chen,
ajournal = "Bull. Amer. Math. Soc.",
author = "Chen, Kuo-Tsai",
journal = "Bull. Amer. Math. Soc.",
number = "5",
pages = "831--879",
publisher = "American Mathematical Society",
title = "Iterated path integrals",
url = "http://projecteuclid.org/euclid.bams/1183539443",
volume = "83",
year = "1977"
}

@article{Remiddi:1999ew,
      author         = "Remiddi, E. and Vermaseren, J.A.M.",
      title          = "{Harmonic polylogarithms}",
      journal        = "Int.J.Mod.Phys.",
      volume         = "A15",
      pages          = "725-754",
      doi            = "10.1142/S0217751X00000367",
      year           = "2000",
      eprint         = "hep-ph/9905237",
      archivePrefix  = "arXiv",
      primaryClass   = "hep-ph",
      reportNumber   = "NIKHEF-99-005, TTP-99-08",
      SLACcitation   = "%%CITATION = HEP-PH/9905237;%%",
}

@article{Borwein:1999js,
      author         = "Borwein, Jonathan M. and Bradley, David M. and
                        Broadhurst, David J. and Lisonek, Petr",
      title          = "{Special values of multiple polylogarithms}",
      journal        = "Trans. Am. Math. Soc.",
      volume         = "353",
      year           = "2001",
      pages          = "907-941",
      doi            = "10.1090/S0002-9947-00-02616-7",
      eprint         = "math/9910045",
      archivePrefix  = "arXiv",
      primaryClass   = "math-ca",
      SLACcitation   = "%%CITATION = MATH/9910045;%%"
}

@article{Moch:2001zr,
      author         = "Moch, Sven and Uwer, Peter and Weinzierl, Stefan",
      title          = "{Nested sums, expansion of transcendental functions and
                        multiscale multiloop integrals}",
      journal        = "J.Math.Phys.",
      volume         = "43",
      pages          = "3363-3386",
      doi            = "10.1063/1.1471366",
      year           = "2002",
      eprint         = "hep-ph/0110083",
      archivePrefix  = "arXiv",
      primaryClass   = "hep-ph",
      reportNumber   = "TTP-01-25, UPRF-2001-21",
      SLACcitation   = "%%CITATION = HEP-PH/0110083;%%",
}

@article{Brown:2011ik,
      author         = "Brown, Francis",
      title          = "{On the decomposition of motivic multiple zeta values}",
      journal = "{Adv. Studies in Pure Math.}",
      year = "2012",
      volume = "63",
      pages = "31--58",
      eprint         = "1102.1310",
      archivePrefix  = "arXiv",
      primaryClass   = "math.NT",
      SLACcitation   = "%%CITATION = ARXIV:1102.1310;%%",
}

@article{Goncharov:2010jf,
      author         = "Goncharov, Alexander B. and Spradlin, Marcus and Vergu,
                        C. and Volovich, Anastasia",
      title          = "{Classical Polylogarithms for Amplitudes and Wilson
                        Loops}",
      journal        = "Phys.Rev.Lett.",
      volume         = "105",
      pages          = "151605",
      doi            = "10.1103/PhysRevLett.105.151605",
      year           = "2010",
      eprint         = "1006.5703",
      archivePrefix  = "arXiv",
      primaryClass   = "hep-th",
      reportNumber   = "BROWN-HET-1602",
      SLACcitation   = "%%CITATION = ARXIV:1006.5703;%%",
}

@article{Dixon:2022rse,
    author = "Dixon, Lance J. and Gurdogan, Omer and McLeod, Andrew J. and Wilhelm, Matthias",
    title = "{Bootstrapping a stress-tensor form factor through eight loops}",
    eprint = "2204.11901",
    archivePrefix = "arXiv",
    primaryClass = "hep-th",
    reportNumber = "SLAC-PUB-17653, CERN-TH-2022-039",
    doi = "10.1007/JHEP07(2022)153",
    journal = "JHEP",
    volume = "07",
    pages = "153",
    year = "2022"
}

@article{Caron-Huot:2018dsv,
    author = "Caron-Huot, Simon and Dixon, Lance J. and von Hippel, Matt and McLeod, Andrew J. and Papathanasiou, Georgios",
    title = "{The Double Pentaladder Integral to All Orders}",
    eprint = "1806.01361",
    archivePrefix = "arXiv",
    primaryClass = "hep-th",
    reportNumber = "SLAC-PUB-17228, DESY-18-041",
    doi = "10.1007/JHEP07(2018)170",
    journal = "JHEP",
    volume = "07",
    pages = "170",
    year = "2018"
}

@article{DelDuca:2010zg,
    author = "Del Duca, Vittorio and Duhr, Claude and Smirnov, Vladimir A.",
    title = "{The Two-Loop Hexagon Wilson Loop in N = 4 SYM}",
    eprint = "1003.1702",
    archivePrefix = "arXiv",
    primaryClass = "hep-th",
    reportNumber = "IPPP-10-21, DCPT-10-42, CERN-PH-TH-2010-059",
    doi = "10.1007/JHEP05(2010)084",
    journal = "JHEP",
    volume = "05",
    pages = "084",
    year = "2010"
}

@article{Abreu:2017enx,
    author = "Abreu, Samuel and Britto, Ruth and Duhr, Claude and Gardi, Einan",
    title = "{Algebraic Structure of Cut Feynman Integrals and the Diagrammatic Coaction}",
    eprint = "1703.05064",
    archivePrefix = "arXiv",
    primaryClass = "hep-th",
    reportNumber = "CERN-TH-2017-056",
    doi = "10.1103/PhysRevLett.119.051601",
    journal = "Phys. Rev. Lett.",
    volume = "119",
    number = "5",
    pages = "051601",
    year = "2017"
}

@article{Broadhurst:1998rz,
    author = "Broadhurst, David J.",
    title = "{Massive three - loop Feynman diagrams reducible to SC* primitives of algebras of the sixth root of unity}",
    eprint = "hep-th/9803091",
    archivePrefix = "arXiv",
    reportNumber = "OUT-4102-72",
    doi = "10.1007/s100529900935",
    journal = "Eur. Phys. J. C",
    volume = "8",
    pages = "311--333",
    year = "1999"
}

@article{Broedel:2018iwv,
    author = "Broedel, Johannes and Duhr, Claude and Dulat, Falko and Penante, Brenda and Tancredi, Lorenzo",
    title = "{Elliptic symbol calculus: from elliptic polylogarithms to iterated integrals of Eisenstein series}",
    eprint = "1803.10256",
    archivePrefix = "arXiv",
    primaryClass = "hep-th",
    reportNumber = "CP3-18-24, CERN-TH-2018-057, HU-Mathematik-2018-03, HU-EP-18/09, SLAC-PUB-17240",
    doi = "10.1007/JHEP08(2018)014",
    journal = "JHEP",
    volume = "08",
    pages = "014",
    year = "2018"
}

@article{DelDuca:2009au,
    author = "Del Duca, Vittorio and Duhr, Claude and Smirnov, Vladimir A.",
    title = "{An Analytic Result for the Two-Loop Hexagon Wilson Loop in N = 4 SYM}",
    eprint = "0911.5332",
    archivePrefix = "arXiv",
    primaryClass = "hep-ph",
    reportNumber = "IPPP-09-92, DCPT-09-184",
    doi = "10.1007/JHEP03(2010)099",
    journal = "JHEP",
    volume = "03",
    pages = "099",
    year = "2010"
}

@article{Duhr:2012fh,
      author         = "Duhr, Claude",
      title          = "{Hopf algebras, coproducts and symbols: an application to
                        Higgs boson amplitudes}",
      journal        = "JHEP",
      volume         = "1208",
      pages          = "043",
      doi            = "10.1007/JHEP08(2012)043",
      year           = "2012",
      eprint         = "1203.0454",
      archivePrefix  = "arXiv",
      primaryClass   = "hep-ph",
      SLACcitation   = "%%CITATION = ARXIV:1203.0454;%%",
}

@article{Gonch2,
ajournal = "Duke Math. J.",
author = "Goncharov, A. B.",
doi = "10.1215/S0012-7094-04-12822-2",
journal = "Duke Math. J.",
month = "06",
number = "2",
pages = "209--284",
publisher = "Duke University Press",
title = "Galois symmetries of fundamental groupoids and noncommutative geometry",
volume = "128",
year = "2005",
eprint         = "math/0208144",
      archivePrefix  = "arXiv",
      primaryClass   = "math.AG",
}

@article{Abreu:2019wzk,
    author = "Abreu, Samuel and Britto, Ruth and Duhr, Claude and Gardi, Einan and Matthew, James",
    title = "{From positive geometries to a coaction on hypergeometric functions}",
    eprint = "1910.08358",
    archivePrefix = "arXiv",
    primaryClass = "hep-th",
    reportNumber = "CERN-TH-2019-168",
    doi = "10.1007/JHEP02(2020)122",
    journal = "JHEP",
    volume = "02",
    pages = "122",
    year = "2020"
}

@article{Abreu:2018sat,
    author = "Abreu, Samuel and Britto, Ruth and Duhr, Claude and Gardi, Einan",
    editor = "Hoang, Andre and Schneider, Carsten",
    title = "{The diagrammatic coaction and the algebraic structure of cut Feynman integrals}",
    eprint = "1803.05894",
    archivePrefix = "arXiv",
    primaryClass = "hep-th",
    reportNumber = "CERN-TH-2018-002, CP3-18-01, Edinburgh 2018/1, FR-PHENO-2018-001, EDINBURGH-2018-1",
    doi = "10.22323/1.290.0002",
    journal = "PoS",
    volume = "RADCOR2017",
    pages = "002",
    year = "2018"
}

@article{Abreu:2017mtm,
    author = "Abreu, Samuel and Britto, Ruth and Duhr, Claude and Gardi, Einan",
    title = "{Diagrammatic Hopf algebra of cut Feynman integrals: the one-loop case}",
    eprint = "1704.07931",
    archivePrefix = "arXiv",
    primaryClass = "hep-th",
    reportNumber = "CERN-TH-2017-092, CP3-17-11, EDINBURGH-2017-09, FR-PHENO-2017-010, TCDMATH-17-09",
    doi = "10.1007/JHEP12(2017)090",
    journal = "JHEP",
    volume = "12",
    pages = "090",
    year = "2017"
}

@article{Abreu:2017ptx,
    author = "Abreu, Samuel and Britto, Ruth and Duhr, Claude and Gardi, Einan",
    title = "{Cuts from residues: the one-loop case}",
    eprint = "1702.03163",
    archivePrefix = "arXiv",
    primaryClass = "hep-th",
    reportNumber = "CERN-TH-2017-033, CP3-17-05, Edinburgh-2017-05, FR-PHENO-2017-001",
    doi = "10.1007/JHEP06(2017)114",
    journal = "JHEP",
    volume = "06",
    pages = "114",
    year = "2017"
}

@article{Bogner:2007mn,
    author = "Bogner, Christian and Weinzierl, Stefan",
    title = "{Periods and Feynman integrals}",
    eprint = "0711.4863",
    archivePrefix = "arXiv",
    primaryClass = "hep-th",
    reportNumber = "MZ-TH-07-19",
    doi = "10.1063/1.3106041",
    journal = "J. Math. Phys.",
    volume = "50",
    pages = "042302",
    year = "2009"
}

@inproceedings{Bourjaily:2022bwx,
    author = "Bourjaily, Jacob L. and others",
    title = "{Functions Beyond Multiple Polylogarithms for Precision Collider Physics}",
    booktitle = "{Snowmass 2021}",
    eprint = "2203.07088",
    archivePrefix = "arXiv",
    primaryClass = "hep-ph",
    reportNumber = "BONN-TH-2022-05, UUITP-11/22, CERN-TH-2022-029, TUM-HEP-1391/22,
  HU-EP-22/08, MITP-22-022",
    month = "3",
    year = "2022"
}

@article{Landau:1959fi,
	author = "Landau, L.D.",
	title = "{On analytic properties of vertex parts in quantum field theory}",
	doi = "10.1016/B978-0-08-010586-4.50103-6",
	journal = "Nucl. Phys.",
	volume = "13",
	number = "1",
	pages = "181--192",
	year = "1960"
}

@article{Steinmann,
	Author = {Steinmann, O},
	Doi = {10.3929/ethz-a-000107369},
	Journal = {Helv. Physica Acta},
	Pages = {257},
	Title = {{\"Uber den Zusammenhang Zwischen den Wightmanfunktionen und der Retardierten Kommutatoren}},
	Volume = {33},
	Year = {1960}}

@article{Steinmann2,
	Author = {Steinmann, O},
	Journal = {Helv. Physica Acta},
	Note = {\url{www.e-periodica.ch/cntmng?pid=hpa-001:1960:33::1079}},
	Pages = {347},
	Title = {{Wightman-Funktionen und Retardierte Kommutatoren. II}},
	Volume = {33},
	Year = {1960}}

@article{pham,
     author = {Pham, Fr\'ed\'eric},
     title = {Singularit\'es des processus de diffusion multiple},
     journal = {Annales de l'I.H.P. Physique th\'eorique},
     pages = {89--204},
     publisher = {Gauthier-Villars},
     volume = {6},
     number = {2},
     year = {1967},
     zbl = {0154.46102},
     mrnumber = {214341},
     language = {fr},
     url = {www.numdam.org/item/AIHPA_1967__6_2_89_0/}
}

@book{Arkani-Hamed:2012zlh,
    author = "Arkani-Hamed, Nima and Bourjaily, Jacob L. and Cachazo, Freddy and Goncharov, Alexander B. and Postnikov, Alexander and Trnka, Jaroslav",
    title = "{Grassmannian Geometry of Scattering Amplitudes}",
    eprint = "1212.5605",
    archivePrefix = "arXiv",
    primaryClass = "hep-th",
    reportNumber = "PUPT-2435",
    doi = "10.1017/CBO9781316091548",
    isbn = "978-1-107-08658-6, 978-1-316-57296-2",
    publisher = "Cambridge University Press",
    month = "4",
    year = "2016"
}

@article{Schlotterer:2012ny,
      author         = "Schlotterer, O. and Stieberger, S.",
      title          = "{Motivic Multiple Zeta Values and Superstring
                        Amplitudes}",
      journal        = "J. Phys.",
      volume         = "A46",
      year           = "2013",
      pages          = "475401",
      doi            = "10.1088/1751-8113/46/47/475401",
      eprint         = "1205.1516",
      archivePrefix  = "arXiv",
      primaryClass   = "hep-th",
      reportNumber   = "AEI-2012-039, MPP-2012-859",
      SLACcitation   = "%%CITATION = ARXIV:1205.1516;%%"
}

@article{Brown:2015fyf,
      author         = "Brown, Francis",
      title          = "{Feynman amplitudes, coaction principle, and cosmic
                        Galois group}",
      journal        = "Commun. Num. Theor. Phys.",
      volume         = "11",
      year           = "2017",
      pages          = "453-556",
      doi            = "10.4310/CNTP.2017.v11.n3.a1",
      eprint         = "1512.06409",
      archivePrefix  = "arXiv",
      primaryClass   = "math-ph",
      SLACcitation   = "%%CITATION = ARXIV:1512.06409;%%"
}

@article{Panzer:2016snt,
      author         = "Panzer, Erik and Schnetz, Oliver",
      title          = "{The Galois coaction on $\phi^4$ periods}",
      journal        = "Commun. Num. Theor. Phys.",
      volume         = "11",
      year           = "2017",
      pages          = "657-705",
      doi            = "10.4310/CNTP.2017.v11.n3.a3",
      eprint         = "1603.04289",
      archivePrefix  = "arXiv",
      primaryClass   = "hep-th",
      SLACcitation   = "%%CITATION = ARXIV:1603.04289;%%"
}

@article{Schnetz:2017bko,
      author         = "Schnetz, Oliver",
      title          = "{The Galois coaction on the electron anomalous magnetic
                        moment}",
      journal        = "Commun. Num. Theor. Phys.",
      volume         = "12",
      year           = "2018",
      pages          = "335-354",
      doi            = "10.4310/CNTP.2018.v12.n2.a4",
      eprint         = "1711.05118",
      archivePrefix  = "arXiv",
      primaryClass   = "math-ph",
      SLACcitation   = "%%CITATION = ARXIV:1711.05118;%%"
}

@article{Duhr:2011zq,
    author = "Duhr, Claude and Gangl, Herbert and Rhodes, John R.",
    title = "{From polygons and symbols to polylogarithmic functions}",
    eprint = "1110.0458",
    archivePrefix = "arXiv",
    primaryClass = "math-ph",
    reportNumber = "IPPP-11-56, DCPT-11-112",
    doi = "10.1007/JHEP10(2012)075",
    journal = "JHEP",
    volume = "10",
    pages = "075",
    year = "2012"
}

@article{He:2025tyv,
    author = "He, Song and Jiang, Xuhang and Li, Xiang and Liu, Jiahao",
    title = "{Heptagon Symbols at Five Loops and All-Loop Sequences}",
    eprint = "2511.09669",
    archivePrefix = "arXiv",
    primaryClass = "hep-th",
    month = "11",
    year = "2025"
}

@article{Gurdogan:2020ppd,
    author = {G\"urdo\u{g}an, \"Omer},
    title = "{From integrability to the Galois coaction on Feynman periods}",
    eprint = "2011.04781",
    archivePrefix = "arXiv",
    primaryClass = "hep-th",
    doi = "10.1103/PhysRevD.103.L081703",
    journal = "Phys. Rev. D",
    volume = "103",
    number = "8",
    pages = "L081703",
    year = "2021"
}

@article{Dixon:2021tdw,
    author = "Dixon, Lance J. and Gurdogan, Omer and McLeod, Andrew J. and Wilhelm, Matthias",
    title = "{Folding Amplitudes into Form Factors: An Antipodal Duality}",
    eprint = "2112.06243",
    archivePrefix = "arXiv",
    primaryClass = "hep-th",
    reportNumber = "SLAC-PUB-17637",
    doi = "10.1103/PhysRevLett.128.111602",
    journal = "Phys. Rev. Lett.",
    volume = "128",
    number = "11",
    pages = "111602",
    year = "2022"
}

@article{Dixon:2022xqh,
    author = {Dixon, Lance J. and G\"urdo\u{g}an, \"Omer and Liu, Yu-Ting and McLeod, Andrew J. and Wilhelm, Matthias},
    title = "{Antipodal Self-Duality for a Four-Particle Form Factor}",
    eprint = "2212.02410",
    archivePrefix = "arXiv",
    primaryClass = "hep-th",
    reportNumber = "CERN-TH-2022-190, SLAC-PUB-17711",
    doi = "10.1103/PhysRevLett.130.111601",
    journal = "Phys. Rev. Lett.",
    volume = "130",
    number = "11",
    pages = "111601",
    year = "2023"
}

@article{Hannesdottir:2024cnn,
    author = "Hannesdottir, Holmfridur S. and Lippstreu, Luke and McLeod, Andrew J. and Polackova, Maria",
    title = "{Minimal Cuts and Genealogical Constraints on Feynman Integrals}",
    eprint = "2406.05943",
    archivePrefix = "arXiv",
    primaryClass = "hep-th",
    month = "6",
    year = "2024"
}

@article{Basso:2024hlx,
    author = "Basso, Benjamin and Dixon, Lance J. and Tumanov, Alexander G.",
    title = "{The three-point form factor of Tr {\ensuremath{\phi}}$^{3}$ to six loops}",
    eprint = "2410.22402",
    archivePrefix = "arXiv",
    primaryClass = "hep-th",
    doi = "10.1007/JHEP02(2025)034",
    journal = "JHEP",
    volume = "02",
    pages = "034",
    year = "2025"
}

@article{Frost:2023stm,
    author = "Frost, Hadleigh and Hidding, Martijn and Kamlesh, Deepak and Rodriguez, Carlos and Schlotterer, Oliver and Verbeek, Bram",
    title = "{Motivic coaction and single-valued map of polylogarithms from zeta generators}",
    eprint = "2312.00697",
    archivePrefix = "arXiv",
    primaryClass = "hep-th",
    doi = "10.1088/1751-8121/ad5edf",
    journal = "J. Phys. A",
    volume = "57",
    number = "31",
    pages = "31LT01",
    year = "2024"
}

@article{Frost:2025lre,
    author = "Frost, Hadleigh and Hidding, Martijn and Kamlesh, Deepak and Rodriguez, Carlos and Schlotterer, Oliver and Verbeek, Bram",
    title = "{Deriving motivic coactions and single-valued maps at genus zero from zeta generators}",
    eprint = "2503.02096",
    archivePrefix = "arXiv",
    primaryClass = "hep-th",
    reportNumber = "MPIM-Bonn-2025, UUITP{\textendash}09/25",
    month = "3",
    year = "2025"
}

@article{Kleinschmidt:2025dtk,
    author = "Kleinschmidt, Axel and Porkert, Franziska and Schlotterer, Oliver",
    title = "{Towards Motivic Coactions at Genus One from Zeta Generators}",
    eprint = "2508.02800",
    archivePrefix = "arXiv",
    primaryClass = "hep-th",
    month = "8",
    year = "2025"
}

@article{Brown:2019jng,
    author = "Brown, Francis and Dupont, Cl{\'e}ment",
    title = "{Lauricella Hypergeometric Functions, Unipotent Fundamental Groups of the Punctured Riemann Sphere, and Their Motivic Coactions}",
    eprint = "1907.06603",
    archivePrefix = "arXiv",
    primaryClass = "math.AG",
    doi = "10.1017/nmj.2022.27",
    journal = "Nagoya Math. J.",
    volume = "249",
    pages = "148--220",
    year = "2023"
}

@article{Dixon:2023kop,
    author = "Dixon, Lance J. and Liu, Yu-Ting",
    title = "{An eight loop amplitude via antipodal duality}",
    eprint = "2308.08199",
    archivePrefix = "arXiv",
    primaryClass = "hep-th",
    reportNumber = "SLAC-PUB-17693",
    doi = "10.1007/JHEP09(2023)098",
    journal = "JHEP",
    volume = "09",
    pages = "098",
    year = "2023"
}

@article{Abreu:2014cla,
    author = "Abreu, Samuel and Britto, Ruth and Duhr, Claude and Gardi, Einan",
    title = "{From multiple unitarity cuts to the coproduct of Feynman integrals}",
    eprint = "1401.3546",
    archivePrefix = "arXiv",
    primaryClass = "hep-th",
    doi = "10.1007/JHEP10(2014)125",
    journal = "JHEP",
    volume = "10",
    pages = "125",
    year = "2014"
}

@article{De:2023xue,
    author = "De, Shounak and Pokraka, Andrzej",
    title = "{Cosmology meets cohomology}",
    eprint = "2308.03753",
    archivePrefix = "arXiv",
    primaryClass = "hep-th",
    doi = "10.1007/JHEP03(2024)156",
    journal = "JHEP",
    volume = "03",
    pages = "156",
    year = "2024"
}

@article{Arkani-Hamed:2023kig,
    author = "Arkani-Hamed, Nima and Baumann, Daniel and Hillman, Aaron and Joyce, Austin and Lee, Hayden and Pimentel, Guilherme L.",
    title = "{Differential Equations for Cosmological Correlators}",
    eprint = "2312.05303",
    archivePrefix = "arXiv",
    primaryClass = "hep-th",
    month = "12",
    year = "2023"
}

@article{tapuskovic2023cosmic,
      title="{The cosmic Galois group, the sunrise Feynman integral, and the relative completion of $\Gamma_1(6)$}", 
      author={Matija Tapušković},
      year={2023},
      eprint={2303.17534},
      archivePrefix={arXiv},
      primaryClass={math.AG}
}

@article{He:2024olr,
    author = "He, Song and Jiang, Xuhang and Liu, Jiahao and Yang, Qinglin and Zhang, Yao-Qi",
    title = "{Differential equations and recursive solutions for cosmological amplitudes}",
    eprint = "2407.17715",
    archivePrefix = "arXiv",
    primaryClass = "hep-th",
    month = "7",
    year = "2024"
}

@article{Arkani-Hamed:2017fdk,
    author = "Arkani-Hamed, Nima and Benincasa, Paolo and Postnikov, Alexander",
    title = "{Cosmological Polytopes and the Wavefunction of the Universe}",
    eprint = "1709.02813",
    archivePrefix = "arXiv",
    primaryClass = "hep-th",
    month = "9",
    year = "2017"
}
\end{document}